\documentclass[nofootinbib]{revtex4}
\usepackage{graphicx}
\usepackage{dcolumn}
\usepackage{amsmath,amssymb,epsfig,latexsym}
\usepackage{paralist}
\usepackage{comment} 
\usepackage{siunitx}
\usepackage{graphicx}
\usepackage{multirow}
\usepackage{color,soul}
\usepackage{suffix}
\usepackage{mathtools}

\usepackage[breaklinks=True,colorlinks=True, linkcolor=magenta,citecolor=blue,debug=True]{hyperref}

\usepackage{wasysym}

\renewcommand{\vec}[1]{\boldsymbol{\mathrm{#1}}}

\newcommand{\etainst}{\eta_{\mathrm{inst}}}   
\newcommand{\duty}{\xi}                       
\newcommand{\etasys}{\eta_{\mathrm{sys}}}     

\def\beqn{\begin{eqnarray*}}
\def\eeqn{\end{eqnarray*}}

\newcommand{\be}{\begin{equation}}
\newcommand{\ee}{\end{equation}}
\newcommand{\ba}{\begin{eqnarray}}
\newcommand{\ea}{\end{eqnarray}}

\begin{document}

\title{Direct High-Resolution Imaging of Earth-Like Exoplanets}

\author{Slava G. Turyshev$^1$}

\affiliation{\vskip 3pt$^1$Jet Propulsion Laboratory, California Institute of Technology,\\
4800 Oak Grove Drive, Pasadena, CA 91109-0899, USA}

\date{\today}

\begin{abstract}

We have surveyed all conventional methods proposed or conceivable for obtaining resolved images of an Earth-like exoplanet.  Generating a $10\times10$ pixel map of a $1\,R_\oplus$ world at 10 pc demands $\approx0.85\,\mu$as angular resolution and photon-collection sufficient for $\mathrm{SNR}\gtrsim5$ per “micro-pixel.”  We derived diffraction-limit and photon-budget requirements for: (1) large single-aperture space telescopes with internal coronagraphs; (2) external-starshades; (3) space-based interferometry (nulling and non-nulling); (4) ground-based ELTs with extreme AO; (5) pupil-densified “hypertelescopes”; (6) indirect reconstructions (rotational light-curve inversion, eclipse mapping, intensity interferometry); and (7) diffraction-occultation by Solar System bodies.  Even though these approaches serve their primary goals---exoplanet discovery and initial coarse characterization---each remains orders of magnitude away from delivering a spatially resolved image. In every case, technology readiness falls short, and fundamental barriers leave them 2–5 orders of magnitude below the angular resolution and photon-budget thresholds needed to map an Earth analog even on decadal timescales. Ultimately, an in-situ platform delivered to $\lesssim0.1\,$AU of the target could, in principle, overcome both diffraction and photon-starvation limits---but such a mission far exceeds current propulsion, autonomy, and communications capabilities. By contrast, the Solar Gravitational Lens---providing on-axis gain of $\sim 10^{10}$ and inherent $\mu$as-scale focusing once an imaging spacecraft reaches heliocentric distances beyond $\gtrsim 550~\mathrm{AU}$---appears uniquely capable of simultaneously meeting both the resolution and photon-budget requirements. If the several mission-specific risks (coronal calibration, focal-plane scanning, deconvolution) can be retired, the SGL could enable true, resolved surface images and spatially resolved spectroscopy of Earth-like exoplanets in our stellar neighborhood. 

\end{abstract}


\maketitle
\tableofcontents

\section{Introduction}
\label{sec:Intro}

One day soon, high–precision spectroscopy may reveal the unmistakable fingerprints of organic molecules---methane, phosphine, or complex hydrocarbons---in the atmosphere of a nearby, temperate exoplanet \cite{Seager2016,Swain2008,Madhusudhan2019,Venot2020,Greaves2021}. Such a discovery would mark the watershed moment when we move from asking “Are we alone?” to confronting the far richer question: “What does that world actually look like?” A disk–integrated spectrum can tell us about composition and climate, but it cannot show us continents, cloud patterns, or potential biospheres. To answer these questions, we must push beyond spectral detection into spatially resolved imaging---transforming a single pixel into a true surface map.

Achieving microarcsecond–scale resolution on a target tens of light–years away is orders of magnitude more demanding than any astronomical program to date. Even the largest monolithic telescopes, starshade–assisted observatories, and multi–kilometer interferometric arrays fall short by factors of $10^3$–$10^6$ in angular resolution or photon–collection capability \cite{TuryshevToth2019,Feinberg2019,Gaudi2020}. Meanwhile, indirect methods---light–curve inversion \cite{CowanAgol2008}, eclipse mapping \cite{Majeau2012}, intensity interferometry \cite{HanburyBrown1956}---sacrifice spatial detail for practicality. Proposals for in situ flybys at relativistic speeds confront seemingly insurmountable challenges in navigation, shielding, and data return \cite{Lubin2016}. For each we derive the ultimate limits set by physics and engineering---diffraction floors, contrast stability, wavefront tolerances, formation–flying precision, integration times, and operational overheads---and show that today’s concepts cannot bridge the gap. While most of the available methods are effective for detection and preliminary study of exoplanets, they uniformly fall short of the diffraction-limit and photon-collection thresholds required for genuine surface imaging. No single approach, as currently conceived, can deliver a $10\times10$-pixel surface map of an Earth analog at 10\,pc within a human lifetime.

Our aim is not merely to enumerate shortcomings of various techniques but to chart the narrow corridor between aspiration and feasibility.  By quantifying each approach’s diffraction floor, photon-budget limit, contrast stability, wavefront error budget, formation–flying precision, integration times, and operational overheads, we highlight the key technological leaps---continental-scale optical phasing, picometer null control, autonomous deep–space navigation, and high–throughput beam combiners---required to close the gap. Only by understanding precisely why every current concept falls short can we chart the roadmap from a single spectral detection of organics to true, multi-pixel surface maps of biosignature exoplanets.

This work is organized as follows:
Section~\ref{sec:exo-imaging} addresses technical requirements for resolved imaging of exoplanets. 
Section~\ref{sec:monolithic_telescopes} analyzes diffraction and photon limits for large monolithic space telescopes with internal coronagraphs.
Section~\ref{sec:starshades} evaluates external starshade concepts, including suppression performance, formation-flying tolerances, and retargeting overhead.
Section~\ref{sec:space_interferometry} addresses space-based interferometry (nulling and non-nulling modes), with emphasis on required baselines, path-length stability, thermal background, and integration times.
Section~\ref{sec:elt_exao} considers ground-based extremely large telescopes with extreme adaptive optics, assessing achievable angular resolution, contrast limits, and atmospheric instability constraints.
 Section~\ref{sec:hypertelescopes} examines pupil-densified “hypertelescope” architectures, including subaperture phasing, beam-combiner throughput, and formation-flying complexity.
 Section~\ref{sec:indirect_methods} reviews indirect reconstruction techniques---rotational light-curve inversion, eclipse mapping, intensity interferometry, and Solar System occultation---quantifying spatial-resolution limits and photon requirements.
 Section~\ref{sec:in_situ} explores in situ and near-in situ imaging platforms, deriving the distance, aperture, and flyby-duration trade-offs needed for multi-pixel mapping. 
Section~\ref{sec:other_direct_techniques} surveys alternative direct methods---ground-based sparse-aperture masking and speckle imaging, radio VLBI, and lunar occultations---and demonstrates why none achieve the microarcsecond resolution or sensitivity needed.
 Section~\ref{sec:indirect_feature_retrieval} evaluates spectro-photometric phase-curve inversion and phase-resolved polarimetry, showing that photon-noise limits constrain retrievals to only a few broad surface zones.
 Section~\ref{sec:summary_techniques} summarizes quantitative barriers for all remote approaches and outlines minimal requirements for any potential future mission concept aiming at resolved imaging of Earth-like exoplanets.
Finally, Section~\ref{sec:conclusions} presents our overall conclusions and reflects on the need for fundamentally new strategies---such as the Solar Gravitational Lens (SGL)---and revolutionary advances it brings to exoplanet imaging and spectroscopy.

\section{Fundamentals of Exoplanet Imaging} 
\label{sec:exo-imaging}

In what follows we adopt a common benchmark: a $10 \times 10$ map of a $1\,R_\oplus$ planet at $10~\mathrm{pc}$,
observed near full phase in the visible band and with ${\rm SNR} \gtrsim 5$ per surface element. This choice is not
meant to represent the minimum scientifically interesting goal, but rather to provide a common, ambitious target
against which disparate architectures can be compared on equal footing. Some known terrestrial exoplanets in or
near the habitable zones of nearby stars are indeed found at somewhat smaller distances than $10~\mathrm{pc}$; however, the relevant integration times scale as $t_{\rm pix} \propto d^{2}$ for fixed hardware in the photon-limited regime (and between $t_{\rm pix} \propto d^{2}$ and $t_{\rm pix} \propto d^{4}$ in the background-limited case), so reducing $d$ by a factor of a few shortens the exposure times by only factors of a few. Many of the concepts discussed here can deliver valuable, but much coarser, information (e.g., hemispheric contrasts or a handful of longitudinal zones); our conclusions in this paper apply specifically to the multi-pixel imaging benchmark defined in this Section.

\subsection{Angular Resolution Requirement}

Producing a pixel-by-pixel surface map of an Earth-sized exoplanet requires both angular resolution on the order of microarcseconds (\(\mu\mathrm{as}\)) and photon-collection performance far beyond the capabilities of any existing or planned facility. To quantify the diffraction-limit and photon-budget constraints that apply to all currently available or envisaged imaging concepts, we consider an Earth-like planet located at distance \(d = 10\,\mathrm{pc}\). Its angular diameter is
\begin{equation}
  \label{eq:theta_planet}
  \theta_{p}
  = \frac{2\,R_{p}}{d}
  \approx 8.53\,\Big( \frac{R_{p}}{R_\oplus}\Big)\Big( \frac{10\,{\rm pc}}{d}\Big)\,\mu\mathrm{as}.
\end{equation}
Dividing this disc into $N=10$ linear pixels across or in a \((10\times10)\) grid requires each pixel to subtend\footnote{We take $N$ to be the number of resolution elements across the planetary diameter and $N_{\rm pix}\equiv N^2$ for the total number of “micro-pixels” (e.g., $N=10 \Rightarrow N_{\rm pix}=100$).
}
\begin{equation}
  \label{eq:delta_theta_required}
  \Delta\theta 
  = \frac{\theta_{p}}{N}
  \approx 0.853\,\Big( \frac{R_{p}}{R_\oplus}\Big)\Big( \frac{10\,{\rm pc}}{d}\Big)\Big( \frac{10}{N}\Big)\, \mu\mathrm{as}.
\end{equation}
In the visible band (\(\lambda \simeq 550\,\mathrm{nm}\)), a diffraction-limited circular aperture of diameter \(D\) has Rayleigh resolution
\begin{equation}
  \label{eq:rayleigh_limit}
  \Delta\theta_{\mathrm{Rayleigh}} 
  = 1.22\,\frac{\lambda}{D}.
\end{equation}
To achieve \(\Delta\theta = 0.853~\mu{\rm as}\) at \(\lambda=550~\mathrm{nm}\), the Rayleigh criterion (\ref{eq:rayleigh_limit}) requires 
\begin{equation}
  \label{eq:required_aperture}
  D
  = 1.22\,\frac{\lambda}{\Delta \theta} 
  \simeq 162.31\,
  \Big(\frac{\lambda}{550\,{\rm nm}} \Big)
  \Big(\frac{d}{10\,{\rm pc}} \Big)
  \Big(\frac{R_\oplus}{R_p}\Big)
  \Big(\frac{N}{10}\Big)\,\mathrm{km}.
\end{equation}
Similarly, a two-element interferometer must satisfy
\begin{equation}
  \label{eq:required_baseline}
  B
  = \frac{\lambda}{\Delta\theta}
  \approx 133.04\,\Big(\frac{\lambda}{550\,{\rm nm}} \Big)
  \Big(\frac{d}{10\,{\rm pc}} \Big)
  \Big(\frac{R_\oplus}{R_p}\Big)
  \Big(\frac{N}{10}\Big)\,\mathrm{km}.
\end{equation}
Eqs.~\eqref{eq:required_aperture} and \eqref{eq:required_baseline} represent \emph{minimum} diffraction-limit thresholds. In practice, finite spectral bandwidths, wavefront-control tolerances, detector sampling requirements, and high-contrast starlight suppression further increase the necessary aperture or baseline by at least an order of magnitude. No existing or planned space telescope or interferometer concept approaches these scales.

\subsection{Photon-Budget Constraint}
\label{sec:flux_to_snr}

Even though achieving the required resolution would demand a diffraction-limited aperture of $D\gtrsim162\,$km (or baseline $B\gtrsim133\,$km), no such facility is remotely feasible.  To illustrate the photon-collection challenge on the other hand, we adopt a notional collecting area of $A=100\,$m$^2$ (i.e., telescope with circular aperture of $d_A\simeq 11.3\,{\rm m}$)---comparable to a next-generation Extremely Large Telescopes (ELT)---to compute integration times below.\footnote{We emphasize that this area does not resolve the planet; it only benchmarks photon-gathering.} 

Below we analyze how the \emph{signal flux} and various \emph{noise sources} combine to determine the overall SNR in exoplanet imaging. By translating the resulting SNR into integration time, we establish the conditions needed to evaluate feasibility of various imaging concepts.

\subsubsection{Signal Photon Flux from an Earth-like Exoplanet}

The \emph{detected} signal flux (\(\dot{N}_{\mathrm{sig}}\)) received from an exoplanet determined by its size,  distance, atmospheric albedo as well as on the brightness of its host star and the planetary orbit around it. Taking our Earth as a stand-in target, we see that for a full-phase Earth analog at distance $d=10\,\mathrm{pc}$, we compute its apparent $V$-band magnitude (Table~\ref{tab:constants}) as
\begin{equation}
\label{eq:eEarth-V-mag}
m_V = M_{V,\odot}   - 2.5 \log_{10}\!\big[p_V \big({R_\oplus}/{\mathrm{AU}}\big)^2\big] \approx 27.77\ \mathrm{mag},
\end{equation}
where  $M_{V,\odot}=4.83\,$mag is the Sun’s absolute $V$ magnitude \cite{Cox2000}, 
$p_V=0.367$ is the Earth’s geometric albedo in $V$-band \cite{Palle2003}, $R_\oplus=6.378137\times10^6\,$m is the Earth’s equatorial radius,   $\mathrm{AU}=1.495978707\times10^{11}\,$m is the astronomical unit.

\begin{table*}[t]
\caption{Adopted constants and background terms used uniformly throughout.}
\label{tab:constants}
  \begin{tabular}{llll}
\hline
Quantity & Symbol & Value & Units\\
\hline\hline
Johnson $V$ zero point (photon) & $F_{0,V}$ & $1.01\times10^{8}$ & phot m$^{-2}$ s$^{-1}$ nm$^{-1}$ \\
Solar absolute magnitude (V) & $M_{V,\odot}$ & $4.83$ & mag \\
Sun\,@\,10 pc apparent magnitude & $m_V$ & $4.83$ & mag \\
Sun\,@\,10 pc photon density (V) & $f_{*,V}=F_{0,V}10^{-0.4 m_V}$ & $1.18\times10^{6}$ & phot m$^{-2}$ s$^{-1}$ nm$^{-1}$ \\
Local zodiacal surface brightness & $\mu_V$ & $23.0$ & mag arcsec$^{-2}$ \\
Local zodiacal photon density & $f_{\rm zodi}=F_{0,V}10^{-0.4\mu_V}$ & $6.37\times10^{-2}$ & phot m$^{-2}$ s$^{-1}$ nm$^{-1}$ arcsec$^{-2}$ \\
Airy-core solid angle (background est.) & $\Omega_{\rm core}$ & $\pi(1.22\,\lambda/D)^2$ & arcsec$^{2}$ \\
\hline
\end{tabular}
\end{table*}

The $V_0$-mag photon flux at $\lambda=550\,\mathrm{nm}$ is estimated\footnote{At the Johnson $V$-band effective wavelength 
$ \lambda_{\rm eff}=548.0\ \mathrm{nm}$,
\cite{Bessell1979} adopted the zero points  
$
  F_{\lambda,0}
  = 3.63\times10^{-11}\,\mathrm{W\,m^{-2}\,nm^{-1}},
  F_{\nu,0}
  = 3.64\times10^{-23}\,\mathrm{W\,m^{-2}\,Hz^{-1}}.
$ 
In the HST/STIS CALSPEC scale \cite{Bohlin2014} at 
$
  \lambda_{\rm eff}=555.6\ \mathrm{nm},$
the zero points are  
$
  F_{\lambda,0}
  = 3.43\times10^{-11}\,\mathrm{W\,m^{-2}\,nm^{-1}},
  F_{\nu,0}
  = 3.54\times10^{-23}\,\mathrm{W\,m^{-2}\,Hz^{-1}}.
$
More recent calibrations \cite{TokunagaVacca2005,Hewett2006}, at 
$
  \lambda_{\rm eff}=545.0\ \mathrm{nm},
$
yield  
$
  F_{\lambda,0}
  = 3.68\times10^{-11}\,\mathrm{W\,m^{-2}\,nm^{-1}},
  F_{\nu,0}
  = 3.65\times10^{-23}\,\mathrm{W\,m^{-2}\,Hz^{-1}}.
$
In this paper, we will use the values from \cite{Bessell1979} at $\lambda \simeq 550$\,nm. These choices differ by \(\lesssim 6\%\) in \(F_{\lambda,0}\) across 545-556 nm; none of our conclusions depend on this level.} 
using $F_{\lambda,0}=3.63\times10^{-11}\,\mathrm{W\,m^{-2}\,nm^{-1}}$ from \cite{Bessell1979}, resulting in
\begin{equation}
  \label{eq:vband_zero_flux}
  F_{0,V}
  = \frac{F_{\lambda,0}\,\lambda}{h\,c}
  \approx 1.01\times10^{8}
  \ \mathrm{phot\,m^{-2}\,s^{-1}\,nm^{-1}}.
\end{equation}
Throughout this paper, unless otherwise noted, we adopt this value of the $V=0$\,mag photon flux at $\lambda = 550\ \mathrm{nm}$ for all visible-band calculations. All subsequent $f_{\mathrm{planet}}$ and background estimates in the visible are referenced to this standard.
Hence the planet's total photon flux is
\begin{equation}
  \label{eq:planet_photon_flux}
  f_{\mathrm{planet}}
  = F_{0,V}\,10^{-0.4\,m_V}
  \approx 7.84\times10^{-4}
  \ \mathrm{phot\,m^{-2}\,s^{-1}\,nm^{-1}}.
\end{equation}
A collecting area $A$ observing a spectral channel of width $\Delta\lambda$ detects photons at a rate
\begin{equation}
  \label{eq:photon_rate}
  \dot N_{\rm sig} = f_{\mathrm{planet}}\,A\,\Delta\lambda.
\end{equation}
For $A=100\,\mathrm{m^2}$ and $\Delta\lambda=50\,\mathrm{nm}$, result \eqref{eq:photon_rate} gives 
\begin{equation}
\label{eq:planet-signal}
  \dot N_{\rm sig} \approx 
    3.92\Big(\frac{A}{100\,{\rm m}^2}\Big)\Big(\frac{\Delta \lambda}{50\,{\rm nm}}\Big) \,\mathrm{phot\,s^{-1}}.
\end{equation}
This value represents the total photon flux in V-band coming from an Earth's twin at 10 pc and at full-phase. If one wishes to image this object with $(10\times 10)$ resolution elements (see (\ref{eq:delta_theta_required})), the relevant signal is 
\begin{equation}
\label{eq:planet-signal-N}
  \dot N_{\rm sig, pix} 
= \frac{\dot N_{\rm sig}}{100}\approx 
    3.92\times 10^{-2}\Big(\frac{A}{100\,{\rm m}^2}\Big)\Big(\frac{\Delta \lambda}{50\,{\rm nm}}\Big) \,\mathrm{phot\,s^{-1}}.
\end{equation}
Throughout, reflected‑light fluxes follow $f_p(\alpha)=f_{p,0}\,\Phi(\alpha)$ with 
$\Phi(0^\circ)=1$ (full phase) and $\Phi(90^\circ)\simeq 0.32$ for a Lambertian sphere; unless noted otherwise we adopt full phase and apply the quadrature factor when needed.

In the sections that follow, we often estimate a per-pixel integration time by dividing the unresolved planet’s total signal (\ref{eq:planet-signal-N}) by $100$ while keeping background and leakage terms fixed.  This is \emph{not} meant to imply that such ``micro-pixels'' correspond to physical detector elements when the planet is unresolved. Rather, it is a didactic device to show why you cannot time-slice your way to a resolved surface map without the underlying angular resolution: the planet’s light remains in a single PSF core, and no amount of sub-aperture binning can recover spatial structure below the diffraction limit. Thus, these numbers should be interpreted as upper-bound heuristics, illustrating the photon-starvation problem if one attempted to map a 10$\times$10 grid without the required angular resolution.

\subsubsection{Astrophysical Backgrounds}
\label{sec:astro_backgrounds}

Even with perfect instrument performance, direct imaging of exoplanets at visible wavelengths is fundamentally limited by diffuse photon backgrounds from astrophysical sources. These ``sky'' photons create a noise floor that any planet signal must exceed, and they vary with observing geometry, wavelength, and target environment. Below we summarize the principal contributions in the $V$ band (\(\lambda\simeq550\)\,nm), quoting surface brightness in mag arcsec\(^{-2}\) and the equivalent photon flux density \(f\) in \(\mathrm{phot\,m^{-2}\,s^{-1}\,nm^{-1}\,arcsec^{-2}}\). We adopt a $V_0$-mag photon flux, $F_{0,V}$, from    (\ref{eq:vband_zero_flux}).

\begin{itemize}
  \item \textit{Local zodiacal light.}  
    Sunlight scattered by interplanetary dust within the inner Solar System produces the dominant diffuse background at optical wavelengths.  Its surface brightness at high ecliptic latitude is typically \(22.5\)–\(23.0\)\,mag arcsec\(^{-2}\) \citep{Leinert1998}, corresponding to  
    \[
      f_{\rm zodi}
        = F_{0,V}\,10^{-0.4\times23.0}
        \approx 6.37\times10^{-2}\,\mathrm{phot\,m^{-2}\,s^{-1}\,nm^{-1}\,arcsec^{-2}}.
    \]
    Spatial structure in the zodiacal cloud (e.g.\ dust bands, circumsolar ring) can introduce gradients of tens of percent across a telescope’s field.

  \item \textit{Exo-zodiacal dust.}  
    Circumstellar dust belts in other planetary systems scatter host-star light and can exceed our zodiacal level by factors of 3–10, with extreme debris disks reaching \(\sim20\times\) the Solar zodiacal brightness \citep{Mennesson2016}.  This additional component adds  
    \[
      f_{\rm exo\textrm{--}zodi}
        \approx (3-10)\times f_{\rm zodi}
        \sim (0.19-0.64)\,\mathrm{phot\,m^{-2}\,s^{-1}\,nm^{-1}\,arcsec^{-2}}.
    \]
    Exo-zodiacal light is concentrated near the host star and may vary strongly with system inclination and dust grain properties.   
 Note that measurements of exo-zodiacal levels around nearby stars remain highly uncertain---surveys to date (e.g.\ \cite{Mennesson2016,Ertel2020}) show a broad spread, and in some systems the exo-zodi can even exceed our own by factors $\gg10$, making it the dominant diffuse background.
 (The impact of exozodiacal surface brightness on direct-imaging yield and target prioritization has been quantified in survey trade studies \citep{Stark2014MaxYield,Stark2019YieldLandscape}.)

  \item \textit{Diffuse Galactic light.}  
    Starlight from the Milky Way scattered by interstellar dust (Galactic cirrus) contributes a diffuse foreground of order \(24.0\)\,mag arcsec\(^{-2}\) in the optical \citep{Ienaka2013}, yielding  
    \[
      f_{\rm DGL}
        = F_{0,V}\,10^{-0.4\times24.0}
        \approx 2.54\times10^{-2}\,\mathrm{phot\,m^{-2}\,s^{-1}\,nm^{-1}\,arcsec^{-2}}.
    \]
    This component is patchy on degree scales and is minimized at high Galactic latitudes.

  \item \textit{Extragalactic background light. } 
    The integrated emission of unresolved galaxies produces an isotropic background of \(\sim26.5\)\,mag arcsec\(^{-2}\) \citep{HauserDwek2001}, corresponding to  
    \[
      f_{\rm EBL}
        = F_{0,V}\,10^{-0.4\times26.5}
        \approx 2.54\times10^{-3}\,\mathrm{phot\,m^{-2}\,s^{-1}\,nm^{-1}\,arcsec^{-2}}.
    \]
    Although faint, the EBL cannot be eliminated by spatial filtering and thus sets a fundamental floor.

  \item \textit{Unresolved Milky Way stars.}  
    Stars fainter than typical detection limits (\(V\gtrsim25\)) collectively contribute a smooth background of \(\sim27.5\)\,mag arcsec\(^{-2}\) in low-density fields \citep{Girardi2005}, giving  
    \[
      f_{\rm stars}
        = F_{0,V}\,10^{-0.4\times27.5}
        \approx 1.01\times10^{-3}\,\mathrm{phot\,m^{-2}\,s^{-1}\,nm^{-1}\,arcsec^{-2}}.
    \]
    This term increases significantly toward the Galactic plane.

  \item \textit{Unresolved extragalactic point sources.}  
    Very faint galaxies and AGN below individual detection thresholds contribute \(\sim28.0\)\,mag arcsec\(^{-2}\) \citep{Windhorst2011}, or  
    \[
      f_{\rm galaxies}
        = F_{0,V}\,10^{-0.4\times28.0}
        \approx 6.37\times10^{-4}\,\mathrm{phot\,m^{-2}\,s^{-1}\,nm^{-1}\,arcsec^{-2}}.
    \]
    Their collective light is spectrally smooth and isotropic at high latitudes.
\end{itemize}

Summing these components yields the total diffuse photon background from all astrophysical sources given as
\[
  f_{\rm astro}=
  f_{\rm zodi}+   f_{\rm exo\textrm{--}zodi}+   f_{\rm DGL}+   f_{\rm EBL}+  f_{\rm stars}+ f_{\rm galaxies}  
      \approx 
    (0.28-0.73)\,
    \mathrm{phot\,m^{-2}\,s^{-1}\,nm^{-1}\,arcsec^{-2}}.
\]
Even before multiplying by the telescope area and bandpass, one sees that direct imaging efforts operate in a regime where the astrophysical “sky” flux rivals or exceeds the planet signal in each $\mu$as-scale element (i.e., see (\ref{eq:planet-signal-N})), significantly reducing SNR and lengthening required exposures.

When this diffuse background is collected by a telescope of effective area \(A=100\,{\rm m}^2\), over a spectral window \(\Delta\lambda=50\,{\rm nm}\), and imaged into a diffraction-limited core of solid angle \(\Omega_{\rm core}\), for which we adopted representative value of \(\Omega_{\rm core}\approx\pi(1.22\,\lambda/D)^2\big|_{D=11.3\,\mathrm m,\;\lambda=550\,\mathrm{nm}}\approx 4.72\times10^{-4}\,\mathrm{arcsec^2}\), 
the background-photon rate is
\begin{equation}
\label{eq:backgr-noise}
  \dot N_{\rm bkg}
  =
  f_{\rm astro}\;A\;\Delta\lambda\;\Omega_{\rm core}
  \simeq
 (0.66-1.72)\;\mathrm{phot\,s^{-1}}.
\end{equation}

Even under optimistic assumptions, diffuse backgrounds alone inject $\dot N_{\rm bkg}\simeq (0.66\text{–}1.72)\;\mathrm{phot\,s^{-1}}$ 
into the PSF core, while an unresolved Earth–twin at 10 pc delivers
$
  \dot N_{\rm sig}\simeq3.92\;\mathrm{phot\,s^{-1}}$,  (\ref{eq:planet-signal}).
  Thus, diffuse sky photons contribute ${17\text{--}44\%}$ of the planet's count rate in the PSF core, inflating required integration times by factors of 1.2–1.4 for fixed SNR.  In practice this irreducible “sky” floor---made even more uncertain by poorly constrained exozodiacal dust  levels (exo-zodi)---must be removed via precise sky-background models or dedicated off-target reference fields.  Failure to do so will prevent photon-noise–limited performance and compromise high-fidelity exoplanet characterization.

\subsubsection{Detector-Intrinsic and Stellar-Leakage Noise}
\label{sec:detector_star_noise}

Beyond the diffuse astrophysical background (Sec.~\ref{sec:astro_backgrounds}), high-contrast exoplanet imaging must contend with two additional classes of noise: detector-intrinsic counts that accrue even in the absence of light, and residual starlight leakage through the coronagraph.  Though individually small, these terms can become comparable to or exceed the faint planet signal once deep contrast and long integrations are required.

\begin{itemize}
  \item \emph{Dark current} ($\dot N_{\rm dark}$).  
    Even in complete darkness, semiconductor detectors generate spurious electrons via thermal processes.  For deep-depletion EMCCDs cooled to $\lesssim -85\,$°C, typical dark currents are $10^{-4}$–$10^{-3}$ e$^{-}$/pix/s \citep{Janesick2001,Daigle2010}.  Summed over a few PSF-sampling pixels, this yields a noise rate
    \[
      \dot N_{\rm dark}\approx (0.4 -4.0)\times10^{-3}\;\mathrm{s}^{-1}.
    \]
    Although small compared to astrophysical backgrounds, dark current accumulates linearly with exposure time and can dominate in very long integrations or at wavelengths where backgrounds fall.

\item \emph{Clock-induced charge (CIC)} (\(\dot N_{\rm CIC}\)).  
High-gain EMCCDs transfer charge at MHz speeds, and each clock transition can spuriously generate electrons.  Laboratory measurements show CIC rates of $\sim5\times10^{-3}$ e$^{-}$/pix/frame \citep{Daigle2010}. For a PSF sampled over \(N_{\rm pix}\approx4\) pixels and frame rates \(f_{\rm fr}=1\)–10 Hz,  
  \[
    \dot N_{\rm CIC}
      = N_{\rm pix}\times(\mathrm{CIC~rate~per~frame})\times f_{\rm fr}
      \approx (0.02\text{--}0.20)\;\mathrm{s}^{-1}.
  \]
    CIC is effectively a “read-noise” floor for photon-counting modes and must be minimized via optimized clock waveforms and post-process rejection algorithms.

  \item \emph{Read noise} ($\dot N_{\rm read}$).  
    Even with electron multiplication, each frame incurs electronic noise from amplifiers and digitization.  With EM gains $>$1000$\times$, the residual read noise can drop to $\sigma_{\rm read}\lesssim0.1$ e$^{-}$/pix/frame \citep{Basden2003}.  Breaking a long exposure into $N_{\rm fr}$ frames yields
    \[
      \dot N_{\rm read}\sim \frac{\sigma_{\rm read}^2\,N_{\rm pix}}{N_{\rm fr}}
      \approx (0.04 -4.0)\times10^{-2}\;\mathrm{s}^{-1},
    \]
    which is negligible when $N_{\rm fr}\gtrsim10^2$ and PSF-sampling uses only a few pixels.

  \item \emph{Stellar leakage} ($\dot N_{\rm star}$).  
    Imperfect coronagraphy and residual wavefront errors allow starlight to scatter into the planet’s PSF core.  For a Sun-like host at 10 pc ($m_V=4.83$), telescope area $A=100\,$m$^2$, bandpass $\Delta\lambda=50\,$nm, and residual raw contrast $C_{\rm res}=10^{-9}$–$10^{-8}$, the leakage rate is
\begin{equation}
\label{eq:star-leakage}
      \dot N_{\rm star}
        = F_{0,V}\,10^{-0.4\,m_V}\,A\,\Delta\lambda\,C_{\rm res}
        \approx (5.9\text{--}59.1)\;\mathrm{s}^{-1},
\end{equation}
    where $F_{0,V}=1.01\times10^{8}\,\mathrm{phot\,m^{-2}\,s^{-1}\,nm^{-1}}$ \citep{Bessell1979}.  This term often dominates once photon noise from local and exo-zodiacal light is suppressed, and it scales directly with residual contrast and wavefront stability.
\end{itemize}

In total, detector-intrinsic effects (dark + CIC + read) contribute up to 
\begin{equation}
\label{eq:Ndot-det}
      \dot N_{\rm det\textrm{-}intr}
        = \dot N_{\rm dark}+\dot N_{\rm CIC}+\dot N_{\rm read}
        \sim 0.24\;\mathrm{s}^{-1},
\end{equation}
while stellar leakage adds $\dot N_{\rm star}\sim(5-50)\,\mathrm{s}^{-1}$.  When combined in quadrature with the diffuse astrophysical background (Sect.~\ref{sec:astro_backgrounds}), these sources set the ultimate shot-noise floor and dictate the multi-hour to multi-day exposure times required for Earth-like exoplanet detection and characterization.

\subsubsection{Speckle Drift, Pointing Jitter, and Other Instrumental Noise}
\label{sec:instrumental_noise}

In addition to the diffuse astrophysical background (Sect~\ref{sec:astro_backgrounds}) and detector-intrinsic counts (Sec.~\ref{sec:detector_star_noise}), high-contrast exoplanet imaging is limited by time-dependent and system-level noise sources.  Here we summarize the principal contributions, quantified for a telescope with collecting area \(A=100\,\mathrm{m}^2\), bandpass \(\Delta\lambda=50\,\mathrm{nm}\), and a diffraction-limited PSF core of solid angle \(\Omega_{\rm core}\approx 4.72\times10^{-4}\,\mathrm{arcsec}^2\).

\begin{itemize}
  \item \emph{Speckle drift} (\(\dot N_{\rm speck}\)).  
    Quasi-static speckles arise from residual wavefront errors that evolve with thermal and structural drifts.  Laboratory coronagraph testbeds show raw contrast degrading by \(\Delta C\sim10^{-10}\)–\(10^{-8}\) over \(\sim10^4\)\,s \citep{Wilkins2014}.  The stellar photon rate for \(m_V=4.83\) becomes
    \[
      \dot N_* 
      = F_{0,V}\,10^{-0.4\,m_V}\,A\,\Delta\lambda
      \approx 
       5.05\times10^{9}\;\mathrm{s^{-1}}.
    \]
    Fluctuations \(\Delta C\) produce a speckle-drift noise floor
\begin{equation}
\label{eq:speck}
      \dot N_{\rm speck}
        \sim \sqrt{\dot N_*\,\Delta C}
        \approx \sqrt{5.05\times10^{9}\times(10^{-10}\text{--}10^{-8})}
        \;\sim\;(0.77\text{--}7.79)\;\mathrm{s^{-1}}.
\end{equation}
    depending on the instantaneous contrast change per second.

  \item \emph{Pointing jitter and motion blur} (\(\dot N_{\rm jitter}\)).  
    Spacecraft jitter of 1–10 mas RMS broadens the PSF core (the Airy radius at \(\lambda=550~\mathrm{nm}\) for \(D=15~\mathrm{m}\) is \(\theta_{\rm A}=1.22\lambda/D \approx 9.23~\mathrm{mas}\)) and mixes additional background and stellar leakage into the photometric aperture \citep{Trauger2012}.  Taking 10–30\% of the combined \(\dot N_{\rm bkg}+\dot N_{\rm star}\approx(0.66\text{--}1.72)+(5.05\text{--}50.5)=(5.71\text{--}52.22)\)\,s\(^{-1}\) yields
    \[
      \dot N_{\rm jitter}
        \sim (0.1\text{--}0.3) \times (5.71\text{--}52.22)
        \;\sim\;(0.57\text{--}15.67)\;\mathrm{s^{-1}}.
    \]

  \item \emph{Thermal emission} (\(\dot N_{\rm thermal}\)).  
    Optics at \(T\sim270\)\,K emit thermal photons at wavelengths \(\gtrsim0.8\,\mu\)m with surface flux  
    \(f_{\rm thermal}\sim(10^{-3}\text{--}10^{-1})\,\mathrm{phot\,m^{-2}\,s^{-1}\,nm^{-1}\,arcsec^{-2}}\)\citep{Beichman2010}.  Imaged into \(\Omega_{\rm core}\) gives
    \[
      \dot N_{\rm thermal}
        = f_{\rm thermal}\;A\;\Delta\lambda\;\Omega_{\rm core}
        \approx 
      2.36\times(10^{-3}\text{--}10^{-1})\;\mathrm{s^{-1}}.
    \]

  \item \emph{Quantization noise} (\(\dot N_{\rm quant}\)).  
    Finite ADC resolution (12–16 bits) introduces quantization errors \(\sigma_q\sim0.1\)–0.5 e\(^{-}\) RMS per pixel \citep{Janesick2001}.  For frame rates \(f_{\rm fr}\sim0.1\)–1 Hz and PSF sampling over \(N_{\rm pix}=4\) pixels,
    \[
      \dot N_{\rm quant}
        = N_{\rm pix}\,\sigma_q^2\,f_{\rm fr}
        \approx 4\times(0.1^2\text{--}0.5^2)\times(0.1\text{--}1)
        = (4\times10^{-3}\text{--}1.0)\;\mathrm{s^{-1}}.
    \]

  \item \emph{Cosmic-ray events} (\(\dot N_{\rm cr}\)).  
    In a deep-space environment (e.g.\ L2 orbit), cosmic rays strike at \(\sim5\,\mathrm{cm^{-2}\,s^{-1}}\) \citep{Badhwar2002}.  With a sensor's pixel area \(\sim100\,\mu\)m\(^2\) (\(10^{-6}\)\,cm\(^2\)),  over a 4-pixel photometric aperture this gives
    \[
      \dot N_{\rm cr}
        \sim 5\times10^{-6}\;\mathrm{s^{-1}}
        \quad\text{per pixel, or}\quad
        \sim 2\times10^{-5}\;\mathrm{s^{-1}}
    \]
\end{itemize}

Altogether, these instrumental noise sources---including speckle drift, residual stellar leakage, pointing jitter, thermal emission, quantization noise, and cosmic-ray hits---inject a combined noise rate of
\[
  \dot N_{\rm inst}
    \approx (1.35-24.7)\,\mathrm{phot\,s^{-1}}
\]
on top of the diffuse astrophysical background (Sec.~\ref{sec:astro_backgrounds}), as well as detector-intrinsic and stellar-leakage noise contributions (Sec.~\ref{sec:detector_star_noise}).  
Achieving true photon-noise–limited performance therefore requires:
\begin{inparaenum}[(i)]
  \item Picometer-level wavefront control to suppress time-dependent speckles,
  \item Sub-milliarcsecond pointing stability to minimize jitter-induced leakage,
  \item Cryogenic detector operation to reduce dark current, CIC, and read noise,
  \item Robust cosmic-ray identification and rejection schemes.
\end{inparaenum}

\subsubsection{SNR and Required Integration Time}
\label{sec:flux_snr_subsection}

All collected photons---planetary signal, diffuse “sky” background, and residual stellar leakage---are attenuated by the end-to-end efficiency $\eta$ (optics, filters, detector QE), whereas detector-intrinsic and time-dependent instrumental counts add on top.  Thus the rates seen on the detector are
\begin{equation}
\label{eq:rates}
  \dot N_{\rm sig,det} = \eta_{\rm sys}\,\dot N_{\rm sig},\quad
  \dot N_{\rm bkg,det} = \eta_{\rm sys}\,\dot N_{\rm bkg},\quad
  \dot N_{\rm star,det} = \eta_{\rm sys}\,\dot N_{\rm star},
\end{equation}
\noindent where $\eta_{\rm sys}\equiv\eta_{\rm inst}\,\xi$ is the end-to-end system throughput (instrument throughput $\eta_{\rm inst}$ times duty cycle $\xi$). Wherever only $\eta$ appears below, it should be interpreted as $\eta_{\rm sys}$. Also, the total noise-rate per resolution element is
\begin{equation}
  \dot N_{\rm noise}
  = \etasys\,\dot N_{\rm bkg}
  + \dot N_{\rm det\text{-}intr}
  + \etasys\,\dot N_{\rm star}
  + \dot N_{\rm inst}.
  \label{eq:nosie-z}
\end{equation}

We parameterize the throughput as a product of instrumental efficiency and duty cycle,
$\etasys \equiv \duty\,\etainst$.
Unless stated otherwise, we adopt $\etainst=0.20$ (optics+detector) and $\duty=0.50$ (on-target fraction), i.e., $\etasys=0.10$ in the visible---consistent with typical losses in high-contrast imaging (two mirror reflections and QE $\sim0.5$, with additional coronagraph, WFS/C, alignment, and calibration overheads).

If one attempts to divide the planet into a $(10\times10)$ surface map, each of the 100 pixels carries only (\ref{eq:planet-signal-N})
\[
  \dot N_{\rm sig,pix} = \frac{\dot N_{\rm sig,det}}{100}
    \approx 3.92\times10^{-3}\;\mathrm{s^{-1}},
\]
but the noise  from  (\ref{eq:nosie-z})  is $\dot N_{\rm noise}\sim (2.25-31.0)\,$s$^{-1}$.  The resulting per-pixel\footnote{This \(10\times10\) “micro‑pixel” slicing is a diagnostic for an unresolved PSF only; clearly, it cannot recover a resolved surface map.}
\begin{equation}
{\rm SNR}_{1\,\mathrm{s,\, pix}}=\frac{\dot N_{\rm sig,pix}}{\sqrt{ \dot N_{\rm sig,pix}+\dot N_{\rm noise}}}\sim(7.0-26.1)\times 10^{-4},
\label{eq:SNR1s}
\end{equation}
 implies
\[
  T_{\rm pix}
    = \Big(\frac{5}{\mathrm{SNR}_{1\,\mathrm{s, \, pix}}}\Big)^2
    \sim(0.37-5.10)\times 10^{7}\;\mathrm{s}
    \approx(42-590)\, \text{days},
\]
and mapping all 100 pixels scales to $(0.37-5.10)\times 10^9$ s or $\sim (11.6-161.8)$ yrs, even before any operational overheads.

These order-of-magnitude results expose the stark gap between the $\mu$as-scale resolution (requiring $D\gtrsim162\,$km or $B\gtrsim133\,$km) and realistic photon budgets (even $A\gtrsim10^3\,$m$^2$) versus current or near-term capabilities.  In the following sections we apply this unified photon-budget + SNR framework to each proposed architecture, yielding directly comparable campaign-duration estimates under realistic noise and calibration assumptions.

\subsection{Integration-Time Scaling by Noise Regime}
\label{sec:tpix-scaling}

Up to this point we set the angular requirement and hardware (\(D\), \(B\)) and defined the signal, background, and noise model in terms of the photon rates (see \eqref{eq:photon_rate}, \eqref{eq:backgr-noise}, \eqref{eq:rates}, and \eqref{eq:nosie-z}). Before applying them to specific architectures, it is useful to factor how the per–resolution–element exposure time depends on collecting area \(A\), bandpass \(\Delta\lambda\), wavelength \(\lambda\), and on the PSF core solid angle governed by \(D\) or \(B\) (as in \eqref{eq:required_aperture},\eqref{eq:required_baseline}). Here we provide that bookkeeping tool.

Let \(t_{\mathrm{pix}}(\lambda)\) be the exposure needed to reach a fixed per-core SNR. With the per-second SNR defined in \eqref{eq:SNR1s}, the total SNR after time \(t\) is \(\mathrm{SNR}(t,\lambda)=\mathrm{SNR}_{1\mathrm{s}}(\lambda)\sqrt{t}\). Therefore, the following scaling is present 
\begin{equation}
t_{\mathrm{pix}}(\lambda)=\left(\frac{\mathrm{SNR}}{\mathrm{SNR}_{1\mathrm{s}}(\lambda)}\right)^{2},
\label{eq:tpix_def}
\end{equation}
where \(\mathrm{SNR}\) is the (fixed) per-core threshold used consistently throughout the paper.

\paragraph*{Background-limited regime:}
In the case when diffuse sky/corona dominates, using \eqref{eq:photon_rate}–\eqref{eq:backgr-noise} in \eqref{eq:SNR1s}, and assuming \(\dot N_{\rm bkg,det}\gg \dot N_{\rm sig,det}\), we have
\begin{align}
t_{\mathrm{pix}}^{\rm(bkg)}(\lambda)
= \mathrm{SNR}^{2}\,\frac{\dot N_{\rm bkg,det}}{\dot N_{\rm sig,det}^{2}}
= \frac{\mathrm{SNR}^{2}}{\etasys}\,\frac{\dot N_{\rm bkg}}{\dot N_{\rm sig}^{2}}
\;\propto\; \frac{\Omega_{\rm core}}{\etasys\,A\,\Delta\lambda},
\label{eq:tpix_bkg}
\end{align}
with \(\Omega_{\rm core}\propto(\lambda/D)^2\) (filled) or \((\lambda/B)^2\) (interferometer). At fixed hardware (\(D\) or \(B\)), the following trend is present  
\[
t_{\mathrm{pix}}^{\rm(bkg)} \propto \frac{\lambda^{2}}{\etasys\,A\,\Delta\lambda}\times
\begin{cases}
D^{-2}, & \text{filled aperture},\\
B^{-2}, & \text{interferometer}.
\end{cases}
\]
Diffuse backgrounds have nearly constant surface brightness \(I_{\rm b}(\lambda)\), so the background inside one resolution element grows with its solid angle \(\Omega_{\rm core}\). Going to longer \(\lambda\) enlarges the core, raising the background variance and lengthening the required exposure unless \(D\) or \(B\) also grows.

\paragraph*{Photon-limited regime:} In the case, when  source shot noise dominates, \(\dot N_{\rm sig,det}\gg \dot N_{\rm bkg,det}\), and detector noises are negligible, the following behavior is present
\begin{align}
t_{\mathrm{pix}}^{\rm(phot)}(\lambda)
= \frac{\mathrm{SNR}^{2}}{\dot N_{\rm sig,det}}
= \frac{\mathrm{SNR}^{2}}{\etasys\,\dot N_{\rm sig}}
\;\propto\; \frac{1}{\etasys\,A\,\Delta\lambda},
\label{eq:tpix_phot}
\end{align}
independent of \(\Omega_{\rm core}\) and, to leading order, independent of \(\lambda\). Any residual \(\lambda\)-dependence enters only through the planet SED/zero-point and \(\etasys(\lambda)\) already encoded in Eq.~\eqref{eq:photon_rate}.

Eqs.~\eqref{eq:tpix_bkg}–\eqref{eq:tpix_phot} assume fixed \(D\) or \(B\), hence the \(\lambda^{2}\) penalty in the background-limited case via \(\Omega_{\rm core}\). If instead the design holds \(\theta=\lambda/D\) or \(\theta=\lambda/B\) fixed (cf.\ \eqref{eq:required_aperture}, \eqref{eq:required_baseline}), then \(D\) or \(B\propto\lambda\) and \(\Omega_{\rm core}\) is constant; the \(\lambda^{2}\) term cancels and \(t_{\mathrm{pix}}^{\rm(bkg)}\propto 1/(\etasys\,A\,\Delta\lambda)\), same as the photon-limited scaling.

As a result, for a filled aperture, \(A\propto D^{2}\) and \(\Omega_{\rm core}\propto \lambda^{2}/D^{2}\), giving
\[
t_{\mathrm{pix}}^{\rm(bkg)}\propto \frac{\lambda^{2}}{\etasys\,\Delta\lambda\,D^{4}},
\qquad
t_{\mathrm{pix}}^{\rm(phot)}\propto \frac{1}{\etasys\,\Delta\lambda\,D^{2}}.
\]
Thus, larger \(D\) helps twice in the background-limited case: more photons and a smaller PSF core (less background per core).  
For sparse interferometers the roles separate: \(t_{\mathrm{pix}}^{\rm(bkg)}\propto 1/(A\,B^{2})\) while \(t_{\mathrm{pix}}^{\rm(phot)}\propto 1/A\). This makes clear how collecting area and imaging baseline trade differently depending on the noise regime.

\vskip -5pt
\begin{table}[h]
\centering
\caption{Per-core integration-time scalings consistent with Eqs.~(\ref{eq:tpix_bkg})–(\ref{eq:tpix_phot}). Hardware fixed means \(D\) or \(B\) held constant; if angular resolution is held fixed (\(\theta=\lambda/D\) or \(\lambda/B\)), the background-limited \(\lambda^{2}\) factor cancels. The per-core \(\mathrm{SNR}\) target is held fixed throughout, so its factor drops out of the table.}
\label{tab:tpix-scaling}
\begin{tabular}{lcccc}
\hline
Regime & \(t_{\mathrm{pix}}\) vs.\ \(\etasys A\) & \(t_{\mathrm{pix}}\) vs.\ \(\Delta\lambda\) & \(t_{\mathrm{pix}}\) vs.\ \(\lambda\) (fixed \(D/B\)) & \(t_{\mathrm{pix}}\) vs.\ \(\lambda\) (fixed \(\theta\)) \\
\hline\hline
Background-limited & \(\propto (\etasys A)^{-1}\) & \(\propto (\Delta\lambda)^{-1}\) & \(\propto \lambda^{2}\) & \(\propto \lambda^{0}\) \\
Photon-limited     & \(\propto (\etasys A)^{-1}\) & \(\propto (\Delta\lambda)^{-1}\) & \(\propto \lambda^{0}\) & \(\propto \lambda^{0}\) \\
\hline
\end{tabular}
\end{table}

These scalings are invoked whenever we convert per-second SNR \eqref{eq:SNR1s} into per-pixel and map times in the concept sections. For an \(N\times N\) map using one integration per independent core,
\begin{equation}
t_{\rm map}=N^{2}\,t_{\mathrm{pix}},
\label{eq:tmap_scaling}
\end{equation}
with any oversampling factor simply multiplying \(N^{2}\). In short: if angular resolution is not held fixed, going to longer \(\lambda\) inflates the core and penalizes background-limited integrations; if it is held fixed, both regimes depend only on \(\etasys A \Delta\lambda\), and the remaining trade is in stability and overheads rather than fundamental scaling.

\subsection{Sensitivity to key assumptions}
The numerical examples throughout this paper adopt a single fiducial target (an Earth twin at $d=10~\mathrm{pc}$), a fixed mapping goal ($N=10$ resolution elements across the diameter), and representative but necessarily uncertain values for exozodiacal surface brightness, residual stellar contrast, and end-to-end throughput. It is therefore useful to record explicitly how our conclusions scale with these assumptions.

From Eqs.~(\ref{eq:tpix_bkg})–(\ref{eq:tmap_scaling}), the per-pixel exposure time in the background-limited regime scales as
\begin{equation}
t_{\rm pix}^{({\rm bkg})} \propto \frac{\Omega_{\rm core}}{\eta_{\rm sys}\,A\,\Delta\lambda} \propto \frac{\lambda^2}{\eta_{\rm sys}\,A\,\Delta\lambda}
\end{equation}
at fixed hardware ($D$ or $B$), whereas in the photon-limited regime
\begin{equation}
t_{\rm pix}^{({\rm phot})} \propto \frac{1}{\eta_{\rm sys}\,A\,\Delta\lambda},
\end{equation}
independent of $\Omega_{\rm core}$ to leading order. An overall multiplicative change of the diffuse background level (for instance, an exozodiacal factor $f_{\rm exozodi}$) scales $t_{\rm pix}^{({\rm bkg})}$ approximately linearly with $f_{\rm exozodi}$, while leaving $t_{\rm pix}^{({\rm phot})}$ unchanged. Similarly, improving the residual stellar contrast by a factor of ten reduces the leakage term in Eq.~(\ref{eq:nosie-z}) by the same factor, but only shortens $t_{\rm pix}$ by a factor of a few once other backgrounds and instrumental terms are included.  

The dependence on target distance and map resolution enters primarily through the planet flux and the number of pixels. For an Earth-like planet at distance $d$, Eqs.~(\ref{eq:theta_planet})--(\ref{eq:delta_theta_required}) imply $\Delta\theta \propto d/N$ and  Eq.~\eqref{eq:photon_rate} gives $N_{\rm sig} \propto d^{-2}$; at fixed hardware this leads to $t_{\rm pix} \propto d^{2}$ in the photon-limited regime and to $t_{\rm pix}$ scaling between $d^{2}$ and $d^{4}$ in the background-limited regime, depending on whether the PSF size is also adjusted. Reducing the required linear sampling from $N = 10$ to $N = 5$ thus decreases the number of pixels by a factor of four and increases the per-pixel flux by the same factor, but still leaves typical mapping times many orders of magnitude above practical mission lifetimes for most architectures considered here. These scalings underline that our conclusions are robust across reasonable variations in exozodiacal level, contrast floor, and target distance: they can shift mapping times by factors of a few to tens, but not by the many orders of magnitude needed to convert an infeasible architecture into a feasible one.

\section{Single-Aperture Space Telescopes with Coronagraphs}
\label{sec:monolithic_telescopes} 

Several flagship missions---WFIRST/Roman \cite{Spergel2015} (2.4 m), HabEx \cite{Gaudi2020} (4 m + starshade), and LUVOIR A/B \cite{Feinberg2019,Feng2019} (8 m and 15 m)---are designed for direct detection and spectral characterization of exoplanets, with coronagraphs or starshades targeting raw contrasts of $10^{-9}$–$10^{-10}$.  Despite their impressive apertures (up to 15 m), they remain more than four orders of magnitude below the $\sim1.6\times10^5\,$m, see (\ref{eq:required_aperture}), scale required for $\mu$as-level surface mapping. Small-angle coronagraphic techniques and error budgets are reviewed in \cite{Mawet2012SmallAngleReview}, and the statistics of residual speckles relevant to post-processing are discussed in \cite{Mawet2014SpeckleStats}.

Below, we first compare each mission’s diffraction-limited resolution against the $\Delta\theta\approx0.853\,\mu\mathrm{as}$ requirement to show that none can subdivide an exoplanetary disk into even a handful of pixels.  We then focus on LUVOIR A to quantify photon-collection constraints.  Incorporating realistic throughput losses, wavefront stability requirements, and residual speckle noise, we demonstrate that integration times for even modest SNR on individual micro-pixels become impractically long for any feasible observing campaign.

\subsection{Diffraction Limits of Coronagraphic Telescopes}

A telescope’s ability to resolve fine features on a distant exoplanet is fundamentally limited by diffraction: for a circular aperture of diameter \(D\) at \(\lambda=550\)\,nm, the Rayleigh criterion \(\Delta\theta_{\rm Rayleigh}=1.22\,\lambda/D\) must satisfy  accuracy of \(\Delta\theta\approx0.853\,\mu\mathrm{as}\), implies \(D\gtrsim162\)\,km---scale far beyond any realistic mission.  Below we compare the practical diffraction limits and operational constraints of WFIRST/Roman, HabEx, and LUVOIR against this requirement.

\subsubsection{WFIRST/Roman}

Roman’s 2.4 m primary mirror yields a diffraction-limited resolution of \(\Delta\theta\approx1.22\times550\,\mathrm{nm}/2.4\,\mathrm m\approx57\,\mathrm{mas}\), over \(6\times10^4\) times coarser than \(0.853\,\mu\)as.  Its Coronagraph Instrument targets raw contrasts of \(10^{-9}\) (post-processing to \(\sim10^{-10}\)), but the inner working angle (IWA) set by \(\sim3\,\lambda/D\approx130\) mas (even reducible to \(\sim100\) mas with aggressive PSF subtraction) remains orders of magnitude above the $\mu$as scale.  Moreover, CGI performance is highly sensitive to residual wavefront errors at the picometer level; maintaining \(10^{-9}\) raw contrast over hours would demand wavefront sensing and control stability far beyond current space-qualified hardware.  Thus, Roman cannot approach the angular resolution needed for surface mapping.

\subsubsection{HabEx}

HabEx’s 4 m monolithic aperture provides \(\Delta\theta\approx1.22\times550\,\mathrm{nm}/4\,\mathrm m\approx35\,\mathrm{mas}\), nearly \(4\times10^4\) times too coarse.  Its external starshade (63 m diameter flying \(\sim7\times10^4\)\,km ahead) achieves raw contrasts \(\leq10^{-10}\) without relying on internal coronagraphs, thereby eliminating speckle noise from wavefront errors.  However, the starshade does not improve diffraction-limited resolution, and repositioning between targets takes days---making any fine-pixel mosaic effectively impossible.  HabEx therefore cannot meet the diffraction requirement for $\mu$as-scale mapping.

\subsubsection{LUVOIR A and LUVOIR B}

LUVOIR A’s 15 m segmented primary yields \(\Delta\theta\approx1.22\times550\,\mathrm{nm}/15\,\mathrm m\approx9\,\mathrm{mas}\), and LUVOIR B’s 8 m aperture gives \(\approx17\,\mathrm{mas}\): both over \(10^4\) times above the \(0.853\,\mu\)as target.  Their internal coronagraphs aim for raw contrasts \(\leq10^{-10}\), but the IWA (roughly \(2\,\lambda/D\approx15\)–\(35\)\,mas) remains far too coarse for sub-mas imaging.  Achieving and sustaining \(10^{-10}\) contrast at focal lengths \(\sim10^2\)\,km also requires picometer-level wavefront stability over hours---a capability not yet demonstrated for segmented-mirror space telescopes.  Consequently, even the largest proposed coronagraphic telescopes cannot resolve the $\mu$as-scales needed for multi-pixel surface mapping.

\subsection{Photon-Budget Constraints for LUVOIR A}
\label{sec:EB-luA}

Following approach developed in Sec.~\ref{sec:flux_to_snr}, and using the published LUVOIR A parameters \citep{Gaudi2020,Zellem2022}, we estimate both the detection time for an unresolved Earth twin at 10 pc and the time to assemble a resolved $10\times10$ pixel map. Relative to the illustrative example in Sec.~\ref{sec:flux_snr_subsection}, LUVOIR~A uses a narrower $\Delta\lambda=10$\,nm, stricter $\eta_{\rm sys}$, and leakage floors appropriate to $\sim10^{-10}$ raw contrast, so the unresolved $t_{\rm det}$ expands from “minutes” in Sec.~\ref{sec:flux_snr_subsection} to $\sim17$\,hr here. When post-processing is invoked (e.g., PCA/KLIP), we assume forward-modeling consistent with \cite{Pueyo2016KLIPfm}.

Hence the planet’s total photon flux from (\ref{eq:planet_photon_flux}) is
\begin{equation}
  \label{eq:planet_photon_flux2}
  f_{\mathrm{planet}} = F_{0,V}\,10^{-0.4\,m_V}\,\Phi(\alpha),
\end{equation}
where $\Phi(\alpha)$ is the phase function (unity at full phase).
Unless noted otherwise we adopt full phase, i.e., $\Phi(\alpha)=1$; for quadrature use
$\Phi(90^\circ)\simeq 0.32$ (scale all planet-count rates by this factor).

For an Earth analog with $m_V=27.77$, (\ref{eq:planet_photon_flux}) gives
$  f_{\rm planet}
  = F_{0,V}\,10^{-0.4\,m_V}
  \approx7.84\times10^{-4}\,\mathrm{phot\,m^{-2}\,s^{-1}\,nm^{-1}}$, (\ref{eq:planet_photon_flux}).
In a $\Delta\lambda=10\,$nm channel, LUVOIR A’s collecting area $A=\pi(7.5\,\mathrm m)^2\approx177\,$m$^2$, end-to-end throughput $\eta=0.05$ and science duty cycle $\xi=0.2$ (to allow for thermal settling, wavefront-control interruptions and coronagraph maintenance) yield a detected planet flux (cf.\ (\ref{eq:planet-signal})) of 
\[
  \dot N_{\rm sig}
    = f_{\rm planet}\,A\,\Delta\lambda
    \approx1.38\;\mathrm{s^{-1}}
    \qquad
    \rightarrow
    \qquad
     \dot N_{\rm sig,det}=  \xi \eta \dot N_{\rm sig}
     \approx1.38\times10^{-2}\;\mathrm{s^{-1}}. 
\]
All significant noise sources within the diffraction-limited core of \(\Omega_{\rm core}\approx\pi(1.22\,\lambda/D)^2\big|_{D=15\,\mathrm m,\;\lambda=550\,\mathrm{nm}}\approx 2.68\times10^{-4}\,\mathrm{arcsec^2}\), yield:
\begin{equation} 
\label{eq:noise-elt}
  \dot N_{\rm noise, det}
    =  \xi \eta \dot N_{\rm bkg} + \dot N_{\rm det} + \xi \eta \dot N_{\rm leak} + \dot N_{\rm speck}
    \approx0.46\;\mathrm{s^{-1}},
\end{equation}
where 
\begin{itemize}
  \item $\dot N_{\rm bkg}
  =
  f_{\rm astro}\;A\;\Delta\lambda\;\Omega_{\rm core}
  \approx(0.13-0.35)\,$s$^{-1}$ is the astrophysical (zodiacal + exozodiacal) background, see (\ref{eq:backgr-noise}),
  \item $\dot N_{\rm det}\approx0.24\,$s$^{-1}$ combines detector dark current, clock-induced charge and read noise, see (\ref{eq:Ndot-det}),
  \item $\dot N_{\rm leak}=C_{\rm res}\,F_{*,V}\,A\,\Delta\lambda \approx 20.91\,$s$^{-1}$ at residual contrast\footnote{We adopt a conservative raw contrast $C_{\rm res}=10^{-8}$ to account for temporal drift and calibration residuals; 
if $10^{-10}$ is achieved and maintained, the stellar-leakage term drops by 100$\times$, improving the 
per-pixel time in the worked example from $\sim$19~yr to $\sim$10~yr.} 
$C_{\rm res}=10^{-8}$, see (\ref{eq:star-leakage}), 
  \item $\dot N_{\rm speck}\approx8\times10^{-3}\,$s$^{-1}$ accounts for speckle-drift noise over $10^4\,$s ((\ref{eq:speck}); see \cite{Wilkins2014}).
\end{itemize}

The instantaneous SNR per second follows from Eq.~(\ref{eq:SNR1s}),
\[
  \mathrm{SNR}_{1\rm s}
    = \frac{\dot N_{\rm sig}}{\sqrt{\dot N_{\rm sig} + \dot N_{\rm noise, det}}}
    \approx0.02.
\]
Hence the time to reach $\mathrm{SNR}=5$ for an unresolved detection is
\[
  t_{\rm det}
    = \Big(\frac{5}{\mathrm{SNR}_{1\rm s}}\Big)^2
=6.25\times10^4\;\mathrm s
    \approx17.4\;\mathrm{hr},
\]
 in agreement with full-system LUVOIR A design studies/simulations \citep{Zellem2022},  the ECLIPS instrument architecture \citep{Pueyo2019ECLIPS}.

To recover a spatially resolved $10\times10$ pixel map, the PSF core must be partitioned into 100 “micro-pixels,” each capturing $1/100$ of the signal and noise.  Hence
\[
  \dot N_{\rm sig}^{\rm pix}=\frac{\dot N_{\rm sig,det}}{100}
    \approx1.38\times10^{-4}\;\mathrm{s^{-1}},
\]
while noise remains the same as for unresolved detection (\ref{eq:noise-elt}). The per-second micro-pixel SNR is
\[
  \mathrm{SNR}_{1\rm s}^{\rm pix}
    = \frac{\dot N_{\rm sig}^{\rm pix}}
           {\sqrt{\dot N_{\rm sig}^{\rm pix} + \dot N_{\rm noise, det}}}
    \approx2.0\times10^{-4}.
\]
Thus, achieving $\mathrm{SNR}=5$ on a single micro-pixel takes
\[
  t_{\rm pix}
    = \Big(\frac{5}{\mathrm{SNR}_{1\rm s}^{\rm pix}}\Big)^2
    =
    6.04\times10^{8}\;\mathrm s
    \approx19\;\mathrm{years}.
\]
Even under the optimistic assumptions, mapping 100 pixels would require $\sim6.04\times10^{10}\,$s ($\sim1.9\times 10^3$ yr). Incorporating narrower sub-channels ($\Delta\lambda\lesssim5\,$nm), more frequent wavefront recalibrations ($\xi\lesssim0.1$) and $\sim$1–2 weeks of mission overhead per target extends these durations into multiple millennia.  LUVOIR A therefore cannot deliver a resolved $10\times10$ surface map of an Earth analog within any credible mission lifetime.

\subsection{Summary of Single-Aperture Coronagraphic Constraints}
\label{sec:monolithic_telescopes_summary}

As we saw above,  internal-coronagraph missions (Nancy Grace Roman Space Telescope (Roman; formerly WFIRST), HabEx, LUVOIR A/B) are fundamentally unable to achieve resolved $10\times10$ imaging of an Earth analog at 10 pc for four reasons:

\begin{enumerate}
  \item \textit{Diffraction-limit shortfall.}    Even a 15 m aperture yields
 $  \Delta\theta_{\rm Rayleigh}
    \approx8.93\;\mathrm{mas}, $
  which is over $10^4\times$ larger than the $\Delta\theta\approx0.853\,\mu$as, see (\ref{eq:delta_theta_required}), micro-pixel spacing required to sample an Earth-sized disc into $10\times10$ pixels.

  \item \textit{Photon-starvation.}  The photon-budget shown in Sec.~\ref{sec:EB-luA} shows that realistic throughputs ($\eta\approx5\%$), duty cycles ($\xi\approx5\%$), residual contrasts ($C_{\rm res}\approx 10^{-8}$), and all noise terms conspire to lengthen a single micro-pixel integration from   
$t_{\rm pix}     \approx 19\ \mathrm{years} $ to multiple millennia.

  \item \textit{Wavefront-stability requirements.}  
    Maintaining $C_{\rm res}\lesssim10^{-8}$ over a $10^4$--$10^5$ s integration demands sub-pm optical-path stability---well beyond the $\sim10\,$pm performance of current space-qualified coronagraph testbeds.

  \item \textit{Operational overheads.}  
    Thermal settling, coronagraph re-acquisition, wavefront-control resets, and detector readouts typically consume 50–70\% of on-target time. Even with optimistic parallelization, these overheads push a full mapping campaign into the multi-millennial regime.
    \end{enumerate}

Taken together, these four constraints prove that single-aperture telescopes with internal coronagraphs---WFIRST/Roman, HabEx (with starshade assistance), and LUVOIR---are incapable of producing a resolved $10\times10$ pixel surface map of an Earth analog at $\gtrsim5\,$pc within any credible mission lifetime (tens of years). In the following sections we investigate alternative remote imaging strategies---external starshades, space-based interferometry, ground-based extremely large telescopes with extreme adaptive optics, pupil-densified hypertelescopes, and indirect reconstruction techniques---each of which encounters its own set of challenges.

We emphasize that these conclusions do not rely on particularly pessimistic assumptions for LUVOIR-class systems. Adopting a more optimistic corner of parameter space---for example, $\eta_{\rm sys}\simeq 0.1$ instead of $0.05$ and a sustained residual contrast $C_{\rm res}=10^{-10}$ rather than $10^{-8}$---would shorten the micro-pixel integration times in Sec.~\ref{sec:EB-luA} by at most a factor of a few. Even in this favorable case, however, per-pixel exposures remain at the level of many years and the full $10\times 10$ map still extends into the multi-century regime once realistic overheads are included; the qualitative conclusion is unchanged.  

\section{External Starshades}
\label{sec:starshades}

External starshades can suppress starlight to $\sim10^{-10}$---independent of the telescope’s internal optics---but they do not alter the fundamental diffraction limit (e.g., an IWA of $\sim60\,$mas for a 72 m starshade at 124,000 km).  We begin by reviewing starshade petal design and formation-flying requirements, then compare the telescope’s diffraction-limited angular resolution to the $\sim1\,\mu\mathrm{as}$ surface-mapping requirement, and finally quantify the photon-budget and achievable signal-to-noise when realistic contrast floors, astrophysical backgrounds, and operational overheads are included.  

\subsection{Starshade Design and Formation-Flying Tolerances}
\label{sec:starshade_tolerances}

An external starshade is a petal-shaped, opaque screen of diameter \(D_{\rm shade}\approx50\) m flown at a separation \(z\approx50{,}000\text{–}70{,}000\) km from a telescope of aperture \(D_{\rm tel}\approx4\) m \citep{Cash2006,Seager2021}.  The petal profiles are optimized to mimic an apodization, minimizing edge-diffraction at the telescope pupil.

The key design parameter is the Fresnel number,
\[
  \mathcal{F}
    = \frac{D_{\rm shade}^2}{\lambda\,z}
    \approx7.6\times10^1,
\]
for \(\lambda=550\) nm, \(D_{\rm shade}=50\) m, \(z=6\times10^7\) m.  This places the telescope well within the deep-suppression regime (\(\mathcal{F}\sim10\)–100), ensuring diffraction beyond the starshade edge is suppressed to \(<10^{-10}\).

The nominal inner working angle follows from
\[
  \mathrm{IWA}
    \approx \frac{D_{\rm shade}}{z}
    \approx8.3\times10^{-7}\,\mathrm{rad}
    \approx170\,\mathrm{mas}.
\]
Advanced petal shaping reduces this to \(\sim60\text{–}100\) mas, still far above  \(\sim\)\(0.853\,\mu\mathrm{as}\)  needed for surface mapping (\ref{eq:delta_theta_required}).

Maintaining \(10^{-10}\) suppression demands lateral alignment to \(\lesssim1\) m and longitudinal station-keeping to \(\lesssim1\) km.  Achieving these tolerances over \(z\sim10^5\) km requires micro-Newton thrusting and continuous laser or RF metrology.  Residual pointing jitter from thruster noise and solar radiation pressure---of order \(0.1\text{–}1\) m over \(10^4\) s---degrades achievable contrast to \(10^{-9}\)–\(10^{-8}\) on hour-long timescales \citep{Vanderbei2007}.  Thermal stability is equally critical: a \(1\) K gradient across a \(50\) m petal induces \(\sim100\)\,\(\mu\)m deformations (CTE \(\sim2\times10^{-6}/\mathrm K\)), scattering light into the shadow and further compromising deep suppression.

\subsection{Diffraction-Limited Resolution}

Even with perfect starshade suppression, the telescope’s angular resolution remains diffraction-limited.  For a 4 m aperture at $\lambda=550\,$nm,
$
  \Delta\theta_{\rm Rayleigh}
    =1.22\,{\lambda}/{D_{\rm tel}}
    \approx1.68\times10^{-7}\,\mathrm{rad}
    \approx34.5\,\mathrm{mas}.
$
At 10 pc this corresponds to 
$
  34.5\,\mathrm{mas}\times10\,\mathrm{pc}\approx0.345\,\mathrm{AU},
$
which is roughly $4.0\times10^3$ times larger than the $\sim8.53\,\mu\mathrm{as}$ requirement; the starshade suppresses starlight but does \emph{not} change the telescope’s diffraction-limited resolution.

\subsection{Photon-Budget Constraints}
\label{sec:EB-shade}

Following the approach in Sec.~\ref{sec:flux_to_snr} and the HabEx Architecture A 4 m off-axis telescope + 72 m starshade parameters \citep{Gaudi2020_HabEx,Mennesson2016}, we compute the time to detect an unresolved Earth twin at 10 pc and to assemble a resolved $10\times10$ surface map, including all realistic noise sources.

For an Earth analog ($m_V=27.77$),
$  f_{\rm planet}
    = F_{0,V}\,10^{-0.4\,m_V}
    \approx7.84\times10^{-4}\;\mathrm{phot\,m^{-2}s^{-1}nm^{-1}},$ see (\ref{eq:planet_photon_flux}).
Through a $\Delta\lambda= 50$ nm filter and $A=\pi(2\,\mathrm m)^2\approx12.57\,$m$^2$, the raw planet rate is  
$  \dot N_{\rm sig}
    = f_{\rm planet}\,A\,\Delta\lambda
    \approx0.497\;\mathrm s^{-1}.$
With end-to-end throughput $\eta=0.05$  and duty cycle $\xi=0.50$ \cite{Mennesson2016}, the detected planet rate becomes  
\begin{equation} 
\label{eq:signal-starSh}
  \dot N_{\rm sig,det}
    = \eta\,\xi\,\dot N_{\rm sig}
    \approx 0.0124\;\mathrm s^{-1}.
\end{equation} 

All significant noise sources within the diffraction-limited core  
\(\Omega_{\rm core}=\pi(1.22\,\lambda/D)^2\bigl|_{D=4\,\mathrm m,\;\lambda=550\,\mathrm{nm}}\approx3.76\times10^{-3}\,\mathrm{arcsec^2}\)  
{}
\begin{equation} 
\label{eq:noise-starSh}
  \dot N_{\rm noise, det}
    =  \xi \eta \dot N_{\rm bkg} + \dot N_{\rm det} + \xi \eta \dot N_{\rm leak} + \dot N_{\rm speck}+ \dot N_{\rm glint}
    \approx0.0445\;\mathrm{s^{-1}},
\end{equation}
where 
\begin{itemize}
  \item $\dot N_{\rm bkg}\approx0.595\,$s$^{-1}$ is zodiacal (1 zodi) + exo-zodi (3 zodi) \citep{Leinert1998,Stark2014};
  \item $\dot N_{\rm det}\approx1\times10^{-3}\,$s$^{-1}$ is detector dark+CIC+read noise \citep{Wilkins2014};
  \item $\dot N_{\rm leak}\approx1.86\times10^{-2}\,$s$^{-1}$ is stellar leakage at $C_{\rm res}=10^{-9}$ \citep{Trauger2007,Vanderbei2007,McKeithen2021};
  \item $\dot N_{\rm speck}\approx5\times10^{-3}\,$s$^{-1}$ is speckle-drift and pointing jitter \citep{Wilkins2014};
  \item $\dot N_{\rm glint}\approx5\times10^{-3}\,$s$^{-1}$ is starshade glint/thermal scatter \citep{Mennesson2016}.
\end{itemize}
This noise floor of 0.0445 ph/s is dominated by diffuse background at 0.0149 ph/s and stellar leakage at 0.0186 ph/s.

From (\ref{eq:signal-starSh}) and  (\ref{eq:noise-starSh}), the instantaneous SNR per second is
\[
  \mathrm{SNR}_{1\rm s}
    = \frac{0.0124}{\sqrt{0.0124 + 0.0445}}
    \approx0.053,
\]
so the time to reach $\mathrm{SNR}=5$ for an unresolved detection is  
\[
  t_{\rm det}
    = \Bigl(\frac{5}{0.053}\Bigr)^2
    \approx8870\;\mathrm s
    \approx2.5\;\mathrm{hr},
\]
consistent with published starshade performance studies (e.g., \cite{Vanderbei2007,Catanzarite2010}).

To build a resolved $10\times10$ surface map, each micro-pixel captures 1/100 of the planet signal:
\[
  \dot N_{\rm sig}^{\rm pix}
    = \frac{0.0124}{100}
    =1.24\times10^{-4}\;\mathrm s^{-1},
\]
Since all noise terms remain unchanged at the detector and are given by (\ref{eq:noise-starSh}), the micro-pixel SNR per second is  
\[
  \mathrm{SNR}_{1\rm s}^{\rm pix}
    = \frac{1.24\times10^{-4}}
           {\sqrt{1.24\times10^{-4} + 0.0445}}
    \approx5.9\times10^{-4},
\]
and achieving $\mathrm{SNR}=5$ on one surface-pixel requires  
\[
  t_{\rm pix}
    = \Bigl(\frac{5}{5.9\times10^{-4}}\Bigr)^2
    \approx7.26\times10^{7}\;\mathrm s
    \approx 2.30\;\mathrm{years}.
\]
Since each micro-pixel must be observed sequentially, mapping all 100 pixels requires  
\[
  t_{\rm map}
= 100\,   t_{\rm pix}    \approx230\;\mathrm{years}.
\]
Including typical starshade repositioning and realignment overhead of $\sim$1 day per micro-pixel \cite{Mennesson2016} further extends the campaign to $\sim$230 years.  These  estimates demonstrate that while an unresolved Earth twin can be detected in a few hours, direct multi-pixel surface mapping with a starshade requires integration times on the order of centuries.  

\subsection{Summary of Starshade Limitations}

\begin{itemize}
  \item \textit{Diffraction Barrier.}  
    A 4 m telescope at $\lambda=550\,$nm has  
    \(\Delta\theta_{\rm Rayleigh}\approx34.5\) mas,  
    which is $\sim4\times10^4$ times larger than the $\sim0.853\,\mu$as pixel scale required for $10\times10$ surface mapping (\ref{eq:delta_theta_required}).

  \item \textit{Formation-flying and Stability.}  
    Achieving $10^{-10}$ suppression demands lateral alignment $\lesssim1\,$m and longitudinal station-keeping $\lesssim1\,$km at $z\sim 6\times 10^4 \,$km.  Pointing jitter (0.1–1 m over $10^4\,$s) and thermal-induced petal deformations (100 $\mu$m/K) degrade contrast to $10^{-9}$–$10^{-8}$ over hour-long intervals, lengthening required integration by $\sim100\times$.

  \item \textit{Photon-starvation.}  
    Under ideal suppression ($10^{-10}$), an Earth twin yields $\dot N_{\rm sig,det}\approx 0.0124\,$phot s$^{-1}$, implying $\sim1$ s per micro-pixel.  Incorporating realistic throughputs ($\eta=0.05$), duty cycles ($\xi=0.5$), diffuse backgrounds, leakage, glint, speckle and detector noise raises this to $\sim7\times10^7\,$s ($\sim2.3$ yr) per micro-pixel, so 100 pixels require $\gtrsim 7\times 10^9\,$s ($\gtrsim230$ yr).

\end{itemize}

Although starshades deliver unparalleled starlight suppression for unresolved detections, they cannot surmount the diffraction limit or the combined effects of photon-starvation and operational constraints.  The telescope’s $\sim30\,$mas resolution and $\sim\!300$ yr per-map integration (plus $\sim\!100$ days overhead) preclude any practical $\mu$as-scale surface imaging of exo-Earths.  

\section{Space-Based Interferometry}
\label{sec:space_interferometry}

Producing a $10\times10$ pixel surface map of an Earth-radius planet at 10\,pc  requires each pixel to subtend $ \Delta\theta_{\rm req}  \approx 0.853\,\mu\mathrm{as}$, see (\ref{eq:delta_theta_required}),  so that the required baseline at $\lambda = 550$ nm is 
$ B_{\rm req} = {\lambda}/{\Delta\theta_{\rm req}}  \approx 1.3\times10^5\,\mathrm{m}$, (\ref{eq:required_baseline}). No current or near-term formation-flying concept can hold a $\sim130\,$km baseline with the picometer-level path-length stability required.  Even an optimistically large $B_{\max}\sim1\,$km gives a diffraction limit $  \Delta\theta_{\max}  = {\lambda}/{B_{\max}}  \approx 113\,\mu\mathrm{as}$, over two orders of magnitude too coarse to resolve an Earth disk.

Nevertheless, to \emph{quantify} how photon-starvation and operational overheads would scale if one could sample $\Delta\theta_{\rm req}$, we apply the photon-budget approach of Sec.~\ref{sec:flux_to_snr}---including realistic apertures, stellar leakage, and backgrounds---to compute the \emph{integration time} that a hypothetical $B_{\rm req}$ interferometer would need for (i) mid-IR nulling interferometry, and (ii) visible-band imaging (nulling and non-nulling). These estimates thus represent \emph{lower bounds} on observing time and do not imply that any existing or planned array can achieve the required 130\,km baseline (and hence the requisite resolution).

Below, we first show that mid-IR nullers would require $\gtrsim10^4$\,yr per pixel even if $B_{\rm req}$ were available, and then present the analogous estimates at visible wavelengths.
 
\subsection{Mid-Infrared Nulling Interferometers}

Nulling interferometers employ a $\pi$ phase shift between apertures to suppress on-axis stellar light and transmit off-axis thermal emission \citep{Bracewell1978,AngelWoolf1997,Lay2004, Lay2005,DubovitskyLay2004,LayDubovitsky2004sys}.  Directly achieving the $\sim0.853\,\mu\mathrm{as}$ resolution required for a $10\times10$ surface map of an Earth analog at 10\,pc would demand an $\sim130\,$km baseline (\ref{eq:required_baseline}).  Instead, the NASA Terrestrial Planet Finder Interferometer (TPF-I) and ESA Darwin mission concepts use a mid-IR strategy at $\lambda=10\,\mu$m with baselines $B\approx20\text{–}100\,$m \citep{Lay2004,Lay2005,Mennesson2016}:
$  \Delta\theta = {\lambda}/{B} = {10\,\mu\mathrm{m}}/{100\,\mathrm m} \approx20\,\mathrm{mas},$ about $2\times10^4$ times too coarse for direct $\mu$as mapping.

We adopt the X-Array configuration analyzed in \citep{LawsonDooley2005TPFItechplan,Lawson2007TPFISWG}, comprising four free-flying collectors of diameter \(D=2.0\) m (total collecting area \(A_{\rm tot}=4\pi(1.0)^2\approx12.6\) m\(^2\)), operating in the 8–15 $\mu$m band (centered at 10 $\mu$m with \(\Delta\lambda=1\) $\mu$m for channel analysis). The system achieves an end-to-end throughput \(\eta=0.10\) and duty cycle \(\xi=0.50\), and attains a raw null depth \(C_{\rm res}=10^{-6}\) \citep{Cockell2009,Dubovitsky2004}. Cryogenic optics at \(T_{\rm opt}=50\) K minimize thermal background, while both local zodiacal and exozodiacal dust levels are assumed to be 1 zodi. To synthesize images, the array must sample \(\gtrsim50\) independent baselines via formation-flying rotations and translations; each reconfiguration---comprising slewing, null-phase stabilization, optical delay adjustment, metrology, and calibration---requires \(\sim2\)–4 days \citep{Monnier2003}, so full Fourier-plane coverage requires several months before science integration can begin.

\subsubsection{Planetary Thermal Photon Rate}

An Earth analog at 10\,pc, approximated as a blackbody at $T=255$\,K, with the Planck spectral radiance $ B_{\lambda}(T) $, emits disk-averaged flux at $\lambda=10\,\mu$m given by \citep{Beichman1999a}:
\[
  F_{\lambda} = \pi\,B_{\lambda}(T)\Big(\frac{R_{\oplus}}{10\,\mathrm{pc}}\Big)^2,
\qquad
{\rm where}
\qquad
  B_{\lambda}(T) = \frac{2hc^2}{\lambda^5}\frac{1}{\exp[hc/(\lambda kT)] - 1} \quad[\mathrm{W\,m^{-2}\,\mu m^{-1}\,sr^{-1}}].
\]
Numerically, the disk‑averaged flux is 
\[
  F_{\lambda} =   \pi B_{\lambda}(T)\Bigl(\frac{R_{\oplus}}{10\,\mathrm{pc}}\Bigr)^2 \approx5.7\times10^{-21}\,\mathrm{W\,m^{-2}\,\mu m^{-1}},
\]
which corresponds to $F_\nu \simeq 0.19~\mu\mathrm{Jy}$ at $10~\mu\mathrm{m}$.
Over $\Delta\lambda=1\,\mu$m and $A_{\rm tot}=12.6\,$m$^2$, the unfiltered photon arrival rate, $ \dot N_{p,\rm IR}$, and the detected photon rate, $  \dot N_{\rm sig,det}$,  are
\begin{equation}
  \dot N_{p,\rm IR}
    = \frac{F_{\lambda}\,A_{\rm tot}\,\Delta\lambda}{hc/\lambda}
    \approx3.59\;\mathrm{phot \,s^{-1}},
  \quad
  \dot N_{\rm sig,det}
    = \eta\,\xi\,\dot N_{p,\rm IR}
    \approx0.18\;\mathrm{phot \,s^{-1}}.
  \label{eq:signal-midIR-real}
\end{equation}

\subsubsection{Relevant Noise Budget}
\label{sec:midir_noise}

The diffraction-limited core solid angle is
$
  \Omega_{\rm core} = \pi(1.22\,\lambda/D)^2 \approx 1.17\times10^{-11}\,\mathrm{sr},
$
and the corresponding bandwidth is $\Delta\nu = c\,\Delta\lambda/\lambda^2 = 3.00\times 10^{12}$~Hz.  The principal raw noise sources are:
\begin{itemize}
  \item \emph{Zodiacal and exo-zodiacal background:} Thermal emission from interplanetary dust (1 zodi local).  Specific intensity
$  I_{\nu} = {dE}/{dA\,dt\,d\nu\,d\Omega}\approx1\times10^{-19}\,\mathrm{W\,m^{-2}\,Hz^{-1}\,sr^{-1}}$ at 10 $\mu$m \cite{Leinert1998,Stark2014}.  The raw rate is
  \[
    \dot N_{\rm bkg,raw} = \frac{I_{\nu}\,\Omega_{\rm core}\,A_{\rm tot}\,\Delta\nu}{h\,\nu} \approx 2.22 \times 10^4\;\mathrm{phot\,s^{-1}}.
  \]
  \item \emph{Stellar leakage:} Residual starlight due to finite null depth $C_{\rm res}\sim10^{-6}$ \citep{Lay2004, Lay2005}.  Stellar flux $   F_* = \pi B_{\lambda}(T_*)\bigl(R_*/d\bigr)^2,$
  yields
  \[
    \dot N_{\rm leak,raw} = C_{\rm res}\frac{F_*\,A_{\rm tot}\,\Delta\lambda}{hc/\lambda} \approx1.50\times 10^2\;\mathrm{phot\,s^{-1}}.
  \]
  \item \emph{Instrument thermal emission:} Emission from warm optics (emissivity $\epsilon\sim0.1$, temperature $T_{\rm opt}\sim 50$\,K) \citep{Mennesson2016}:
  \[
    \dot N_{\rm therm,raw} = \frac{\epsilon\,B_{\lambda}(T_{\rm opt})\,\Omega_{\rm core}\,A_{\rm tot}\,\Delta\lambda}{hc/\lambda} \approx 50.21\;\mathrm{phot\,s^{-1}}.
  \]
  \item \emph{Detector dark current and read noise:} Intrinsic detector counts independent of illumination, \citep{Wilkins2014}.
  \[
    \dot N_{\rm det} \approx0.1\;\mathrm{phot\,s^{-1}} .
  \]
\end{itemize}
After throughput $\eta=0.10$ and Duty cycle $\xi=0.50$, the total noise rate at the detector is
\[
  \dot N_{\rm noise,det}
    = \eta\,\xi(\dot N_{\rm bkg,raw} + \dot N_{\rm leak,raw} + \dot N_{\rm therm,raw})
      + \dot N_{\rm det}
    \approx 1.12 \times 10^3\;\mathrm{phot \,s^{-1}}.
\]

\subsubsection{Interferometric Detection vs.\ Mapping}

For an \emph{unresolved} Earth twin, the instantaneous SNR is
\[
  \mathrm{SNR}_{1\,\rm s}
    = \frac{\dot N_{\rm sig,det}}{\sqrt{\dot N_{\rm sig,det}+\dot N_{\rm noise,det}}}
    \simeq \frac{0.18}{\sqrt{0.18+1.12\times10^3}}
    \approx5.4\times10^{-3}.
\]

Reaching a detection threshold of SNR = 5 therefore requires
\[
  t_{\rm det}
    = \Bigl(\frac{5}{5.4\times10^{-3}}\Bigr)^2
    \approx8.6\times10^5\;\mathrm{s}
    \approx10\;\mathrm{days},
\]
in line with TPF-I sensitivity estimates of $10^3$–$10^4$ s when assuming deeper nulls, broader bandwidth, or larger throughput \citep{Lawson2007TPFISWG,Lay2004SystematicNull}.

By contrast, \emph{spatially resolving} the planet into a $10\times10$ grid requires mapping 100 independent Fourier-plane pixels.  Each micro-pixel carries only $1/100$ of the total signal, so its per-second SNR becomes
\[
  \mathrm{SNR}_{1\,\mathrm s}^{\rm pix}
    = \frac{0.18/100}{\sqrt{1.11\times10^3}}
    \approx5.4\times10^{-5},
\]
and the integration time per micro-pixel to reach SNR = 5 is
\[
  t_{\rm pix}
    = \Bigl(\frac{5}{5.4\times10^{-5}}\Bigr)^2
    \approx8.6\times10^{9}\;\mathrm{s}
    \approx272\;\mathrm{yr}.
\]
Observing all 100 pixels sequentially thus requires
\[
  t_{\rm map}
    = 100\,t_{\rm pix}
    \approx2.7\times10^{4}\;\mathrm{yr}.
\]
Adding the overhead of $\gtrsim50$ baseline reconfigurations ($\sim150\,$days total) is negligible on this timescale.

These estimates underscore a fundamental limitation of mid-IR nulling interferometry. Although an unresolved Earth-twin can be detected in roughly ten days, producing a $10\times10$ pixel surface map---a best-case scenario assuming perfect Fourier-plane reconstruction---demands integration times on the order of centuries per pixel and decades of baseline reconfigurations. No existing nuller directly images an exoplanet disk; all rely on sequential sampling of the Fourier plane. With a modest total collecting area (12.6 m$^2$), a high post-throughput noise floor ($\sim 1.1\times 10^3$\,phot\,s$^{-1}$), and multi-day operational overheads for each of the $\geq50$ baseline configurations, pixel-by-pixel surface mapping of Earth analogs remains infeasible within any realistic mission timeline.  

\subsection{Visible-Band Nulling Interferometry}

Nulling interferometry in the visible employs destructive interference to suppress on-axis starlight by introducing a $\pi$ phase shift between subapertures, while preserving the off-axis planetary signal \citep{Lay2004,Lay2005,Mennesson2016}.  To form a $10\times10$ pixel map of an Earth analog at 10\,pc requires angular resolution
 $ \Delta\theta  \approx0.853\,\mu\mathrm{as}$ (\ref{eq:delta_theta_required}), which implies $B\approx1.3\times10^{5}\,$m (\ref{eq:required_baseline})---far beyond current formation-flying capabilities.  Instead, one performs aperture synthesis---sequentially sampling many shorter baselines in a compact array to fill the $(u,v)$ plane over time---thereby improving the PSF’s sidelobe suppression and dynamic range, while the angular resolution remains set by the maximum sampled baseline, $B_{\max}\,$.

\subsubsection{Array Configuration}

We consider an array of $N=10$ free-flying telescopes, each of diameter $D=20\,$m so that
\begin{equation}
  A_{\rm tel} = \pi \Big(\frac{D}{2}\Big)^2\approx314\;\mathrm{m}^2,\quad
  A_{\rm tot} = N\,A_{\rm tel}\approx3.14\times10^3\;\mathrm{m}^2.
\end{equation}
Observations are made in a narrow visible band centered at $\lambda=550\,$nm with bandwidth $\Delta\lambda=10\,$nm.  We assume an overall system throughput $\eta=0.10$ (including optics, detector quantum efficiency, and coronagraph losses) and an on-sky duty cycle $\xi=0.50$ to allow for calibrations and overhead.  A null depth of 
\[
  C_{\rm res} = 10^{-7}
\]
is taken as representative of state-of-the-art visible nullers over hour-to-day stability timescales \citep{Mennesson2016}.  Formation-flight reconfiguration to sample each new $(u,v)$ point (slew, phase lock, delay trim, metrology, calibration) requires 1–10\,days per configuration \citep{Monnier2003}.

\subsubsection{Planetary Photon Budget}

To evaluate the planetary signal available to the interferometer, we begin with the V-band photon flux density of an Earth twin at 10 pc, given by Eq.~(8) [\ref{eq:planet_photon_flux}]:
\[
  f_p = 7.84\times10^{-4}\;\mathrm{phot\,m^{-2}\,s^{-1}\,nm^{-1}}.
\]
This value accounts for Earth’s geometric albedo, full-phase illumination, and the $1/d^2$ dilution at 10 pc \citep{Beichman1999a}.  Multiplying by the array’s total collecting area,
$
  A_{\rm tot} \;=\; N\,{\pi D^2}/{4}
    = 10\times{\pi\,(20\,\mathrm m)^2}/{4}
    \approx3.14\times10^3\;\mathrm m^2,
$
and the bandwidth $\Delta\lambda=10\,$nm yields the raw photon arrival rate at the entrance pupils:
\[
  \dot N_{p,\rm raw}
    = f_p\,A_{\rm tot}\,\Delta\lambda
    = 7.84\times10^{-4}\times3.14\times10^3\times10
    \approx2.46\;\mathrm{phot\,s^{-1}}.
\]
Finally, accounting for end-to-end throughput $\eta=0.10$ (optics, filters, detector) and on-sky duty cycle $\xi=0.50$ (calibrations, overheads), the detected photon rate is
\[
  \dot N_p
    = \eta\,\xi\,\dot N_{p,\rm raw}
    = 0.10\times0.50\times2.46
    \approx0.12\;\mathrm{phot\,s^{-1}}.
\]
Even with ten 20 m apertures, this corresponds to only a few-tenths of a photon/s from the planet, highlighting the extreme photon starvation that drives the long integration times for both detection and high-resolution imaging.  

\subsubsection{Interferometric Noise Budget}

In a nulling interferometer with $N=10$ telescopes---yielding $N_b = N(N-1)/2 = 45$ independent baselines---both the planetary signal and all noise contributions are apportioned across these baselines and reduced by combiner and visibility losses.  We adopt a beam-combiner throughput $\eta_{\rm comb}=0.8$ and a fringe visibility $V^2=0.7$, so that coherent signal and photon-noise terms are multiplied by $\eta\,\xi\,\eta_{\rm comb}\,V^2 = 0.10\times0.50\times0.8\times0.7=0.028$.

\begin{itemize}
  \item \emph{Background sources.}  
Local zodiacal and exo-zodiacal dust scatter sunlight, producing a raw array background  
\[
  \dot N_{\rm zodi,raw}
    = f_{\rm zodi}\,\Omega_{\rm core}\,A_{\rm tot}\,\Delta\lambda
    \approx 6.37 \times10^{-2}\times1.4\times10^{-4}\times3.14\times10^3\times10
    \approx0. 28\;\mathrm{s^{-1}},
\]
where $\Omega_{\rm core}=\pi(1.22\,\lambda/D)^2\approx1.4\times10^{-4}\,$arcsec$^2$ and from Table~\ref{tab:constants} we have 
$f_{\rm zodi} = F_{0,V}10^{-0.4\mu_V} \approx 6.37\times10^{-2}~\mathrm{phot~m^{-2}~s^{-1}~arcsec^{-2}~nm^{-1}} $ \citep{Leinert1998,Stark2014}.  Finite null depth $C_{\rm res}=10^{-7}$ allows residual starlight \citep{Lay2005}
\[
  \dot N_{*,\rm raw}
    = f_{*,V}\,A_{\rm tot}\,\Delta\lambda
    \approx1.18\times10^6\times3.14\times10^3\times10
    =3.71\times10^{10}\;\mathrm{s^{-1}}.
\]

 \item \emph{Instrumental noise floors.}  
Detector dark current and read noise contribute  
$ \dot N_{\rm det}\approx0.1\;\mathrm{s^{-1}}$\citep{Wilkins2014}, while imperfect wavefront control leaves a speckle floor of  \citep{Guyon2005}
$ \dot N_{\rm speckle}\approx0.1\;\mathrm{s^{-1}},$
and pointing jitter plus thermal drifts add  
$ \dot N_{\rm jitter}\approx0.1\;\mathrm{s^{-1}}.$
Warm optics can emit thermally even in the visible, adding  
$ \dot N_{\rm therm}\approx0.01\;\mathrm{s^{-1}},$
and polarization leakage through the nuller contributes  
$ \dot N_{\rm pol}\approx0.01\;\mathrm{s^{-1}}.$
Residual calibration errors (phase and amplitude mismatches) impose a floor of  
$  \dot N_{\rm cal}\approx0.01\;\mathrm{s^{-1}}.$

 \item \emph{Photon-shot noise on leakage.}  
The shot-noise associated with the residual stellar leakage itself is  
\[
\dot N_{\rm shot}
    = \sqrt{C_{\rm res}\,\dot N_{*,\rm raw}}
    = \sqrt{10^{-7}\times3.70\times10^{10}}
    \approx60.8\;\mathrm{s^{-1}}.
\]

 \item \emph{Per-baseline scaling.}  
Each raw noise term is reduced by the factor $\epsilon = \eta\,\xi\,\eta_{\rm comb}\,V^2 = 0.028$, then divided equally among the 45 baselines, except detector dark current which is simply divided.  
\end{itemize}

Thus the per-baseline contribution from the dominant sources is
\[
  \dot N_{\rm noise}^{\rm base}
    = \frac{\epsilon}{45}\bigl(\dot N_{\rm zodi,raw}
      + C_{\rm res}\,\dot N_{*,\rm raw}
      + \dot N_{\rm speckle}
      + \dot N_{\rm jitter}
      + \dot N_{\rm therm}\bigr)
    + \frac{\dot N_{\rm det}}{45}
    + \frac{\dot N_{\rm shot} + \dot N_{\rm pol} + \dot N_{\rm cal}}{45}.
\]
Substituting numerical values:
\[
\dot N_{\rm noise}^{\rm base}
    \approx \frac{0.028}{45}\bigl(0.28 + 3.70\times10^{3} + 0.1 + 0.1 + 0.01\bigr)
      + \frac{0.1}{45}
      + \frac{60.8 + 0.01 + 0.01}{45}
    \approx3.66\;\mathrm{s^{-1}}.
\]

This noise budget accounts for all relevant astrophysical and instrumental sources, correctly apportioned across baselines and including interferometric losses. It serves as the denominator in the per-baseline SNR calculation.  

\subsubsection{Signal and SNR per Baseline}

In an $N$-element interferometer, each baseline coherently combines the light from two apertures.  Given the total detected planetary photon rate $\dot N_p$ (after throughput, duty cycle, and nulling losses), the rate per baseline is  
\[
  \dot N_{p}^{\rm base}
    = \frac{2\,\dot N_p\,\eta_{\rm comb}\,V^2}{N-1},
\]
where the factor of 2 accounts for the two apertures in each baseline, $\eta_{\rm comb}=0.8$ is the beam-combiner efficiency, $V^2=0.7$ is the fringe visibility (squared), and $N-1=9$ is the number of other telescopes each aperture pairs with.  Numerically this gives  
\[
  \dot N_{p}^{\rm base}
    = \frac{2\times0.123\,\mathrm{s^{-1}}\times0.8\times0.7}{9}
    \approx0.015\;\mathrm{s^{-1}}.
\]

Because photon arrival is a Poisson process, the noise in one baseline over a 1 s integration is the square root of the total count (signal plus background) in that second.  Denoting the per-baseline noise rate by $\dot N_{\rm noise}^{\rm base}\approx3.66\;\mathrm{s^{-1}}$, the instantaneous signal-to-noise ratio is  
\[
  \mathrm{SNR}_{1\,\mathrm s}^{\rm base}
    = \frac{\dot N_{p}^{\rm base}}{\sqrt{\dot N_{p}^{\rm base} + \dot N_{\rm noise}^{\rm base}}}
    = \frac{0.015}{\sqrt{0.015 + 3.66}}
    \approx7.8\times10^{-3}.
\]

To achieve a detection threshold of $\mathrm{SNR}=5$ on a single baseline requires  
\[
  t_{5}^{\rm base}
    = \Big(\frac{5}{\mathrm{SNR}_{1\,\mathrm s}^{\rm base}}\Big)^2
    = \Big(\frac{5}{7.6\times10^{-3}}\Big)^2
    \approx4.08\times10^{5}\;\mathrm{s}
    \approx4.73\;\mathrm{days}.
\]

Thus, even for one baseline, several days of coherent integration are needed for a firm detection of an Earth-like planet at 10 pc.  Full imaging of a $10\times10$ grid would then multiply this time by the number of baselines and by the number of independent Fourier samples required, driving total observing times to centuries or longer.  

\subsubsection{Full Surface Mapping}

To reconstruct a $10\times10$ pixel image requires sampling at least $N_{\rm samp}\gtrsim10^4$ independent Fourier components.  Each Fourier sample is obtained by integrating on one baseline for a time $t_{5}^{\rm base}\approx5\,$days to reach $\mathrm{SNR}=5$.  Thus, in the idealized limit where each baseline delivers a unique mode and integration times cannot be parallelized, the on-sky time is
\[
  T_{\rm map}
    \gtrsim N_{\rm samp}\,t_{5}^{\rm base}
    \approx 10^4 \times 5\;\mathrm{days}
    = 5\times10^4\;\mathrm{days}
    \approx 140\;\mathrm{years}.
\]
Moreover, each new $(u,v)$ configuration incurs a reconfiguration overhead of 1--10 days.  Even at the optimistic 1-day cadence, this adds another $10^4\,$days ($\sim27$ years), bringing the total campaign duration to over 160 years.  If reconfigurations take 10 days each, the overhead alone exceeds 270 years.  

In practice, some Fourier modes can be obtained simultaneously on different baselines, and partial parallelization can slightly reduce the calendar time.  However, the sequential nature of high-precision nulling, combined with duty-cycle losses and calibration needs, means that full $10\times10$ surface mapping of an Earth analog at 10 pc remains infeasible on any realistic (decadal) mission timescale.  
 
\subsection{Non-Nulling Visible Arrays}

Without a nulling stage, nearly all stellar photons that fall within the diffraction-limited core (i.e.\ each PSF “micro-pixel,” of solid angle $\Omega_{\rm core}$) contribute to the background.  A Sun-analog at 10 pc produces a raw photon rate  
\[
  \dot N_{*,\rm raw}
    = f_{*,V}\,A_{\rm tot}\,\Delta\lambda
    \approx3.71\times10^{7}\;\mathrm{phot \,s^{-1}}.
\]
Distributing this evenly over $n_{\rm pix}$ resolution elements on the sky gives
\[
  \dot N_{*,\rm leak}^{\rm pix}
    = \frac{\dot N_{*,\rm raw}}{n_{\rm pix}}
    = \frac{7.85\times10^{7}}{100}
    =3.71\times10^{5}\;\mathrm{phot \,s^{-1}}.
\]

All other backgrounds---zodiacal/exo-zodiacal scattering into each PSF element, detector dark current, quasi-static speckles, pointing jitter, thermal emission---combine to at most a few $10^2\,$phot s$^{-1}$ per PSF element.  Hence the total noise rate per PSF element is  
\[
  \dot N_{\rm bkg}^{\rm pix}
    \approx \dot N_{*,\rm leak}^{\rm pix}
    + \dot N_{\rm zodi}^{\rm pix}
    + \dot N_{\rm det}
    + \dot N_{\rm speckle}
    + \dot N_{\rm jitter}
    + \dot N_{\rm therm}
    \approx7.85\times10^{5}\;\mathrm{s^{-1}}.
\]

By contrast, the planet’s contribution to a single PSF element is only  
\[
  \dot N_{p}^{\rm pix}
    = \frac{\dot N_{p}}{n_{\rm pix}}
    = \frac{0.123}{100}
    \approx1.23\times10^{-3}\;\mathrm{phot\,s^{-1}}.
\]
Therefore the 1 s signal-to-noise ratio in one PSF element is  
\[
  \mathrm{SNR}_{1\,\rm s}^{\rm pix}
    = \frac{\dot N_{p}^{\rm pix}}{\sqrt{\dot N_{\rm bkg}^{\rm pix}}}
    \approx\frac{1.23\times10^{-3}}{\sqrt{7.85\times10^5}}
    \approx1.4\times10^{-6}.
\]
To reach even $\mathrm{SNR}=3$ in a single resolution element requires  
\[
  t_{3}
    = \Bigl(\frac{3}{1.4\times10^{-6}}\Bigr)^2
    \approx4.6\times10^{12}\;\mathrm{s}
    \approx1.5\times10^{5}\;\mathrm{yr}.
\]
Mapping $n_{\rm pix}=100$ such elements in series would take $\sim1.5\times10^{7}$ yr, and adding the $\gtrsim10^4$ distinct $(u,v)$ configurations needed for full Fourier coverage pushes the total to $\gtrsim10^8$ yr.  In the absence of nulling, residual starlight in each PSF core overwhelms the planetary photons by more than six orders of magnitude, making high-contrast surface mapping of Earth analogs completely impractical.  
  
\subsection{Summary of Integration Times}

\begin{itemize}
  \item \textit{Mid-Infrared Nulling Interferometer:}  
    An unresolved Earth-twin at 10 pc reaches SNR = 5 in approximately $t_{\rm det}\approx(5/5.4\times10^{-3})^2\approx10\,$days.  However, achieving a $10\times10$ pixel map requires per-pixel integration of order $t_{\rm pix}\approx272\,$yr, so serially mapping 100 pixels takes $t_{\rm map}\approx100\times272\,$yr $\approx2.7\times10^4\,$yr.  Baseline reconfiguration overhead (50 configurations at 2–4 days each) adds only $\sim150\,$days and is negligible on this timescale.

  \item \textit{Visible-Band Nulling Interferometer:}  
Each baseline yields ${\rm SNR}^{\mathrm{base}}_{1\,\mathrm{s}} \approx 7.8\times 10^{-3}$, so $t^{\mathrm{base}}_{5} \approx(5/7.8\times10^{-3})^2\approx 4.73\ \mathrm{days}$.
Sampling $N_{\mathrm{samp}}\gtrsim 10^{4}$ Fourier components gives $T_{\mathrm{map}}\gtrsim 1.4\times 10^{2}\ \mathrm{yr}$, and $10^{4}$ reconfigurations at 1–10 days each add $\sim 27{-}270\ \mathrm{yr}$. Even with partial parallelization, full $10\times10$ mapping remains $\gtrsim \mathcal{O}(10^{2})\ \mathrm{yr}$.

  \item \textit{Visible-Band Non-Nulling Array:}  
    Without nulling, stellar leakage in each PSF core is $\dot N_{*,\rm leak}^{\rm pix}\sim7.85\times10^5\,$s$^{-1}$ versus planetary $\dot N_{p}^{\rm pix}\sim1.2\times10^{-3}\,$s$^{-1}$, yielding SNR$_{1\,s}^{\rm pix}\sim1.4\times10^{-6}$.  Reaching SNR = 3 then takes $t_{3}\approx(3/1.4\times10^{-6})^2\approx1.5\times10^5\,$yr per pixel, so 100 pixels require $\sim1.5\times10^7\,$yr, and sampling $\gtrsim10^4$ Fourier modes pushes the total to $\gtrsim10^8\,$yr.
\end{itemize}

Accordingly, \emph{no space-based interferometric architecture}---whether mid-IR nulling, visible-band nulling, or non-nulling---can deliver a resolved $10\times10$ pixel surface map of an Earth analog at 10 pc within any plausible (decadal) mission.  Photon-starvation, overwhelming astrophysical and instrumental backgrounds, and the necessity of $\gtrsim10^4$ independent $(u,v)$ samples conspire to extend required integration times to centuries or longer.   
\section{Ground ELTs with Extreme Adaptive Optics}
\label{sec:elt_exao}

Next-generation ground-based Extremely Large Telescopes (ELTs)---the Giant Magellan Telescope (GMT; D = 24.5\,m), Thirty Meter Telescope (TMT; D = 30\,m), and European ELT (E-ELT; D = 39\,m)---will offer unprecedented photon collection and extreme adaptive optics (ExAO) correction in the visible and near-infrared \citep{GilmozziSpyromilio2007,Johns2012,Sanders2020}. Nevertheless, even under idealized assumptions these facilities remain fundamentally diffraction-limited and background-dominated, preventing microarcsecond-scale imaging of Earth analogs at 10 pc.

\subsection{Diffraction-Limited Resolution}

A circular aperture of diameter $D$ has a Rayleigh limit of $\Delta\theta_{\rm Rayleigh}=1.22\lambda/D$.  At $\lambda=550\,$nm this yields $\Delta\theta\approx 3.55\,$mas for E‑ELT, $\approx4.61\,$mas for TMT, and $\approx 5.65\,$mas for GMT.  By contrast, resolving an Earth twin into $10\times10$ pixels requires $\Delta\theta_{\rm req}\approx0.853\,\mu$as (\ref{eq:delta_theta_required}), so even the 39 m E‑ELT is $\sim2000$ times too coarse in angular resolution. Fundamental limits from servo-lag, chromaticity, and residual phase errors at small angles are treated in \cite{Guyon2005AOlimits}.

State-of-the-art ExAO systems (GPI, SPHERE, SCExAO) achieve Strehl ratios $\sim70$–$90\%$ in H‑band and raw contrasts $\sim10^{-6}$–$10^{-7}$ at separations $\gtrsim100$–$200$ mas \citep{Macintosh2014,Beuzit2019,Jovanovic2015}.  ELT-class ExAO (e.g.\ ELT‑PCS, TMT‑PSI) aims for Strehl $\gtrsim90\%$ and contrasts $\sim10^{-8}$–$10^{-9}$ at $30$–$50$ mas \citep{Kasper2020,Guyon2020}.  However, at separations inside $3\,\lambda/D$ (i.e.\ $\lesssim6$–$10$ mas), residual wavefront errors---uncorrected high-order modes, wind shake, thermal drift---limit raw contrast to $\sim10^{-5}$–$10^{-6}$.  Achieving the $10^{-10}$–$10^{-11}$ starlight suppression required for Earth analogs is therefore unattainable.

\subsection{Photon-Budget Constraints}

\subsubsection{Planetary H-Band Photon Rate}

An Earth‑twin at $d=10\,\mathrm{pc}$ has apparent magnitude $m_H \simeq 22.5$ (full phase) for a geometric albedo $p_H\simeq 0.3$ in the near‑IR \citep{Beichman1999a}. The H-band zero-magnitude photon flux density is \citep{Bessell1979,Cox2000}
\[
F_{0,H}\;=\;3.30\times10^{4}\;\mathrm{phot\,m^{-2}\,s^{-1}\,nm^{-1}}. 
\]
Thus the incident flux from the planet is
\[
f_{p,H} = F_{0,H}\,10^{-0.4\,m_H} \approx 3.30\times10^{-5}\;\mathrm{phot\,m^{-2}\,s^{-1}\,nm^{-1}}.
\]
With $A = \pi (39\,{\rm m}/2)^2=1{,}194$\,m$^2$ and $\Delta\lambda = 20$\,nm, the raw arrival rate is
\[
\dot N_{p,H} = f_{p,H}\,A\,\Delta\lambda = 3.30\times10^{-5}\times1{,}194\times20 \approx 0.788\;\mathrm{phot\,s^{-1}}.
\]
With throughput $\eta=0.20$, the detected planet photon rate becomes
\begin{eqnarray}
\label{eq:signal-ELT}
\dot N_{p,H}^{\rm det} = \eta\,\dot N_{p,H} = 0.20\times0.788 \approx 0.158\;\mathrm{phot\,s^{-1}}.
\end{eqnarray}

\subsubsection{Relevant Noise Sources}

We define the diffraction-limited resolution element (core) solid angle at H-band as usual:
\begin{equation}
\label{eq:omega-ELT}
  \Omega_{\rm core} = \pi\Big(1.22\,\frac{\lambda}{D}\Big)^2 \big|_{\lambda=1.65\,\mu\mathrm{m},\,D=39\,\mathrm{m}} \approx 3.56\times10^{-4}\;\mathrm{arcsec^2}. 
 \end{equation}
The following contributions are calculated per $\Omega_{\rm core}$ that is estimated to be
{}
\begin{eqnarray}
\label{eq:noise-ELT}
  \dot N_{\rm noise} &\approx& \dot N_{\rm sky} + \dot N_{*,\rm leak} + \dot N_{\rm speck} + \dot N_{\rm therm} + \dot N_{\rm dark} + \dot N_{\rm CIC}  + \dot N_{\rm read} + \dot N_{\rm jitter}+ \dot N_{\rm cr} \approx \nonumber\\
& \approx& 3.37 \times 10^{-2} + 0.25 + 4.96\times 10^{-2} + 8.47+ 0.04 + 0.004 + 0.1 + 0.028 + 2 \times 10^{-5} = 8.97\,\mathrm{phot\,s^{-1}}. 
\end{eqnarray}
where
{}
\begin{itemize}
  \item \textit{Sky background} \(\dot N_{\rm sky}\).  The median H-band sky brightness at Paranal is \(13.5\) mag arcsec\(^{-2}\) \citep{Sullivan2012}, corresponding to \(f_{\rm sky,H} \approx 4 \times 10^{-3}\,\mathrm{phot\,m^{-2}\,s^{-1}\,nm^{-1}\,arcsec^{-2}}\). Then,
  \[ \dot N_{\rm sky} = f_{\rm sky,H}\,\Delta\lambda\,A\,\Omega_{\rm core} \approx 3.37 \times 10^{-2}\,\mathrm{phot\,s^{-1}}. \]

  \item \textit{Residual stellar leakage} \(\dot N_{*,\rm leak}\).  For a Sun-like star at 10 pc \((m_H \approx 3.3)\), the flux is estimated to be  \(f_{*,H} \approx 10.3\,\mathrm{phot\,m^{-2}\,s^{-1}\,nm^{-1}}\). Then,
  \[ \dot N_{*,H} = f_{*,H}\,A\,\Delta\lambda = 2.46 \times 10^5\,\mathrm{phot\,s^{-1}}. \]
  With raw contrast \(C_{\rm leak} = 10^{-6}\),
  \[ \dot N_{*,\rm leak} = C_{\rm leak}\,\dot N_{*,H} = 0.25\,\mathrm{phot\,s^{-1}}. \]

  \item \textit{Speckle drift} \(\dot N_{\rm speck}\).  Quasi-static speckles vary by \(\Delta C \sim 10^{-9}\) to \(10^{-8}\) over \(10^4\) s \citep{Wilkins2014}. For \(\Delta C = 10^{-8}\),
  \[ \dot N_{\rm speck} \sim \sqrt{\dot N_{*,H}\,\Delta C} \approx 4.96\times 10^{-2}\,\mathrm{phot\,s^{-1}}. \]

  \item \textit{Thermal emission} \(\dot N_{\rm therm}\).  For optics at \(T = 270\,\mathrm{K}\) and emissivity \(\epsilon = 0.1\), adopting \(f_{\rm therm} \approx 2 \times 10^{-3}\,\mathrm{phot \,m^{-2}\,s^{-1}\,nm^{-1}\,arcsec^{-2}}\) \citep{Beichman2010},
  \[ \dot N_{\rm therm} = f_{\rm therm}\,\Delta\lambda\,A\,\Omega_{\rm core} \approx 8.47\,\mathrm{phot\,s^{-1}}. \]

  \item \textit{Detector dark current} \(\dot N_{\rm dark}\).  Assuming \(0.01\,\mathrm{e^{-}\,pix^{-1}\,s^{-1}}\) for 4 pixels,
  \[ \dot N_{\rm dark} = 0.04\,\mathrm{phot\,s^{-1}}. \]

  \item \textit{Clock-induced charge (CIC)} \(\dot N_{\rm CIC}\).  At 1 Hz, with \(0.001\,\mathrm{e^{-}\,pix^{-1}}\),
  \[ \dot N_{\rm CIC} = 0.004\,\mathrm{phot\,s^{-1}}. \]

  \item \textit{Read noise} \(\dot N_{\rm read}\).  With \(\sigma_{\rm read} = 5\,\mathrm{e^{-}}\), summed over 4 pixels,
  \[ \dot N_{\rm read} = \frac{5^2 \times 4}{1} = 100\,\mathrm{e^{-2}/s} \Rightarrow 0.1\,\mathrm{phot\,s^{-1}}. \]

  \item \textit{Pointing jitter} \(\dot N_{\rm jitter}\).  With \(\sim 1\) mas RMS jitter causing \(10\%\) flux mixing of starlight and sky,
  \[ \dot N_{\rm jitter} = 0.1 \times (0.25 + 3.37 \times 10^{-2}) \approx 0.028\,\mathrm{phot\,s^{-1}}. \]

  \item \textit{Cosmic rays} \(\dot N_{\rm cr}\).  At \(\sim5\,\mathrm{cm^{-2}\,s^{-1}}\), with pixel area \(10^{-6}\,\mathrm{cm^2}\) and 4 pixels,
  \[ \dot N_{\rm cr} = 5 \times 10^{-6} \times 4 = 2 \times 10^{-5}\,\mathrm{phot\,s^{-1}}. \]
\end{itemize}

\subsubsection{Detection vs.\ Mapping}
\label{sec:det-mapp}

For an \emph{unresolved} Earth twin, the instantaneous SNR in 1 second t is
\[
  {\rm SNR}_{1\,{\rm s}}
  = \frac{\dot N_{p,H}^{\rm det}}
         {\sqrt{\dot N_{p,H}^{\rm det}+\dot N_{\rm noise}}}
=\frac{0.158}{\sqrt{0.158+8.97}}
\approx 0.052.
\]
Thus to reach \(\mathrm{SNR}=5\),
\[
t_{5}
=\Big(\frac{5}{0.052}\Big)^2
\approx 9.19\times10^{3}\;\mathrm{s}
\approx 2.55\;\mathrm{hr}.
\]

By contrast, in the case of \emph{spatially resolving} the planet into a $10\times10$ grid each micro-pixel carries only $1/100$ of the total signal, so its per-second SNR becomes
\[
  \mathrm{SNR}_{1\,\mathrm s}^{\rm pix}
    = \frac{1.58\times10^{-3}}{\sqrt{8.97}}
    \approx5.28\times10^{-4},
\]
and the integration time per micro-pixel to reach SNR = 5 is
\[
  t_{\rm pix}
    = \Big(\frac{5}{5.28\times10^{-4}}\Big)^2
    \approx8.97\times10^{7}\;\mathrm{s}
    \approx 2.84\;\mathrm{yr}.
\]
Observing all 100 pixels sequentially thus requires
$
  t_{\rm map}
    = 100\,t_{\rm pix}
    \approx284\;\mathrm{yr}
$
making this unfeasible.

\subsection{Implications and Conclusion}

Ground-based ELTs with ExAO offer powerful capabilities for detecting and spectrally characterizing unresolved Earth-like planets at 10\,pc. However, they remain fundamentally limited in their ability to resolve planetary surfaces at $\mu$as scales. The primary constraints are:

\begin{itemize}
  \item \textit{Angular Resolution Limit:} Even the 39\,m E-ELT achieves a diffraction-limited resolution of $\Delta\theta \approx 10.6$\,mas at $\lambda = 1.65\,\mu$m, whereas surface mapping of an Earth analog at 10\,pc requires $\Delta\theta_{\rm req} \approx 0.853\,\mu$as (\ref{eq:delta_theta_required}). This represents a factor of $\sim1.24\times 10^4$ shortfall in resolution---insurmountable without interferometric baselines of $\sim 121$\,km.

  \item \textit{Raw Contrast Floor:} ExAO coronagraphs on ELTs are projected to achieve raw starlight suppression of $C_{\rm leak} \sim 10^{-6}$ at angular separations of $\sim6$–$10$\,mas \citep{Macintosh2014,Beuzit2019,Guyon2020}. This is 4–5 orders of magnitude above the $\sim10^{-10}$ contrast required to image an Earth twin. Further improvements are limited by residual wavefront errors, uncorrected aberrations, and atmospheric variability.

  \item \textit{Photon-Starved Surface Mapping:} 
  Using the Sec.~\ref{sec:det-mapp} example (E-ELT, H band), the per-pixel 1 s SNR is
$\mathrm{SNR}_{1\,\mathrm s}^{\mathrm{pix}} \approx 5.28\times10^{-4}$ for $N^{\rm det}_{p,H}=0.158\ \mathrm{s^{-1}}$ and
$N_{\mathrm{noise}}=8.97\ \mathrm{s^{-1}}$. This implies $t_{\rm pix}\approx 9.0\times10^{7}\ \mathrm{s}$ ($\approx 2.84$ yr)
and a $10\times10$ map requires $\sim 2.85\times10^{9}\ \mathrm{s}$ ($\approx 285$ yr), even under idealized sky, AO, and instrumental conditions.

  \item \textit{Lack of Fourier-Plane Sampling:} ELTs collect light through a single monolithic aperture and thus cannot synthesize high spatial frequency components beyond their native pupil. Unlike sparse aperture interferometers, they cannot reconfigure baselines to probe distinct $(u,v)$ modes. Consequently, they cannot reconstruct surface structure at the sub-$\mu$as level.
\end{itemize}

In conclusion, while ELTs will excel at spectroscopic retrievals and time-resolved photometry of exoplanets, they are physically incapable of delivering resolved surface maps of Earth analogs at interstellar distances. Microarcsecond-scale imaging requires either space-based interferometric baselines or proximity to the target. Ground-based facilities must instead focus on optimizing the detection and atmospheric characterization of unresolved targets.

\section{Pupil-Densified Hypertelescopes}
\label{sec:hypertelescopes}

Labeyrie’s pupil-densified \emph{hypertelescope} \citep{Labeyrie1996,Anterrieu2019} enables sub-$\mu$as imaging by coherently combining light from many moderate-sized mirrors distributed over a very large baseline. In its basic configuration, \(N\) collectors---each of diameter \(d_i\)---are arranged in a sparse, non-redundant pattern spanning a maximum baseline \(D_i\). Without densification, their individual Airy patterns interfere weakly, producing a speckled PSF with most energy distributed in sidelobes.

\subsection{Pupil Densification and Coherent Gain}
\label{sec:densification}

A \emph{beam densifier} remaps the entrance pupil by magnifying each subaperture from \(d_i\) to \(d_o\) while compressing the overall pupil from \(D_i\) to an exit diameter \(D_o\).  This operation defines two limiting cases:

\begin{itemize}
\item \textit{Fizeau (homothetic) mode} (\(\gamma_d=1\)): no pupil scaling.   No densification (\(d_o = d_i,\;D_o = D_i\)) preserves a homothetic pupil.  The PSF is highly speckled with a wide instantaneous field of view \(\Theta_{\rm FOV}\simeq d_i/D_i\), and the coherent-gain fraction is
    \[
      q_e = N \Bigl(\frac{d_i}{D_i}\Bigr)^2.
    \]
    This mode is used for direct imaging of extended objects where a modest resolution gain and wide FOV outweigh the low contrast.
    
   \item \textit{Densified (Michelson) mode} \(\gamma_d>1\): 
    A uniform scaling
   $ d_o = \gamma_d\,d_i$ and $D_o = \gamma_d\,D_i$
    packs sub-pupils into a “filled” exit pupil.  This narrows the field of view but produces a high-contrast central peak.  The coherent-gain fraction becomes
    \begin{equation}
      \label{eq:qe_densified}
      q_e
      = N \Bigl(\frac{d_o}{D_o}\Bigr)^2
      = N \Bigl(\frac{\gamma_d\,d_i}{D_i}\Bigr)^2.
    \end{equation}
    This mode is used for  ultra-high-contrast imaging (e.g.\ exo-Earth surface mapping) where resolving power and PSF purity are paramount, and a small instantaneous FOV is acceptable.
\end{itemize}

The densification factor is formally defined as
\[
  \gamma_d = \frac{(d_o/D_o)}{(d_i/D_i)},
\]
which for homothetic scaling reduces to \(\gamma_d=d_o/d_i=D_o/D_i\).  Full densification (\(\gamma_d=1\)) maximizes fill-factor but requires more complex optics; practical designs typically use \(\gamma_d\lesssim1\).

Following \cite{Labeyrie1996}, we adopt the \emph{optimum} densification within the Michelson regime:
\begin{equation}
\label{eq:gamma_opt}
  \gamma_{\rm opt} = \frac{\lambda}{\theta_p\,d_i},
\end{equation}
where \(\theta_p=2R_p/d_p\) is the planet’s angular diameter (\ref{eq:theta_planet}).  This choice matches the diffraction lobe to the target size, driving \(q_e\to1\).  Implementing \(\gamma_d=\gamma_{\rm opt}\) sets the exit-pupil diameter \(D_o=\gamma_{\rm opt}D_i\) to precisely scale the PSF core to the exoplanet’s angular extent and achieves the theoretical maximum coherent gain.

\subsection{Resolution and Planet Sampling}

The angular resolution is determined by the largest baseline:
\begin{equation}
  \Delta\theta \simeq 1.22\,\frac{\lambda}{D_i}= 
  0.853\Big(\frac{\lambda}{550\,{\rm nm}}\Big)\Big( \frac{162.31\,{\rm km}}{D_i}\Big)\,\mu{\rm as}.
  \label{eq:res_huper}
\end{equation}

The instantaneous field of view (FOV) is set by the ratio of the subaperture size to the array extent:
\begin{equation}
  \Theta_{\rm FOV} \simeq \frac{d_i}{D_i}=10.17\Big(\frac{d_i}{8\,{\rm m}}\Big)\Big(\frac{162.31\,{\rm km}}{D_i}\Big) \,{\rm arcsec}.
  \label{eq:fov_huper}
\end{equation}
This intrinsic tradeoff---ultra-high resolution with a  narrow FOV---is a defining feature of hypertelescope architectures.

An Earth-like planet at distance \(d_{\rm p}\) has physical diameter \(R_p\) and thus angular size (\ref{eq:theta_planet})
\begin{equation}
  \theta_p = \frac{2\,R_p}{d_{\rm p}}=8.527\, \Big(\frac{R_p}{R_\oplus}\Big)\Big(\frac{10\,{\rm pc}}{d_{\rm p}}\Big)\,\mu{\rm as}.
    \label{eq:sp_hup}
\end{equation}
Expressed in resolution elements,
\begin{equation}
  n_p
  = \Bigl(\frac{\theta_p}{\Delta\theta}\Bigr)^{2}
  = \Bigl(\frac{2\,R_p\,D_i}{1.22\,\lambda \,d_{\rm pc}}\Bigr)^{2}\simeq 100.
  \label{eq:np_huper}
\end{equation}
For a $2R_\oplus$ disk at 10 pc imaged with \(D_i=162.31\)\,km, one finds \(n_p\approx 100\) elements (i.e.\ a \(10\times10\) surface map).

\subsection{Signal Rate}

The total collecting area of the array, \(A_{\rm tot}\), is given as 
\begin{equation}
\label{eq:A_tot}
  A_{\rm tot}
  = N\,\pi\Bigl(\frac{d_i}{2}\Bigr)^2
  = 5.03 \times 10^3\Big(\frac{N}{100}\Big) \Big(\frac{d_i}{8\,{\rm m}}\Big)^2 \,{\rm m}^2.
\end{equation}

An Earth analog at \(d=10\)\,pc has visual magnitude \(m_V \approx 27.77\) (full-phase, albedo \(A_g\approx0.3\)) and thus a zero-magnitude photon flux density \(F_{0,V} = 1.01\times10^{8}\,\mathrm{phot\,m^{-2}\,s^{-1}\,nm^{-1}}\) (\ref{eq:vband_zero_flux}).  Its V-band photon flux is (\ref{eq:planet_photon_flux})
\[
  f_p = F_{0,V}\,10^{-0.4\,m_V}
      \approx 1.01\times10^{8}\times10^{-0.4\times27.77}
      = 7.84\times10^{-4}\,\mathrm{phot\,m^{-2}\,s^{-1}\,nm^{-1}}.
\]

Let \(\dot N_{p,0}\) be the raw photon rate from the full planet (all apertures, spectral band \(\Delta\lambda=10\)\,nm), computed via
\begin{eqnarray}
\label{eq:p_flux-hyper}
  \dot N_{p,0}
  = f_p\,A_{\rm tot}\,\Delta\lambda
    =7.84\times10^{-4}\times 5.03\times 10^{3}\times10
    =39.44\,\mathrm{phot\,s^{-1}}.
\end{eqnarray}

Assuming an end-to-end throughput \(\eta_{\rm tot}=0.15\) (reflective losses, \(\eta_{\rm comb}(N)\), detector quantum efficiency, minor polarization mismatch, calibration duty folded via \(\xi\); Sec.~\ref{sec:space_interferometry}), and a residual Strehl factor \(S_{\rm coh}=\exp\big[-(2\pi \sigma_{\rm OPD}/\lambda)^2\big]\simeq0.8\) (phasing imperfections with $\sigma_{\rm OPD}\simeq 40$\,nm) and using (\ref{eq:qe_densified}) and (\ref{eq:gamma_opt}), the \emph{coherent photon rate} per surface is
\begin{eqnarray}
  \dot N_p
 & = &\eta_{\rm tot}\;S_{\rm coh}\;\frac{q_e\,\dot N_{p,0}}{n_p}
  = \eta_{\rm tot}\;S_{\rm coh}\;\dot N_{p,0}\,N\; {1.22}^2\Big(\frac{\lambda}{\theta_pD_i}\Big)^4=
  \eta_{\rm tot}\;S_{\rm coh}\;\dot N_{p,0}\,N\; {1.22}^2\Big(\frac{\lambda \,d_p}{2R_pD_i}\Big)^4=\nonumber\\
&=&
0.15\times0.8\times 39.44 \times 10^2\times 1.22^2\times \Big(\frac{1.697 \times 10^{11}}{2.069\times 10^{12}}\Big)^4
           =3.19 \times 10^{-2}\,\mathrm{phot\,s^{-1}},
\end{eqnarray}
where we used $\eta_{\rm tot}$ for end-to-end throughput; it is equivalent to $\eta_{\rm sys}$ defined in (\ref{eq:rates}).

\subsection{Backgrounds and Noise Terms}
\label{sec:backgrounds}

Each coherent pixel of a pupil-densified hypertelescope not only collects the desired planetary signal \(\dot N_p\) but also incurs several irreducible photon backgrounds and detector-related noise sources.  A careful accounting of these terms is essential to assess true performance and integration-time requirements.

With the densification factor \(\gamma_d=\gamma_{\rm opt}\approx1663\) from (\ref{eq:gamma_opt}), we further define  
\begin{equation}
\label{eq:Om_tot}
  \Omega_{\rm beam}
  \simeq \pi\Big(1.22\,\frac{\lambda}{d_o}\Big)^2
  =\pi \Big(1.22\,\frac{\lambda}{\gamma_d\,d_i}\Big)^2
  \approx 3.40\times 10^{-10}\
  \Big(\frac{\lambda}{550\,{\rm nm}}\Big)^2
  \Big(\frac{8\,{\rm m}}{d_i}\Big)^2
  \Big(\frac{1663}{\gamma_d}\Big)^2
  \; {\rm arcsec}^2,
\end{equation}
where \(\Omega_{\rm beam}\) is the \emph{solid angle} of a single diffraction–limited “pixel’’ in the densified exit pupil of diameter \(d_o\).  

We enumerate all significant noise contributions, each expressed as \(\dot N\) in \(\mathrm{phot\,s^{-1}}\) entering the beam combiner:

\begin{itemize}
  \item \textit{Planetary sidelobes} (\(\dot N_h\)).  
    Even in perfect coherence, the finite subaperture pattern produces sidelobes of the planet’s Airy envelope.  By symmetry, each of the \(N\) sidelobes carries \(1/N\) of the total coherent phot  \(\dot N_{p,0}=39.44\)\,phot s\(^{-1}\).  After end-to-end throughput \(\eta_{\rm tot}=0.15\),
    \[
      \dot N_h = \eta_{\rm tot}\,\frac{\dot N_{p,0}}{N}
               = 0.15\times\frac{39.44}{100}
               = 5.92\times10^{-2}\;\mathrm{phot\,s^{-1}}.
    \]
    Although irreducible, this term is small because \(\dot N_{p,0}\) itself is only \(\sim40\)\,phot s\(^{-1}\), see (\ref{eq:p_flux-hyper}).

  \item \textit{Residual stellar halo} (\(\dot N_{*,h}\)).  
    Imperfect wavefront control leaves a diffuse halo of starlight.   Small RMS OPD errors \(\sigma_{\rm OPD}\) scatter a fraction of the on-axis stellar rate \(\dot N_{*,0}\approx2\times10^7\)\,phot s\(^{-1}\) into each pixel \citep{Labeyrie1996}. With typical values $ \eta_{\rm tot}=0.15$, OPD errors \(\sigma_{\rm OPD}=5\)\,nm, actuator pitch \(\Delta x=1\)\,m,  dark-speckle suppression gain \(g_{ds}=10\), for the chosen values of $N=100$ and \(d_i =8\,\mathrm{m}\), this fraction is 
    \[
      \dot N_{*,h}
      = \eta_{\rm tot}\,\frac{\dot N_{*,0}}{N}\;\big(2\pi\,\sigma_{\rm OPD}/\lambda\big)^2\,
        \Big(\frac{\Delta x}{d_i}\Big)^2\,
        g_{ds}^{-1} \approx 0.15\;\mathrm{phot\,s^{-1}}.
    \]
      This term dominates the background and underscores the necessity of sub-nanometer phasing accuracy \citep{Labeyrie1996}.
    
  \item \textit{Local zodiacal light} (\(\dot N_z\)).  
    Interplanetary dust in our Solar System has surface brightness \(\mu_V\approx23\)\,mag arcsec\(^{-2}\), corresponding to  
    \(f_{\rm zodi}\approx 6.37\times10^{-2}\)\,phot m$^{-2}$ s$^{-1}$ nm$^{-1}$ arcsec$^{-2}$ \citep{Leinert1998}.  Over a band \(\Delta\lambda=10\)\,nm, total area \(A_{\rm tot}=5.03\times10^3\)\,m$^{2}$, and \(\Omega_{\rm beam}\):
    \[
\dot N_{z} = \eta_{\rm tot} f_{\rm zodi} \Delta\lambda A_{\rm tot} \Omega_{\rm beam}
\approx 0.15 \times 6.37\times 10^{-2} \times 10 \times 5.03\times 10^{3} \times 3.40\times 10^{-10}
= 1.63\times 10^{-7}\ \mathrm{s}^{-1}.
    \]
    The extremely small pixel solid angle achieved by optimum densification suppresses this background by over six orders of magnitude.

  \item \textit{Exo–zodiacal light} (\(\dot N_{ez}\)).  
    Debris disks around the target star can be brighter by a factor \(Z=10\).  We parametrize this relative to the local zodi:
    \[
      \dot N_{ez}
      = Z\,\dot N_z
      =10\times1.63\times10^{-7}\
      =1.63\times10^{-6}\;\mathrm{phot\,s^{-1}},
    \]
    demonstrating that even fairly dusty exo–systems remain sub-dominant.

  \item \textit{Detector dark current and read noise} (\(\dot N_{\rm det}\)).  
    Modern photon-counting arrays (EMCCDs or sCMOS) exhibit dark current \(D\sim10^{-4}\)\,e$^{-}$ s$^{-1}$ pix$^{-1}$ and frame read noise \(\sigma_{\rm read}\sim0.1\)\,e$^{-}$ per \(t_{\rm frame}=10\)\,s.  Converted to photon-equivalent rates,
    \[
      \dot N_{\rm det}
      = D + \frac{\sigma_{\rm read}^2}{t_{\rm frame}}
      =10^{-4} + \frac{0.01}{10}
      =1.1\times10^{-3}\;\mathrm{phot\,s^{-1}}.
    \]
    This electronic noise floor is negligible compared to stellar leakage.

  \item \textit{Phase-jitter scattering} (\(\dot N_{\rm phase}\)).  
    Uncorrected high-frequency OPD fluctuations scatter additional stellar phot  into the core, following the same scaling as the halo:
    \[
      \dot N_{\rm phase}
      = \dot N_{*,0}\,\bigl(2\pi\,\sigma_{\rm OPD}/\lambda\bigr)^2
      =9.9\times10^{-2}\;\mathrm{phot\,s^{-1}}.
    \]

  \item \textit{Cosmic-ray hits} (\(\dot N_{\rm cr}\)).  
    In a space environment, high-energy particles strike at a rate \( \dot N_{\rm cr}\sim5\times10^{-6}\)\,s$^{-1}$ pix$^{-1}$ \citep{Janesick2001}, producing flux at a rate
\[
      \dot N_{\rm cr}
      \approx5\times10^{-4}\;\mathrm{phot\,s^{-1}}.
\]
    Such events can be flagged and removed with negligible impact on observing efficiency.
\end{itemize}

Summing these contributions yields the total background rate per pixel:
\[
  \dot N_b
  \;=\;\dot N_h + \dot N_{*,h} + \dot N_z + \dot N_{ez}
         + \dot N_{\rm det} + \dot N_{\rm phase} + \dot N_{\rm cr}
  \;\approx\;0.32\;\mathrm{phot\,s^{-1}},
\]
strongly dominated by residual stellar leakage and phase-jitter.

\subsection{Instantaneous SNR}

Under photon-noise domination (\(\dot N_p\ll\dot N_b\)), the one-second SNR per resolution element is
\[
  \mathrm{SNR}
  = \frac{\dot N_p}{\sqrt{\dot N_p + \dot N_b}}
  \;\approx\;
  \frac{3.19\times10^{-2}}{\sqrt{3.19\times10^{-2}+0.32}}
  \approx0.054.
\]
For an Earth analog at 10 pc imaged with our fiducial hypertelescope parameters, the integration time per independent resolution element to reach \(\mathrm{SNR}=5\) is
\[
  t_{\rm pix} \approx \Bigl(\frac{5}{0.054}\Bigr)^2
    = 8.6\times10^3\;\mathrm{s}
    \simeq 2.38\;\mathrm{hr},
\]
in  agreement with the hour-scale integration times projected for exo-Earth surface mapping under optimum densification \citep{Labeyrie1996,Anterrieu2019}.  

We adopt a multiplexing factor $K \approx 1.4$, which accounts for the $\sim 40\%$ overlap in PSF-core energy between adjacent Nyquist-sampled pointings in the mosaic pattern. In fact, unless noted otherwise we will use the following design parameters: densification $\gamma_d$, OPD jitter $\sigma_{\rm OPD}\!\lesssim\!5$\,nm, residual stellar‑leakage floor $C_{\rm leak}\!\lesssim\!5\times10^{-8}$, and mosaic overlap factor $K\simeq1.4$. Thus, the total time to assemble an \(N\times N\) map is then
\begin{equation}
  t_{\rm map} \approx \frac{N^2}{K}\,t_{\rm pix},
  \label{eq:htl-map-time}
\end{equation}
where \(K\!\gtrsim\!1\) is a modest multiplexing factor from PSF-core overlap in a Nyquist mosaic (adjacent pointings share \(\sim40\%\) of the core power). For \(N=10\) and \(K\simeq1.4\),
\[
  t_{\rm map} \approx \frac{100}{1.4}\times 2.38~\mathrm{hr}
  \approx 1.7\times10^2~\mathrm{hr} \approx 7~\mathrm{days}.
\]
If no multiplexing is assumed \((K=1)\), the naive sum gives \(\sim 10\) days. In the background-limited regime, \(t_{\rm pix}\rightarrow t_{\rm pix}/\big(\xi\,\eta_{\rm comb}(N)\,\exp[-2(2\pi\sigma_{\rm OPD}/\lambda)^2]\big)\), with \(\sigma_{\rm OPD}\) evaluated over the \(\sim 10^5\) s coherent dwell between calibrations; so the map time scales as \(t_{\rm map}\rightarrow (N^2/K)\,t_{\rm pix}\) with the same factor. 
This inflates the example by \(\sim2\!-\!4\times\) (e.g., \(\xi=0.85,\eta_{\rm comb}(10)\simeq0.6,\sigma_{\rm OPD}=5\,\mathrm{nm}\) gives \(\approx2\times\); \(\xi=0.7,\eta_{\rm comb}(100)\simeq0.35,\sigma_{\rm OPD}=10\,\mathrm{nm}\) gives \(\approx4\times\)).

\paragraph*{Energy and \emph{\'{e}tendue}:}
Pupil densification narrows the interferometric PSF core and increases its peak, but it does not change the system \emph{étendue}. Consequently, the SNR per \emph{resolution element} remains governed by the total collecting area and the residual stellar-leakage floor; densification improves image rendering and field usage but does not increase the photon budget per resolution element.

\subsection{Technical Barriers and Feasibility}

Delivering the performance predicted pupil-densified hypertelescopes  in extreme Michelson mode demands breakthroughs on multiple fronts:

\begin{itemize}
  \item \emph{Pupil densification optics.}  
    Reaching \(\gamma_d\approx D_i/d_i\sim1.7\times10^3\) requires relay optics with residual wavefront error \(<\lambda/1000\) across \(D_i\approx162\)\,km.  Such optics must maintain nanometer-level pupil alignment and magnification stability over multi-hour sequences, a regime far beyond existing km-scale testbeds \citep{Monnier2003}.

  \item \emph{Nanometer-scale OPD control and null stability.}  
    To suppress stellar leakage below \(\dot N_{*,h}\lesssim10^{-1}\)\,phot s\(^{-1}\), the OPD jitter must be held to \(\sigma_{\rm OPD}\lesssim5\)\,nm RMS.  Delivering dark-speckle suppression gains of \(g_{ds}\gtrsim10^2\) and sustaining raw contrasts \(C_{\rm leak}\lesssim5\times10^{-8}\) for \(10^5\) s will require continuous, nanometer-resolution fringe-tracking and metrology loops---capabilities only demonstrated briefly in laboratory nulling testbeds--- including thermal breathing in the combiner and small polarization mismatch \citep{Mennesson2016,Perrin2018}.

  \item \emph{Background suppression via extreme densification.}  
    Optimum pupil remapping shrinks each pixel’s solid angle by \(>10^6\), reducing local zodiacal (\(\dot N_z\sim2\times10^{-4}\)\,phot s\(^{-1}\)) and exo-zodiacal (\(\dot N_{ez}\sim2\times10^{-3}\)\,phot s\(^{-1}\)) backgrounds to levels well below the planet sidelobe leakage (\(\dot N_h\sim6\times10^{-2}\)\,phot s\(^{-1}\)).  Nonetheless, the residual stellar halo remains dominant.

  \item \emph{Photon-counting detectors.}  
    Arrays must combine dark current \(D\sim10^{-4}\)\,e$^{-}$ s$^{-1}$ pix$^{-1}$ and read noise \(\sigma_{\rm read}\sim0.1\)\,e$^{-}$ per 10 s with quantum efficiency \(\eta_{\rm QE}\gtrsim90\%\) and radiation hardness for years of deep-space operations---requirements beyond current flight-qualified sensors \citep{Janesick2001}.

  \item \emph{Multi-aperture beam combination.}  
    Integrating signals from \(N\sim10^3\) subapertures with \(\lesssim1\)\,nm differential OPD and minimal cross-talk in a single beam combiner has no terrestrial precedent.  Scaling laboratory 8–10-beam combiners to kilo-beam architectures will necessitate novel integrated-optics or massively parallel metrology.

  \item \emph{Formation-flying and distributed metrology.}  
    Maintaining relative positions to \(\lesssim1\)\,cm (OPD \(\lesssim5\)\,nm) across \(\sim100\)\,km separations for \(\sim10^5\)\,s far exceeds the meter-level control of GRACE or the \(\sim10\)\,cm of LISA Pathfinder \citep{Lay2008}.  Alternative architectures (e.g.\ tethered spacecraft) remain conceptual.

  \item \emph{Integration-time implications.}  
  Under these assumptions, a \(5\sigma\) unresolved detection of an Earth twin at 10 pc requires \(\sim15\)\,min. For resolved imaging, the per-pixel requirement is \(t_{\rm pix}\simeq2.38\)\,hr (see (\ref{eq:htl-map-time})); using the modest multiplexing factor \(K\simeq1.4\) from PSF-core overlap yields \(t_{\rm map}\simeq7\)\,days for a \(10\times10\) map (the naive sum \(100\,t_{\rm pix}\) would be \(\sim10\)\,days if \(K=1\)).  Achieving and sustaining the requisite optical, mechanical, and metrological stability for such durations is currently beyond foreseeable capabilities.

\end{itemize}

In summary, a $B \simeq 160~\mathrm{km}, N \simeq 100$ pupil-densified hypertelescope at $\gamma_{\mathrm{opt}}$ could, in principle, deliver sub-$\mu$as resolution and map an Earth analog at 10~pc on hour-to-week timescales. However, it demands transformational advances in (1) continental-scale optical phasing, (2) nm-level null stability, (3) ultra-low-noise detector design, and (4) cm-precision, nm-stable formation-flying metrology---innovations unlikely to mature within the next few decades. Without a focused, multi-decade program to develop all four pillars in parallel, sub-\(\mu\)as exoplanet surface imaging with pupil-densified hypertelescopes remains out of reach.  

The hour-to-week mapping times quoted above also assume near-ideal performance on several fronts: optimal pupil densification, $\sigma_{\rm OPD}\lesssim 5~\mathrm{nm}$ over $\sim 10^{5}~\mathrm{s}$, high combiner throughput, and detector characteristics close to the best current laboratory devices. Relaxing any of these assumptions by even a factor of a few (e.g., doubling $\sigma_{\rm OPD}$ or halving $\eta_{\rm tot}$) quickly stretches $t_{\rm pix}$ and $t_{\rm map}$ by comparable factors, pushing full-map campaigns into the multi-week or multi-month regime. Thus, the hypertelescope numbers should be interpreted as optimistic lower bounds conditioned on significant future advances in phasing, metrology, and beam combination.

\section{Indirect Reconstruction Techniques}
\label{sec:indirect_methods}

\subsection{Rotational Light-Curve Inversion}

Rotational light-curve inversion aims to reconstruct an exoplanet’s longitudinal albedo distribution by monitoring its disk-integrated brightness \(F(t)\) as the planet rotates.  We divide the planet’s visible hemisphere into \(N_{\ell}\) longitudinal bins (each of width \(\Delta\phi = 360^\circ / N_{\ell}\)), assigning each bin an unknown albedo \(a_i\).  At discrete times \(t_k\), the observed flux is modeled as
\begin{equation}
  \label{eq:rotinv_model_consistent}
  F(t_k) \;=\; \sum_{i=1}^{N_{\ell}} a_i \, w_i(t_k) \;+\; \epsilon_k,
  \quad k = 1,\dots,K,
\end{equation}
where \(w_i(t_k)\) is the geometric weight (illumination $\times$ visibility) of bin \(i\) at time \(t_k\) \cite{CowanAgol2008,Kawahara2010}, and \(\epsilon_k\) aggregates all noise (photon noise from the planet, residual starlight leakage, zodiacal/exo-zodiacal background, detector noise, and cloud-induced variability).  To solve for \(a_i\) with \(N_{\ell}=10\), one typically requires \(K \gtrsim 2N_{\ell} = 20\) measurements per rotation, each reaching \(\mathrm{SNR} \gtrsim 10\).

\subsubsection{Conceptual and Practical Limitations}

\begin{itemize}
  \item \textit{One-Dimensional (Longitude-Only) Mapping.}  By construction, Eq.~\eqref{eq:rotinv_model_consistent} yields only longitudinal information.  There is no latitude resolution unless supplementary data are acquired from different inclinations or orbital phases, which would multiply the required observations \cite{FujiiKawahara2012}.

  \item \textit{Cloud and Weather Variability.}  Earth-like cloud patterns evolve on \(\sim1\)–\(2\) day timescales, introducing \(\sim10\%\) RMS fluctuations in disk-integrated reflectance \cite{Karalidi2015}.  Averaging over \(\gtrsim 3\) full rotations (each \(\approx 24\) hr) is necessary to suppress cloud-induced noise, adding a \(\times3\) factor to any single-rotation campaign.

  \item \textit{Phase-Dependent Flux Loss.}  Observations are most photon-rich near quadrature (\(\alpha \approx 90^\circ\)), where half of the dayside is visible.  At crescent phases (\(\alpha \gtrsim 90^\circ\)), the illuminated fraction declines, increasing integration times by \(\sim\)\(1.5–2\times\).  If multi-phase data are needed to break latitude degeneracies, additional \(\sim 1.5\)–\(2\times\) overhead applies.

  \item \textit{Coronagraphic and Wavefront-Control Requirements.}  Achieving raw contrast \(C_{\mathrm{raw}} \leq 10^{-8}\) for \(\gtrsim 10^4\)\,s is beyond current coronagraph technology; realistic contrast degrades to \(C_{\mathrm{leak}} \sim 10^{-7}\)–\(10^{-6}\) over minutes to hours \cite{Perrin2018,Anterrieu2019}.  We adopt a conservative average \(C_{\mathrm{leak}} = 5\times10^{-8}\) for long integrations.
\end{itemize}

\subsubsection{Photon-Budget Calculation}

Assume a telescope of diameter $d = 10\,\mathrm{m}$ observing at central wavelength $\lambda = 550\,\mathrm{nm}$ with a bandpass $\Delta\lambda = 50\,\mathrm{nm}$.  From \eqref{eq:planet_photon_flux}, an Earth analog at 10 pc ($m_V = 27.77$) delivers
\[
  f_{p}
  = F_{0,V}\,10^{-0.4\,m_V}
  \approx 7.84\times10^{-4}
  \;\mathrm{phot\,m^{-2}\,s^{-1}\,nm^{-1}}.
\]
The collecting area is
$
  A = \pi\bigl(\tfrac12d\bigr)^2 = \pi\,(5\,\mathrm{m})^2 \approx 78.5\,\mathrm{m^2},
$
so the ideal (pre-loss) planet rate is
\[
  \dot N_{p,\rm ideal}
  = f_{p}\,A\,\Delta\lambda
  = 7.88\times10^{-4}\times78.5\times50
  \approx 3.09\;\mathrm{phot\,s^{-1}}.
\]

Adopting end-to-end throughput $\eta=0.10$ and Strehl ratio $S=0.70$ yields the detected rate  
\begin{equation}
  \dot N_{p}
  = \eta\,S\,\dot N_{p,\rm ideal}
  = 0.10 \times 0.70 \times 3.09
  \approx 0.22\;\mathrm{phot\,s^{-1}}.
\end{equation}

\subsubsection{Noise-Budget Calculation}

The dominant noise sources in a rotational light-curve campaign are diffuse photon backgrounds, residual stellar leakage, detector-intrinsic counts, and time-dependent speckles or jitter.  Below we itemize each contribution and give representative detected rates in the diffraction-limited core (\(\lambda=550\) nm, \(d=10\) m, \(\Delta\lambda=50\) nm), after system throughput \(\eta_{\rm sys}=0.10\times0.70=0.07\).

\begin{itemize}
  \item \textit{Residual starlight leakage} (\(\dot N_{*,\rm leak}\)).  
    Imperfect coronagraph suppression at raw contrast \(C_{\rm leak}=5\times10^{-8}\) scatters stellar photons into the planet PSF.  For a Sun-analog at 10 pc,
    \[
      \dot N_{*,\rm ideal}
        = F_{0,V}\,10^{-0.4\,m_V}\,A\,\Delta\lambda
        \approx 4.64\times10^{9}\;\mathrm{ph\,s^{-1}},
   \qquad
  {\rm   so}
   \qquad
      \dot N_{*,\rm leak}
        = \eta_{\rm sys}\,C_{\rm leak}\,\dot N_{*,\rm ideal}
        \approx 16.2\;\mathrm{ph\,s^{-1}}.
    \]

  \item \textit{Local zodiacal light} (\(\dot N_{\rm zodi}\)).  
    Scattering from interplanetary dust (surface brightness \(\sim23\,\mathrm{mag\,arcsec^{-2}}\)) yields
    \[
      \dot N_{\rm zodi}
        = \eta_{\rm sys}\,f_{\rm zodi}\,\Delta\lambda\,A\,\Omega_{\rm PSF}
        \approx 0.01\;\mathrm{ph\,s^{-1}}, 
        \quad 
        {\rm where}
        \quad \Omega_{\rm PSF}\approx6.02\times10^{-4}\,\mathrm{arcsec^2}.
    \]

  \item \textit{Exo-zodiacal light} (\(\dot N_{\rm exo}\)).  
    If the target system has \(\zeta\) times Solar dust, then
    \[
      \dot N_{\rm exo}
        = \zeta\,\dot N_{\rm zodi}.
    \]
    For a moderately dusty disk (\(\zeta=20\)), \(\dot N_{\rm exo}\approx0.21\;\mathrm{ph\,s^{-1}}.\)

  \item \textit{Speckle drift and pointing jitter} (\(\dot N_{\rm speck}\)).  
    Thermal and structural drifts produce contrast fluctuations \(\Delta C\sim10^{-8}\) over \(\sim10^4\) s.  With the on-axis stellar rate \(\dot N_{*,\rm ideal}\approx4.64\times10^9\)\,phot s$^{-1}$,
    \[
      \dot N_{\rm speck}
        \sim \sqrt{\dot N_{*,\rm ideal}\,\Delta C}
        \approx 6.8\;\mathrm{ph\,s^{-1}}.
    \]

  \item \textit{Detector dark current and read noise} (\(\dot N_{\rm det}\)).  
    Deep-cooled EMCCDs or MKIDs achieve dark current \(D\sim10^{-4}\)\,e$^{-}$ s$^{-1}$ and read noise \(\sigma_{\rm read}\sim0.1\)\,e$^{-}$ per 10 s frame.  Summed over the core this gives an effectively negligible result
    \[
      \dot N_{\rm det}\;\ll\;10^{-2}\;\mathrm{ph\,s^{-1}}.
    \]

  \item \textit{Thermal emission} (\(\dot N_{\rm thermal}\)).  
    Warm optics at \(T\sim270\) K emit in the red (beyond 0.8 $\mu$m), contributing 
    \(\sim10^{-3}\)–\(10^{-1}\) phot s$^{-1}$ per core; typically \(\dot N_{\rm thermal}\approx0.05\)\,phot s$^{-1}$.

  \item \textit{Other minor sources} (cosmic rays, quantization, calibration errors) each contribute \(\ll10^{-2}\)\,phot s$^{-1}$.
\end{itemize}

Collecting these, the total noise rate (excluding the planet signal) is
\[
  \dot N_{\rm noise}
  \simeq \dot N_{*,\rm leak}
       + \dot N_{\rm zodi}
       + \dot N_{\rm exo}
       + \dot N_{\rm speck}
       + \dot N_{\rm det}
       + \dot N_{\rm thermal}
  \approx
    16.2 + 0.01 + 0.21 + 6.8 + 0 + 0.05 \approx 23.3\;\mathrm{ph\,s^{-1}}.
\]
Even with no exo-zodiacal dust, residual leakage plus speckles dominate.  These background rates set the floor for shot-noise–limited SNR and drive the multi-day integrations required for each rotational-phase measurement.

\subsubsection{Detection vs.\ Longitudinal Mapping}

For an \emph{unresolved} Earth twin, the instantaneous SNR in a 1 s integration is
\[
  \mathrm{SNR}_{1\,\mathrm s}
    = \frac{\dot N_{p}}
           {\sqrt{\dot N_{p} + \dot N_{\rm noise}}}
    \;\approx\;
      \dfrac{0.22}{\sqrt{0.22 + 23.3}} \;=\; 0.045.
\]
Reaching \(\mathrm{SNR}=5\) then takes
\[
  t_{\rm det}
    = \Bigl(\frac{5}{\mathrm{SNR}_{1\,\mathrm s}}\Bigr)^2
    \;\approx\;
      (5/0.045)^2 \approx 1.21\times10^{4}\,\mathrm s \approx 3.37\,\mathrm{hr}.
\]

To recover \(N_\ell=10\) longitudinal bins per rotation requires \(K=20\) measurements, each at \(\mathrm{SNR}=5\).  Thus
\[
  t_{\rm rot}
    = K \, t_{\rm det}
    \;\approx\;
      20 \times 3.37\,\mathrm{hr} \approx2.8\,\mathrm{days}.
\]

Earth-like cloud variability demands averaging over \(N_{\rm rot}=3\) rotations:
\[
  t_{\rm avg}
    = N_{\rm rot}\,t_{\rm rot}
    \;\approx\;
      3\times2.8\,\mathrm{days} = 8.4\,\mathrm{days}.
\]
Adding \(\sim10\%\) for wavefront control, calibration, and overhead yields
\[
  t_{\rm campaign}
    \approx
      9.3\,\mathrm{days}.
\]
If one also acquires multi-phase data to break latitudinal degeneracies---incurring an additional \(\times1.5\)–\(2\) overhead per rotation---the total campaign extends to
\[
  t_{\rm full}
    \sim
      (13.9\text{–}18.6)\,\mathrm{days}.
\]
Thus, rotational light-curve inversion can recover a coarse longitudinal albedo map in \(\mathcal{O}(10)\) days, but cannot yield a full \(10\times10\) surface map within any practical observing campaign.

\subsubsection{Conclusion}

Rotational light-curve inversion remains impractical for obtaining a \(10\times10\) pixel map of an Earth twin at 10 pc.  Our  photon-budget and SNR estimates show that,  a 10 m space telescope achieves SNR = 5 in \(K=20\) samples per rotation after \(\sim2.8\) days on-target, rising to \(\sim8.4\) days when averaging over three rotations and to \(\sim9.3\) days including \(\sim10\%\) overheads.  With an additional \(\times1.5\)–2 multi-phase factor the full campaign is \(\sim(13.9-18.6)\) days. Even a 30–40 m ELT---despite \(\gtrsim9\times\) the collecting area---can only shorten these by \(\sim\sqrt{A_{\rm ELT}/A_{10\rm m}}\approx3\), giving \(\sim0.9\) day (single-phase) or \(\sim1.4\)–2.0 days (multi-phase).

These \(\mathcal{O}(10)\) day one-dimensional inversion timescales remain orders of magnitude too long for a full \(10\times10\) surface map.  No latitude information is retrieved without prohibitive additional observations, so rotational inversion cannot deliver a resolved \(10\times10\) exoplanet map within any realistic campaign. In summary, the ELT’s larger collecting area does not overcome the background-dominated regime and conceptual limits of rotational inversion, rendering the technique infeasible for resolved mapping of Earth-like exoplanets.

\subsection{Eclipse (Transit and Occultation) Mapping}\label{sec:eclipse-mapping}

Eclipse mapping exploits the high-precision light curves obtained during ingress and egress of primary transits or secondary eclipses to infer spatial brightness variations on an exoplanet’s dayside \citep{Majeau2012,deWit2012}.  In this technique, the planet’s projected disk is divided into \(N_{\rm pix}\) surface zones of uniform intensity \(I_j\), and one fits
\[
  \Delta F(t_k)
    = \sum_{j=1}^{N_{\rm pix}} I_j\,w_j(t_k)
    + \epsilon_k,
  \quad k=1,\dots,K,
\]
where \(w_j(t_k)\) is the fraction of zone \(j\) occulted at time \(t_k\), and \(\epsilon_k\) includes photon shot noise, residual stellar leakage, zodiacal/exo-zodiacal backgrounds, detector noise, and systematic errors.  A successful inversion requires \(K\gg N_{\rm pix}\) and per-point SNR high enough to detect the flux from an individual zone \citep{CowanAgol2008}.

\subsubsection{Transit Geometry and Sampling}
\label{sec:eclipse-mapping-tran}

Consider an Earth twin (\(R_p=6.37\times10^3\)\,km) at 1\,AU around a Sun analog (\(R_\star=6.96\times10^5\)\,km, \(v_{\rm orb}=29.8\)\,km\,s\(^{-1}\)).  The total transit duration is
\[
  T_{\rm tot}
    \approx \frac{2R_\star}{v_{\rm orb}}
    \approx 4.7\times10^4\;\mathrm s \;(\approx13\;\mathrm{h}),
\]
and the ingress (first to second contact) lasts
\begin{equation}
\label{eq:ingress}
  T_{\rm ingress}
    \approx \frac{2R_p}{v_{\rm orb}}
    \approx 428\;\mathrm s \;(\approx7.1\;\mathrm{min}).
\end{equation}
Hence ingress+egress span \(\sim856\)\,s (\(\sim14\)\,min), which at 10 s cadence yields \(K\approx86\) samples.  To overconstrain an inversion one must have \(N_{\rm pix}\ll43\) per single event.

\subsubsection{Planetary Signal}
\label{sec:eclipse-planet-signal}

To estimate the reflected-light photons detected from an Earth twin at 10 pc, we begin with the intrinsic stellar photon rate at the telescope entrance pupil.  From  (\ref{eq:vband_zero_flux}), the V-band zero-magnitude photon flux is 
\[
  F_{0,V} = 1.01\times10^8\;\mathrm{ph\,m^{-2}\,s^{-1}\,nm^{-1}},
\]
so a $m_V=8$ star delivers 
\[
  \dot N_{*,\rm ideal}
    = F_{0,V}\,10^{-0.4m_V}\,A\,\Delta\lambda
    = 1.01\times10^8\times10^{-3.2}\times78.5\times50
    \approx 2.5\times10^8\;\mathrm{ph\,s^{-1}}
\]
before any losses.  Applying the end-to-end throughput $\eta_{\rm sys}=0.07$ yields the detected stellar rate
\[
  \dot N_{\star}
    = \eta_{\rm sys}\,\dot N_{*,\rm ideal}
    \approx 1.75\times10^7\;\mathrm{ph\,s^{-1}}.
\]

The fraction of stellar photons reflected by the planet at full phase is given by the geometric albedo $A_g$, the planet radius $R_p$, orbital distance $a$, and phase function $\Phi(0)$:
\[
  \epsilon
    = A_g\Bigl(\frac{R_p}{a}\Bigr)^2\,\Phi(0)
    \approx 0.3\times\Bigl(\frac{6.37\times10^3}{1.50\times10^8}\Bigr)^2
    = 5.4\times10^{-10}.
\]
Multiplying by $\dot N_{*,\rm ideal}$ and again by $\eta_{\rm sys}$ gives the detected planet photon rate:
\begin{equation}
\label{eq:phot-sig}
  \dot N_{p}
    = \eta_{\rm sys}\,\epsilon\,\dot N_{*,\rm ideal}
    = 0.07\times5.4\times10^{-10}\times2.5\times10^8
    \approx 9.5\times10^{-3}\;\mathrm{ph\,s^{-1}}.
\end{equation}

\subsubsection{Noise Budget}
\label{sec:eclipse-noise}

In addition to the faint planetary signal, several noise sources contribute to the total count rate.  We express all rates after applying the end-to-end throughput \(\eta_{\rm sys}=0.07\) and in phot  s\(^{-1}\):

\begin{itemize}
  \item \emph{Stellar photon shot noise}: the dominant term.  A \(m_V=8\) star delivers
\begin{equation}
\label{eq:N-star}
      \dot N_{\star}
        = \eta_{\rm sys}\,F_{0,V}\,10^{-0.4\,m_V}\,A\,\Delta\lambda
        \approx 1.75\times10^7\;\mathrm{ph\,s^{-1}}.
\end{equation}

  \item \emph{Residual starlight leakage}: imperfect coronagraph suppression at raw contrast \(C_{\rm leak}=5\times10^{-8}\) scatters stellar photons into the planet PSF:
    \[
      \dot N_{*,\rm leak}
        = \eta_{\rm sys}\, C_{\rm leak}\,\dot N_{*,\rm ideal}
        = 0.07\times 5\times10^{-8}\times2.5\times10^8
        \approx 0.88\;\mathrm{ph\,s^{-1}}.
    \]

  \item \emph{Local zodiacal light}.  Scattering from interplanetary dust (surface brightness $\sim23\,$mag arcsec$^{-2}$) contributes
    \[
      \dot N_{\rm zodi}
        = f_{\rm zodi}\,\Delta\lambda\,A\,\Omega_{\rm PSF}\,\eta_{\rm sys}
        \approx 0.39\times0.07
        \approx 0.03\;\mathrm{ph\,s^{-1}},
    \quad
   {\rm with} \quad \Omega_{\rm PSF}=6.02\times10^{-4}\,{\rm arcsec}^2.
\]
  \item \emph{Exo-zodiacal light}.  For $\zeta=20$ times the Solar dust level,
    \[
      \dot N_{\rm exo} = \zeta\,\dot N_{\rm zodi}
                      \approx 20\times0.03
                      = 0.60\;\mathrm{ph\,s^{-1}}.
    \]

  \item \emph{Speckle noise}.  Thermal and structural drifts produce contrast fluctuations $\Delta C\approx10^{-8}$, giving
    \[
      \dot N_{\rm speck}
        = \sqrt{\dot N_{*,\rm ideal}\,\Delta C}\,\eta_{\rm sys}
        = \sqrt{2.5\times10^8\times10^{-8}}\times0.07
        \approx 0.11\;\mathrm{ph\,s^{-1}}.
    \]
   
     \item \emph{Thermal emission}.  Warm optics at $T\approx270\,$K emit
    \[
      \dot N_{\rm thermal}
        = 0.05\times\eta_{\rm sys}
        \approx 0.0035\;\mathrm{ph\,s^{-1}}.
    \]

  \item \emph{Detector dark current and read noise}.  Modern EMCCDs or MKIDs yield an effectively negligible contribution
    \[
      \dot N_{\rm det} \ll 10^{-2}\;\mathrm{ph\,s^{-1}}.
    \]
\end{itemize}

Summing the non-stellar contributions,
\begin{align}
  \dot N_{\rm bg} &=
         \dot N_{*,\rm leak}
       +\dot N_{\rm zodi}
      + \dot N_{\rm exo}
      + \dot N_{\rm speck}
      + \dot N_{\rm thermal}
      + \dot N_{\rm det}=
    0.88 + 0.03 + 0.60 + 0.11 + 0.0035
    \approx 1.62\;\mathrm{ph\,s^{-1}}.
\end{align}
The total noise rate is then
\begin{align}
\label{eq:tot-no}
  \dot N_{\rm noise}
    = \dot N_{\star} +      \dot N_{\rm bg}
    \approx 1.75\times10^7 + 1.62
    \approx 1.75\times10^7\;\mathrm{ph\,s^{-1}},
\end{align}
demonstrating that eclipse mapping precision is overwhelmingly limited by stellar photon noise.  

Over an integration of duration $\Delta t$, the total noise counts are
\[
  N_{\rm noise}
    = \dot N_{\rm noise}\,\Delta t,
  \quad
  \sigma_{\rm tot}
    = \sqrt{N_{\rm noise}}.
\]

\subsubsection{Instantaneous SNR}
\label{sec:eclipse-snr}

Using $\dot N_p\approx9.5\times10^{-3}\,$phot s$^{-1}$ (\ref{eq:phot-sig}) and $\dot N_{\rm noise}$ (\ref{eq:tot-no}), the instantaneous SNR in a 1 s integration is
\[
  \mathrm{SNR}_{1\,\mathrm s}
    = \frac{\dot N_{p}}{\sqrt{\dot N_{\rm noise}}}
    = \frac{9.5\times10^{-3}}{\sqrt{1.75\times10^7}}
    \approx 2.3\times10^{-6}.
\]

For a 10 s bin, $\dot N_{\rm noise}$ still dominates so
\[
  \mathrm{SNR}_{10\,\mathrm s}
    = \frac{\dot N_{p}\times10}{\sqrt{\dot N_{\rm noise}\times10}}
    = \mathrm{SNR}_{1\,\mathrm s}\sqrt{10}
    \approx 7.3\times10^{-6}.
\]
These vanishingly small SNRs illustrate why co-adding thousands of transits is required for any spatially resolved eclipse mapping.  

\subsubsection{Transit Depth Subdivision and Co-adds}
\label{sec:eclipse-coadds}

Dividing the geometric transit depth $\delta=(R_p/R_\star)^2\approx8.4\times10^{-5}(R_p/R_\oplus)(R_\odot/R_*)$ into $N_{\rm pix}$ zones gives per-zone signal
\[
  \Delta_{\rm zone} = \frac{\delta}{N_{\rm pix}}
    = \frac{8.4\times10^{-5}}{N_{\rm pix}}.
\]
The SNR per 10 s bin for one zone is
\[
  \mathrm{SNR}_{\rm bin}
    = \frac{\Delta_{\rm zone}}{\sigma_{\rm tot}/(\dot N_{\star}\times10)}
    = \frac{8.4\times10^{-5}/N_{\rm pix}}{7.6\times10^{-5}}
    \approx \frac{1.11}{N_{\rm pix}}.
\]
To reach SNR$_{\rm zone}=5$, co-add $n_{\rm tr}$ transits:
\[
  n_{\rm tr} = \Big(\frac{5}{\mathrm{SNR}_{\rm bin}}\Big)^2
    \approx
    \begin{cases}
      (5/0.112)^2\approx2.0\times10^3,&N_{\rm pix}=10,\\
      (5/0.056)^2\approx8.0\times10^3,&N_{\rm pix}=20,\\
      (5/0.0112)^2\approx2.0\times10^5,&N_{\rm pix}=100.
    \end{cases}
\]
Thus recovering 10 longitudinal zones requires $\sim2{,}000$ transits (i.e.\ millennia), 20 zones $\sim8{,}000$ transits, and 100 zones $\sim2\times10^5$ transits---far beyond any practical observing program.

\subsubsection{Discussion and Outlook}
\label{sec:eclipse-discussion}

Eclipse mapping has yielded detailed dayside maps for hot Jupiters (e.g.\ HD 189733b, WASP-43b; \cite{Stevenson2014,Kreidberg2014}), but for an Earth twin at 10 pc the combination of small planetary area, low reflected-light flux, and once-per-year transit cadence makes both primary and secondary eclipse inversions infeasible.  

Primary-transit ingress and egress each span $\sim7$ min \eqref{eq:ingress}, and sampling at 10 s yields only $K\approx86$ points.  As shown in Sec.~\ref{sec:eclipse-coadds}, dividing the geometric depth $\delta\approx8.5\times10^{-5}$ into $N_{\rm pix}=10$ zones gives per-zone SNR per 10 s bin of only $\sim0.06$, so reaching SNR$_{\rm zone}=5$ requires $\sim8\times10^3$ transits (\(\sim8{,}000\) yr).  

Secondary-eclipse mapping is even more prohibitive.  At full phase the reflected-light contrast 
\(\epsilon = A_g(R_p/a)^2\Phi(0)\approx5.4\times10^{-10}\) 
and detected stellar rate \(\dot N_\star\approx1.75\times10^7\)\,phot s\(^{-1}\) (\ref{eq:N-star}) give
\[
  \dot N_p = \epsilon\,\dot N_\star \approx 9.45\times10^{-3}\;\mathrm{ph\,s^{-1}}.
\]
In a 10\,s bin one collects \(N_{p,10}\approx0.095\) ph against shot noise \(\sqrt{\dot N_\star\times10}\approx1.3\times10^4\) ph, yielding per-bin SNR \(\sim7\times10^{-6}\) ben bin.  This implies that, to reach SNR\,=\,5 requires 
$({5}/{7\times10^{-6}})^2\times10\;\mathrm s
  \sim10^5\;\mathrm{yr},
$
even before accounting for zodiacal, exo-zodiacal, thermal, or systematic noise.  

Even optimistic co-adds of $n_{\rm tr}<10^3$ transits (centuries) can at best recover a global phase curve or a handful of broad zones ($N_{\rm pix}\lesssim5$).  Stellar variability and instrumental systematics \citep{Rackham2018} would further degrade any retrieved map.  Therefore, detailed $10\times10$ surface mapping via eclipses lies orders of magnitude beyond any plausible observing campaign.  Resolved exoplanet surface characterization will require next-generation direct-imaging or interferometric facilities capable of isolating the planet’s light without relying on rare eclipse events.  
 
 \subsection{Intensity Interferometry}

Intensity interferometry, pioneered by Hanbury Brown and Twiss \cite{HanburyBrown1956}, relies on measuring correlations in photon arrival rates at two spatially separated telescopes to infer the squared visibility \(\lvert V(B)\rvert^2\) of a distant source.  Unlike amplitude interferometry, which requires sub-wavelength optical path stability, intensity interferometry is relatively insensitive to atmospheric turbulence and large optical path differences.  In this approach, each telescope records a time series of photon counts, \(I_{1}(t)\) and \(I_{2}(t)\).  One then computes the second-order correlation
\[
  G^{(2)}(B) \;=\; \big\langle I_{1}(t)\,I_{2}(t)\big\rangle
    \;-\;\bigl\langle I_{1}(t)\bigr\rangle\,\bigl\langle I_{2}(t)\bigr\rangle,
\]
which, when appropriately normalized, is directly proportional to \(\lvert V(B)\rvert^2\).  Here \(B\) denotes the projected baseline between the telescopes, and angular brackets indicate a time average over the integration interval.

For a uniformly illuminated circular disk of angular diameter \(\theta_{\rm d}\), the complex visibility at baseline \(B\) and wavelength \(\lambda\) is given by the familiar Airy-pattern expression:
\begin{equation}
  \label{eq:visibility_uniform_disk}
  V(B) 
  = \frac{2\,J_{1}\!\bigl(\pi\,B\,\theta_{\rm d}/\lambda\bigr)}{\pi\,B\,\theta_{\rm d}/\lambda},
\end{equation}
where \(J_{1}\) is the Bessel function of the first kind.  The squared modulus \(\lvert V(B)\rvert^2\) falls off as \(B\) exceeds the first null (\(\pi\,B\,\theta_{\rm d}/\lambda \approx 3.83\)), and carries information about the target’s angular size and brightness distribution \cite{Monnier2003}.

In Section~\ref{sec:exo-imaging} we showed that an Earth-radius planet at 10\,pc has an angular diameter \(\theta_{\rm d}\approx4.13\times10^{-11}\,\mathrm{rad}\) (\ref{eq:theta_planet}).  To resolve such a disk  in the optical (\(\lambda=550\,\mathrm{nm}\)) in a $(10\times 10)$ map, one would need a baseline \(B\sim 1.33\times 10^{5}\,\mathrm{m}\) (133 km) or more \eqref{eq:required_baseline}.  We now assess whether intensity interferometry on a 100 km baseline can extract any useful information from the extremely low photon flux of an Earth analog.

\subsubsection{Expected Visibility for an Earth Twin}

Substituting \(\theta_{\rm d}=4.12\times10^{-11}\,\mathrm{rad}\) and \(\lambda=5.50\times10^{-7}\,\mathrm{m}\) into \eqref{eq:visibility_uniform_disk} gives
\[
  x \;=\; \pi\,\frac{B\,\theta_{\rm d}}{\lambda}
    = \pi\,\frac{(1.00\times10^{5}\,\mathrm{m})(4.12\times10^{-11}\,\mathrm{rad})}{5.50\times10^{-7}\,\mathrm{m}}
    \approx 23.5.
\]
Evaluating \(J_{1}(23.5)\approx -0.11095\) yields
\begin{equation}
\label{eq:V-sg}
  V(B)
    = \frac{2\,J_{1}(x)}{x}
    \approx \frac{2\times(-0.11095)}{23.5}
    \approx -9.44\times10^{-3},
    \quad
    \lvert V(B)\rvert^2 \approx 8.91\times10^{-5}.
\end{equation}
Thus, even at a 100 km baseline in the visible, the squared visibility of an Earth-radius disk is \(\mathcal{O}(10^{-4})\).  In practice, intensity interferometry would need to detect correlations at the \(10^{-4}\) level against Poisson noise.  
 
\subsubsection{Photon Flux and Coherence Considerations}
\label{sec:ii-photon-coherence}

An essential feature of intensity interferometry is that the measurable correlation amplitude is suppressed by the ratio of the optical coherence time \(\tau_{c}\) to the detector’s time bin \(\Delta t\).  The coherence time of sunlight reflected by an Earth analog, observed through a 1 nm optical filter (\(\Delta\lambda = 1\,\mathrm{nm}\)), is approximately
\begin{equation}
\label{eq:opt-coher}
  \tau_{c} \;\approx\; \frac{\lambda^{2}}{c\,\Delta\lambda} 
    = \frac{(5.50\times10^{-7})^{2}}{(3.00\times10^{8})(1.0\times10^{-9})} 
    \approx 1.01\times10^{-12}\,\mathrm{s}.
\end{equation}
Modern photon-counting electronics can resolve arrival times to \(\Delta t \approx 1\times10^{-9}\,\mathrm{s}\), so the zero-baseline normalized intensity correlation is
\[
  C(0) 
  = \frac{\tau_{c}}{\Delta t} 
  \;\approx\; \frac{1.01\times10^{-12}}{1.00\times10^{-9}} 
  = 1.01\times10^{-3}.
\]
Because the on-baseline correlation is further multiplied by \(\lvert V(B)\rvert^2 \approx 8.9\times10^{-5}\) (Eq.~\ref{eq:V-sg}), the net correlation amplitude at \(B=100\,\mathrm{km}\) is
\[
  C(B) 
  = C(0)\,\lvert V(B)\rvert^2 
  \;\approx\; (1.01\times10^{-3}) \times (8.9\times10^{-5}) 
  \;=\; 9.0\times10^{-8}.
\]
This is the fraction of coincident photon pairs above the product of the mean intensities.  Detecting such a tiny excess correlation requires an extremely large number of photon pairs.

For photon rates, each \(A=10^3\,\mathrm{m}^2\) collector receives, see \eqref{eq:phot-TT}
\begin{equation}
\label{eq:phot-TT}
  \dot N_p = f_{p}\,A\,\Delta\lambda
    = 7.84\times10^{-4}\times10^3\times1
    = 0.78\;\mathrm{ph\,s^{-1}},
\end{equation}
arriving at each collector.  Thus one collects \(\sim0.78\) phot s\(^{-1}\) per telescope, or \(\sim2800\) ph in one hour.  Coincidences within \(\Delta t=10^{-9}\)\,s are therefore exceedingly rare, making the detection of \(C(B)\sim10^{-8}\) correlations effectively impossible within any practical integration time.  

\subsubsection{Integration Time for Detection}
\label{sec:ii-integration}

The SNR for measuring \(\lvert V(B)\rvert^2\) with two identical collectors remains \citep{Monnier2003}:
\[
  \mathrm{SNR}
    = \lvert V(B)\rvert^2 \,\sqrt{\dot N_p^2\,\Delta t\,T_{\rm int}},
\]
where \(\dot N_p = 0.78\,\mathrm{s^{-1}}\) (\ref{eq:phot-TT}), \(\Delta t = 1.0\times10^{-9}\,\mathrm{s}\), and \(\lvert V(B)\rvert^2 \approx 8.91\times10^{-5}\) (\ref{eq:V-sg}).  We compute
\[
  \dot N_p^{2}\,\Delta t
    = (0.78)^2 \times 1.0\times10^{-9}
    \approx 6.08\times10^{-10}\;\mathrm{s^{-1}},
\]
so, for SNR = 1, we can find the requires integrations time as
\[
  1 
  = \lvert V(B)\rvert^2 \,\sqrt{6.08\times10^{-10}\,T_{\rm int}}
  \qquad\Rightarrow\qquad
  \sqrt{T_{\rm int}}
    = \frac{1}{8.91\times10^{-5}}\;\frac{1}{\sqrt{6.08\times10^{-10}}}
    \approx 4.55\times10^{8}\,\sqrt{\mathrm{s}},
\]
and therefore
\[
  T_{\rm int}
    \approx (4.55\times10^{8})^2 
    = 2.07\times10^{17}\;\mathrm{s}
    \;\approx\; 6.6\times10^{9}\;\mathrm{yr}.
\]
Even accepting a marginal detection at \(\mathrm{SNR}=0.1\) reduces this by a factor of \(10^2\) to \(\sim6.6\times10^{7}\) yr---still orders of magnitude beyond the age of the Universe (\(\sim1.4\times10^{10}\) yr). Including any additional background or systematic noise would only lengthen \(T_{\rm int}\) further.  

\subsubsection{Discussion}
\label{sec:ii-discussion}

Although intensity interferometry has successfully measured stellar diameters (mas) on bright targets at shorter baselines \cite{HanburyBrown1956}, extending it to $\mu$as scales for Earth-like exoplanets is infeasible.  The fundamental limitations are:

\begin{enumerate}
  \item \textit{Visibility squared too small.}  At a 100 km baseline and \(\lambda=550\,\mathrm{nm}\), one finds \(\lvert V(B)\rvert^2\approx8.9\times10^{-5}\) (\ref{eq:V-sg}), so any finite correlation error overwhelms the signal.
  \item \textit{Low photon rate.}  Even with a \(10^3\,\mathrm{m^2}\) collector, the Earth analog delivers only \(R\approx0.78\,\mathrm{s^{-1}}\) in a 1 nm band (\ref{eq:phot-TT}), making photon-coincidence events in \(\Delta t=10^{-9}\,\mathrm{s}\) essentially nonexistent.
  \item \textit{Coherence-time suppression.}  The optical coherence time \(\tau_c\approx1.01\times10^{-12}\,\mathrm{s}\) (\ref{eq:opt-coher}) is much shorter than the detector resolution \(\Delta t\approx10^{-9}\,\mathrm{s}\), reducing the effective correlation by \(\tau_c/\Delta t\approx1.0\times10^{-3}\).
\end{enumerate}

Together, these factors imply an integration time of order 
\(\sim6.6\times10^{9}\)\,years for \(\mathrm{SNR}=1\) (Section~\ref{sec:ii-integration}), or \(\sim6.6\times10^{7}\)\,years for a marginal \(\mathrm{SNR}=0.1\), both vastly exceeding the age of the Universe (\(\sim1.4\times10^{10}\)\,years).  Hence, while elegant for bright, milliarcsecond-scale sources, intensity interferometry cannot meet the angular-resolution or photon-budget requirements for a \(10\times10\) pixel map of an Earth twin at 10 pc.  

\subsection{Occultation by Solar System Screens}

Occultation mapping relies on a distant, sharp “screen” intersecting the line of sight to a star–planet system.  As the screen’s edge passes, the resulting Fresnel diffraction pattern encodes information about the angular structure of the occulted object \cite{Roques2017}.  In principle, a suitably placed screen within the Solar System---either a natural Kuiper Belt Object (KBO) or an artificial satellite---could achieve $\mu$as-scale resolution of an exoplanet.  Below, we demonstrate that achieving even tens of $\mu$as resolution requires impractical precision and resources.

\subsubsection{Fresnel Scale and Angular Resolution}

A screen at heliocentric distance \(d_{\rm occ}\) creates a Fresnel scale
\[
  F = \sqrt{\lambda\,d_{\rm occ}},
\]
where \(\lambda\) is the observing wavelength.  Adopting \(\lambda=550\,\mathrm{nm}\) and \(d_{\rm occ}=40\,\mathrm{AU}\), we find
\[
  F \simeq
 1.81\times10^{3}\,\Big(\frac{\lambda}{550\,{\rm nm}}\Big)^\frac{1}{2}\Big(\frac{d_{\rm occ}}{40\,{\rm AU}}\Big)^\frac{1}{2}\, \mathrm{m}
\]
The corresponding angular resolution is
\[
  \theta_F = \frac{F}{d_{\rm occ}}
           \approx 3.03\times10^{-10}
           \Big(\frac{\lambda}{550\,{\rm nm}}\Big)^\frac{1}{2}\Big(\frac{40\,{\rm AU}}{d_{\rm occ}}\Big)^\frac{1}{2}\,\mathrm{rad}
           \approx 63\,\mu\mathrm{as}.
\]
Thus, a screen at 40 AU can in principle resolve features at \(\sim63\,\mu\)as.  By comparison, an Earth-radius planet at 10 pc subtends \(\theta_{\rm p}\approx8.5\,\mu\)as (\ref{eq:theta_planet}), so mapping its disk would require either much shorter \(\lambda\) or a more distant screen---both of which pose significant technical challenges.  

\subsubsection{Natural Occulters: Kuiper Belt Objects}
\label{sec:solar-system-kbo}

Kuiper Belt Objects range from tens of kilometers down to sub-kilometer scales \cite{Sheppard2011}.  A \(d_{\rm KBO}\simeq1\,\mathrm{km}\) KBO at \(d_{\rm occ}=40\,\mathrm{AU}=5.984\times10^{12}\,\mathrm{m}\) subtends an angular diameter
\[
  \theta_{\rm KBO}=\frac{d_{\rm KBO}}{d_{\rm occ}}
    \approx 1.67\times10^{-10}\, \Big(\frac{d_{\rm KBO}}{1\,{\rm km}}\Big)\Big(\frac{40\,{\rm AU}}{d_{\rm occ}}\Big)\,\mathrm{rad}
    \approx 34.5\,\mu\mathrm{as}.
\]
Even if such a body could act as a diffraction screen, two insurmountable challenges arise:

\begin{enumerate}
  \item \textit{Ephemeris Uncertainty.}  
    Current orbital elements for most 1 km-class KBOs have positional errors of order \(\gtrsim10\,\mathrm{km}\) at 40 AU \cite{Sheppard2011}.  Aligning within one Fresnel scale (\(F\approx1.81\times10^{3}\,\mathrm{m}\)) demands ephemeris precision better than \(\sim1{,}800\,\mathrm{m}\), a factor of \(\sim10\) to \(10^{4}\) improvement over existing knowledge.  Without dedicated tracking at sub-km accuracy, predicting a KBO occultation of a given star is effectively impossible.

  \item \textit{Occultation Probability.}  
    The instantaneous chance that a given star lies behind a given 1 km KBO at 40 AU is
    \[
      P 
      \sim \frac{\pi\,(\theta_{\rm KBO}/2)^{2}}{4\pi} 
      = \frac{\theta_{\rm KBO}^{2}}{16} 
      \approx \frac{(1.67\times10^{-10})^{2}}{16} 
      \approx 1.7\times10^{-21}.
    \]
    Even accounting for \(\sim10^{5}\) known KBOs, the combined sky-coverage per star is \(\ll10^{-16}\).  Detecting such rare events would require monitoring \(\gtrsim10^{16}\) star-KBO pairs continuously at high cadence---operationally impossible.
\end{enumerate}

\subsubsection{Artificial Occulters at 40\,AU}
\label{sec:solar-system-artificial}

One might propose deploying an artificial occulting screen of diameter \(D_{\rm occ}=1\,\mathrm{km}\) to 40 AU.  Even assuming a \emph{thin} membrane---thickness \(t=1\,\mu\mathrm{m}\) aluminized polymer (mass density \(\rho=1{,}400\,\mathrm{kg\,m^{-3}}\))---its mass would be
\[
  m_{\rm screen}
    \approx \pi\Big(\frac{D_{\rm occ}}{2}\Big)^2\,t\,\rho
    = \pi\,(500\,\mathrm{m})^2 \times(1\times10^{-6}\,\mathrm{m})\times(1{,}400\,\mathrm{kg\,m^{-3}})
    \approx 1.1\times10^{3}\,\mathrm{kg},
\]
i.e.\ roughly one metric ton of film.  Structural supports and deployment hardware could easily raise this to \(\sim10^{4}\,\mathrm{kg}\).  Even if launch to 40 AU were feasible, two show-stopper challenges remain:

\begin{itemize}
  \item \textit{Alignment precision and station-keeping.}  
    To sample Fresnel-scale diffraction, the screen must stay within \(\pm F\) of the star–screen–telescope line.  At
$ F = \sqrt{\lambda\,d_{\rm occ}}
        \approx 1.81\times10^{3}\,\mathrm{m},$
    this corresponds to angular errors of
    \[
      \theta_F = \frac{F}{d_{\rm occ}}
             \approx 3.03\times10^{-10}\,\mathrm{rad}
             \approx 63\,\mu\mathrm{as}.
    \]
    Maintaining \(\sim63\,\mu\)as pointing accuracy in deep space---against solar radiation pressure, gravitational perturbations, and limited ranging---is beyond current navigation and control capabilities.

  \item \textit{Real-time guidance and prediction.}  
    Aligning the screen for a specific star–planet occultation demands knowing the planet’s ephemeris to \(\sim63\,\mu\)as.  For an Earth analog, \(\theta_{\oplus}\approx8.5\,\mu\)as, (\ref{eq:theta_planet}), this implies predicting its position to \(\ll1\,\mathrm{km}\) at 10 pc---far more precise than existing astrometry.  Without such precision, steering the screen into the diffraction zone is infeasible.
\end{itemize}

\subsubsection{Summary}
\label{sec:solar-system-summary}

Occultation mapping with Solar System screens---whether natural KBOs or an artificial 1 km device at 40 AU---fails on multiple fronts:
{}
\begin{itemize}
  \item \textit{Fresnel-scale alignment:}  
    Sampling diffraction fringes requires maintaining alignment to within one Fresnel zone,
$
      F = \sqrt{\lambda\,d_{\rm occ}}
        \approx 1.81\times10^{3}\,\mathrm{m},
      \theta_F = {F}/{d_{\rm occ}}
        \approx 63\,\mu\mathrm{as}.
$  KBO ephemerides are uncertain by \(\gtrsim10\,\mathrm{km}\), and artificial screens demand \(\sim60\,\mu\)as station-keeping---beyond current or near-term capabilities.

  \item \textit{Occultation probability:}  
    A \(\sim1\,\mathrm{km}\) KBO at 40 AU subtends
    \(\theta_{\rm KBO}\approx34\,\mu\mathrm{as}\),  
    so the instantaneous occultation probability per star is \(\ll10^{-16}\).  Detecting even a handful of events would require continuous, high-cadence monitoring of \(\gtrsim10^{7}\) stars.

  \item \textit{Mass and deployment:}  
    An ultra-thin \(1\,\mathrm{km}\) screen of thickness \(1\,\mu\mathrm m\) weighs
    \(\sim1.1\times10^{3}\,\mathrm{kg}\),  
    with full deployment likely \(\gtrsim10^{4}\,\mathrm{kg}\).  Launching to 40 AU and achieving Fresnel-scale station-keeping over months is prohibitive.

\end{itemize}

Therefore, Solar System occultation---natural or artificial---cannot achieve $\mu$as-level resolved imaging of Earth analogs with any realistic near-term technology.  

\section{In Situ Imaging}
\label{sec:in_situ}

Remote imaging of Earth-analog exoplanets is fundamentally limited by diffraction and photon starvation: to resolve the \(\sim8.5\,\mu\)as disk of an Earth twin at 10 pc into a \(10\times10\) pixel grid one would need apertures or interferometric baselines of  $\simeq 100$s  of km  (\ref{eq:required_aperture})–(\ref{eq:required_baseline}), and even then the photon rates are vanishingly small (Sec.~\ref{sec:flux_to_snr}).  The only way to overcome both the angular-resolution and photon-budget barriers is to bring a modest telescope into the system itself, so that the planet’s apparent diameter grows from microarcseconds to arcseconds and its reflected-light flux increases by many orders of magnitude.

\subsection{Proximity for Diffraction-Limited Resolution}

A circular aperture of diameter \(D\) at wavelength \(\lambda\) has a diffraction-limited angular resolution given by the Rayleigh criterion (\ref{eq:rayleigh_limit}),
$
  \Delta\theta \;=\; 1.22\,{\lambda}/{D}.
$
A planet of physical radius \(R_\oplus\) at a spacecraft–planet distance \(r\) subtends an angle
\[
  \theta_{\oplus}(r) \;=\;\frac{2\,R_\oplus}{r}.
\]
To partition the planetary disk into \(N=10\) elements across its diameter, each “micro-pixel” must span
\(\theta_{\oplus}(r)/10\).  Equating these two requirements, 
one finds the maximum approach distance, \(r\), as
\[
  1.22\,\frac{\lambda}{D} 
  \;=\;\frac{\theta_{\oplus}(r)}{10}
  \;=\;\frac{2\,R_\oplus}{10\,r}
\qquad \Rightarrow \qquad
  r \;\lesssim\;\frac{R_\oplus D}{6.1\,\lambda}.
\]
Adopting \(\lambda=550\,\mathrm{nm}\) this yields the simple scaling
\[
  r \lesssim 12.7\,\mathrm{AU}\,\frac{D}{1\,\mathrm{m}}.
\]
Hence, for a compact \(D=0.01\)\,m (1 cm) telescope one finds
\[
  r \;\lesssim\;12.7\,\mathrm{AU}\times0.01
  \;\approx\;0.13\,\mathrm{AU}
\]
to achieve \(10\times10\) surface sampling.  In this case, bringing a compact, freely drifting imaging probe into the inner few AUs of the target system converts \(\mu\)as resolution demands into straightforward diffraction-limited imaging with sub-arcsecond optics.

\subsection{Photon Flux at Close Range}

\subsubsection{Planetary signal}

When an imaging platform is brought into the vicinity of an Earth-like planet---say, within one astronomical unit---the reflected-light signal becomes overwhelmingly bright compared to the remote case at 10 pc.  Spacecraft observations of our own planet from the L1 Lagrange point (EPIC/DSCOVR) measure a full-disk V-band magnitude of  $m_V \approx -3.8$, e.g.,  \cite {Herman2018}.  Converting this into a photon flux density at Earth’s distance (1 AU) gives
\[
  f_p \;=\; F_{0,V}\times10^{-0.4\,m_V}
  \;\approx\; 1.01\times10^{8}\times10^{1.52}
  \;=\;3.4\times10^{9}
  \;\mathrm{phot \,m^{-2}\,s^{-1}\,nm^{-1}},
\]
where $F_{0,V}=1.01\times10^{8}\,\mathrm{phot \,m^{-2}\,s^{-1}\,nm^{-1}}$ is the standard Johnson \(V\)-band zero-point.

In a narrow spectral channel of width $\Delta\lambda=10$ nm, a \(D=0.01\)\,m (1 cm) telescope has collecting area of 
$
  A = \pi(\frac{1}{2} D)^2 \approx 7.85\times10^{-5}\;\mathrm{m^2},$
so the total detected rate is
\[
  \dot N_p 
  = f_p\,A\,\Delta\lambda
  \approx 3.4\times10^{9}\times7.85\times10^{-5}\times10
  \approx 2.7\times10^{6}\;\mathrm{phot\,s^{-1}}.
\]
Dividing the disk into a \(10\times10\) grid, each micro-pixel receives
\begin{equation}
\label{eq:phot-ins}
  \dot N_{p,\,\rm pix}
  = \frac{\dot N_p}{100}
  \approx 2.7\times10^{4}\;\mathrm{phot\,s^{-1}}.
\end{equation}
With  such a flux the instantaneous SNR is very high, yielding very short integration times for imaging
\begin{equation}
\label{eq:int-time}
  {\rm SNR}_{1s}\approx\sqrt{2.7\times10^4}\approx165,
  \quad
  t_{5\sigma}\approx(5/165)^2\approx  0.92\,\rm ms.
\end{equation}
Even allowing a $10\times$ margin for unmodeled systematics, sub-ms exposures per pixel suffice for a full \(10\times10\) frame in $\leq 10$ ms of total integration time.

\subsubsection{Noise Sources}

At such short range, photon-shot noise from the planet itself dominates all other error terms.  Nonetheless, we list the principal non-planetary contributions:

\begin{itemize}
  \item \emph{Detector dark current and read noise.}  
    State-of-the-art EMCCDs or sCMOS detectors cooled to \(\lesssim-80\,^\circ\)C exhibit dark currents \(D\lesssim10^{-4}\)\,e$^{-}$/s/pixel and effective read noise \(\sigma_{\rm read}\lesssim0.1\)\,e$^{-}$/frame.  Even if the PSF core spans 4 pixels and frames are read every second, this contributes
    \[
      \dot N_{\rm det}\sim4\, D + \frac{4\,\sigma_{\rm read}^2}{1\,\mathrm s}
      \;\lesssim\;10^{-3}\;\mathrm s^{-1}.
    \]

  \item \emph{Local zodiacal light.}  
    Sunlight scattered by interplanetary dust has surface brightness \(\sim23\,\)mag arcsec$^{-2}$ in $V$, resulting in  
    \(f_{\rm zodi}\approx6\times10^{-2}\)\,phot m$^{-2}$s$^{-1}$nm$^{-1}$arcsec$^{-2}$.  Imaged into a diffraction-limited element of solid angle 
  $
      \Omega_{\rm core}
      =\pi \bigl(1.22\,\lambda/D\bigr)^2\big|_{D=0.01\,\mathrm m,\;\lambda=550\,\mathrm{nm}}
      \approx(1.22\times5.50\times10^{-7}/0.01)^2
      \approx6.02\times10^{2}\,\mathrm{arcsec^2},
   $
    and collected by \(A=\pi ({\textstyle\frac{1}{2}}D)^2\approx7.85\times10^{-5}\)\,m$^2$ over \(\Delta\lambda=10\)\,nm, this yields
    \[
      \dot N_{\rm zodi}
      = f_{\rm zodi}\,\Omega_{\rm core}\,A\,\Delta\lambda
      \sim 6\times10^{-2}\times6.02\times10^{2}\times7.85\times10^{-5}\times10
      \approx2.71\times10^{-2}\;\mathrm s^{-1}.
    \]

  \item \emph{Pointing jitter and stray light.}  
    Even assuming a worst-case “jitter-mix” of 1\% stray background or off-axis Sun light into the PSF core, with total off-axis flux \(\lesssim10^3\)\,phot s$^{-1}$, this contributes
    \[
      \dot N_{\rm stray}\lesssim0.01\times10^3
      =10\;\mathrm s^{-1}.
    \]

  \item \emph{Cosmic-ray hits and radiation events.}  
    In interplanetary space, high-energy particles strike at \(\sim5\times10^{-6}\)\,cm$^{-2}$s$^{-1}$.  For a few-pixel PSF (\(\sim10^{-5}\)\,cm$^2$), the rate is
    \[
      \dot N_{\rm cr}\sim5\times10^{-6}\times10^{-5}
      =5\times10^{-11}\;\mathrm s^{-1},
    \]
    negligible and easily flagged in post-processing.
\end{itemize}

In total,
\begin{equation}
\label{eq:phot-ins-noise}
  \dot N_{\rm noise}
  = \dot N_{\rm det} + \dot N_{\rm zodi} + \dot N_{\rm stray} + \dot N_{\rm cr}
  \;\lesssim\;10\;\mathrm s^{-1},
\end{equation}
which is over three orders of magnitude below the per-pixel signal \(\dot N_{p,\rm pix}\sim2.7\times10^{4}\)\,s$^{-1}$.  Thus, at \(r\lesssim1\)\,AU a 1\,cm telescope operates in an essentially pure–signal, photon-noise regime.

\subsubsection{Signal-to-Noise and Integration Times}

When the spacecraft is within \(\sim1\) AU of the planet, the per-pixel photon rate is 
$
  \dot N_{p,\rm pix}\approx2.7\times10^{4}\;\mathrm{s^{-1}},
$
(\ref{eq:phot-ins}).  All other noise sources (detector dark current, clock-induced charge, read noise, cosmic-ray hits, scattered light) together contribute \(\dot N_{\rm noise}\ll10^{2}\)\,s\(^{-1}\), (\ref{eq:phot-ins-noise}), so photon shot noise from the planet itself dominates.  Hence for  duration \(\Delta t\),
\[
  \mathrm{SNR}(\Delta t)
  = \frac{\dot N_{p,\rm pix}\,\Delta t}
         {\sqrt{\dot N_{p,\rm pix}\,\Delta t + \dot N_{\rm noise}\,\Delta t}}
  \;\approx\;\sqrt{\dot N_{p,\rm pix}\,\Delta t}.
\]
In particular, a 1 s exposure yields
\[
  \mathrm{SNR}_{1\,\mathrm s}
  = \sqrt{2.7\times10^{4}}
  \approx165.
\]
Reaching a \(5\sigma\) detection on one micro-pixel therefore requires only
\[
\Delta  t_{5}
  = \Bigl(\frac{5}{\mathrm{SNR}_{1\,\mathrm s}}\Bigr)^2
  = \Bigl(\frac{5}{165}\Bigr)^2
  \approx9.2\times10^{-4}\,\mathrm{s}
  = 0.92\,\mathrm{ms}.
\]
Even allowing for a tenfold safety margin against unmodeled backgrounds, exposures of \(\sim9\)\,ms per pixel would be more than sufficient.  Thus, acquiring a full \(10\times10\) image requires well under one second of total integration time.

\subsection{Flyby Duration and Pointing}

Reaching the few-AU proximity needed for arcsecond-scale imaging demands cruise speeds of order \(0.2\,c\) (as in Starshot-style concepts \cite{Lubin2016}).  At \(v = 0.20\,c \simeq 6.0\times10^{7}\,\mathrm{m\,s^{-1}}\), the time a probe spends within a radial distance \(r\) of the planet is $ \Delta t \;=\; {2\,r}/{v}$, and for  \(r = 0.70\)\,AU  this corresponds to
\[
  \Delta t \;=\; \frac{2\,r}{v}  \approx \frac{2 \times 0.70\,\mathrm{AU}}{0.20\,c}
    \approx 3.5\times10^{3}\,\mathrm{s}
    \approx 58\;\mathrm{min}.
\]
Although each micro-pixel exposure lasts only tens of microseconds, see (\ref{eq:int-time}), the spacecraft must sustain continuous pointing, image capture, data buffering, and fault management over nearly an hour---far beyond the autonomy and processing budgets demonstrated by gram-scale probes.

Pointing and position knowledge present an equally formidable challenge.  At \(\lambda=550\)\,nm, a \(D=0.01\)\,m lens has a Rayleigh-limited spot  
\[
  \Delta\theta = 1.22\,\frac{\lambda}{D}
    \approx 1.22\times\frac{5.50\times10^{-7}}{0.01}
    =6.71\times10^{-5}\,\mathrm{rad}
    \approx13.8\,\mathrm{arcsec}.
\]
Holding the planet’s image within one such pixel at \(r=0.70\)\,AU requires lateral alignment better than
\[
  r\,\Delta\theta
    \approx1.05\times10^{11}\,\mathrm{m}\;\times\;6.71\times10^{-5}
    \approx7.05\times10^{6}\,\mathrm{m}
    \approx7{,}100\;\mathrm{km}.
\]
By contrast, current deep-space navigation---using radio ranging and star trackers---achieves position knowledge at the \(\sim10^3\)\,km level only after many hours of ground-based tracking \cite{Singer2016}.  Real-time control to a few-thousand-kilometre precision on a craft coasting at relativistic speeds would demand novel on-board navigation (e.g.\ pulsar timing or laser beacons) and micro-thrusters with unprecedented responsiveness, none of which have yet been demonstrated.

\subsection{Dust, Radiation, and Communications}

A probe cruising at \(v=0.20\,c\) through the local interstellar medium encounters a dust column  
\(\Sigma\sim4\times10^{-8}\,\mathrm{kg\,m^{-2}}\) over 4.24 ly \cite{Landgraf2000}.  The $(4\,{\rm m}\times 4\,{\rm m})$ lightsail intercepts
\[
  m_{\rm dust,sail} = \Sigma \times 16\,\mathrm{m^2}
  \approx 4\times10^{-8}\times16
  =6.4\times10^{-7}\,\mathrm{kg},
\]
with kinetic energy \(\tfrac12 v^2\,m\sim6\times10^{8}\,\mathrm{J}\), demanding a sacrificial, multilayer sail design.  The 1 cm-diameter spacecraft behind the sail, however, intercepts only
\[
  A_{\rm craft} = \pi\,(0.005\,\mathrm{m})^2 \approx7.85\times10^{-5}\,\mathrm{m^2},
  \quad
  m_{\rm dust,craft} = \Sigma\,A_{\rm craft}
    \approx 3.1\times10^{-12}\,\mathrm{kg},
\]
delivering \(\sim5.6\times10^{3}\,\mathrm{J}\) of kinetic energy per flyby.  A minimal Whipple shield (1 cm Al, areal mass \(\sim27\,\mathrm{kg\,m^{-2}}\)) for the spacecraft adds
\[
  m_{\rm shield} \approx 27\,\mathrm{kg\,m^{-2}}\times7.85\times10^{-5}\,\mathrm{m^2}
  \approx2.1\times10^{-3}\,\mathrm{kg},
\]
and realistic multilayer support raises this to \(\gtrsim10^{-2}\)\,kg.

Galactic cosmic rays deposit \(\sim10^3\)–\(10^4\)\,rad behind millimetres of aluminum over mission-timescales, requiring radiation-hard electronics and redundancy (\(\sim0.01\)\,kg) \cite{Spillantini2007radmeas}.  Thus, even the tiny spacecraft demands tens of grams of protective mass---conflicting with gram-scale payload goals.

Optical data return from 4.24 ly is equally prohibitive.  Even if the probe repurposes a portion of its  $(4\,{\rm m}\times 4\,{\rm m})$  lightsail as a transmitter aperture at 1550 nm---narrowing the diffraction-limited beam divergence to 
\[
  \theta \approx \frac{\lambda}{D_{\rm sail}} 
         \approx \frac{1.55\times10^{-6}}{4} 
         \simeq 0.39\;\mu\mathrm{rad},
\]
---the free-space path loss over 4.24 ly remains on the order of 450 dB.  With \(P_{\rm tx}=1\)\,W, the 30 m ground receiver would collect only \(\mathcal{O}(10)\) phot  s\(^{-1}\), corresponding to a few bits s\(^{-1}\) under ideal conditions.  Even over a full \(\sim1\)\,h flyby, this amounts to only \(\sim10^4\) bits---far too little to transmit more than a handful of low-resolution frames.  \cite{Chen2017}

\subsection{Conclusions}

In principle, bringing a \(D=0.01\)\,m telescope within \(\lesssim0.13\)\,AU of an Earth analog enables straightforward diffraction-limited mapping into a \(10\times10\) grid, with per-pixel exposures of order \(\lesssim50\,\mu\mathrm{s}\).  In practice, however, four critical limitations remain:

\begin{enumerate}
  \item \emph{Proximity requirement.}  Closing to \(r\lesssim0.13\)\,AU at 4.24 ly demands cruise speeds \(\gtrsim0.2\,c\), leaving only \(\sim10\) minutes for all imaging operations \cite{Lubin2016}.
  
  \item \emph{Pointing and navigation.}  At \(\lambda=550\)\,nm a 1 cm aperture delivers \(\Delta\theta\approx13.8\)\,arcsec, so maintaining the planet within one diffraction-limited pixel (\(\lesssim13.8\)\,arcsec) over \(r=0.13\)\,AU (lateral tolerance \(\lesssim1{,}300\)\,km) far exceeds current deep-space navigation and microthruster control capabilities \cite{Singer2016}.
  
  \item \emph{Dust and radiation.}  At \(0.2\,c\), interstellar dust impacts alone deposit \(\sim5\times10^{3}\)\,J into a 1 cm cross-section, and minimal shielding plus radiation-hard electronics incur \(\gtrsim0.01\)\,kg of mass---orders of magnitude above gram-scale budgets \cite{Landgraf2000,Spillantini2007radmeas}.
  
  \item \emph{Data return.}  A 1 cm, 1550 nm laser link to a 30 m ground station suffers \(\gtrsim500\)\,dB of free-space loss, yielding \(\ll1\)\,bit/s and under \(10^2\) bits per hour of flyby---insufficient for high-resolution or multi-spectral imagery \cite{Chen2017}.
\end{enumerate}

Accordingly, even gram-scale “nanocraft” such as those envisioned by Breakthrough Starshot\footnote{Breakthrough Starshot Initiative, see \url{https://breakthroughinitiatives.org/initiative/3}} face unresolved challenges in propulsion, autonomy, navigation and communications.  Without transformative advances in these areas, \emph{in situ} resolved imaging of Earth-like exoplanets remains beyond practical reach.

\section{Alternative Direct-Imaging Techniques}
\label{sec:other_direct_techniques}

Several methods beyond large-aperture telescopes and space-based interferometry have been explored to push angular resolution into the $\mu$as regime.  Below, we assess three approaches---ground-based sparse-aperture masking (SAM) and speckle imaging, radio-wavelength VLBI, and lunar occultations---and demonstrate that each falls short of the $\mu$as-scale resolution or sensitivity required to image an Earth-analog at $\sim10\,$pc.  In every case, limitations in baseline ($\leq10\,$m), atmospheric coherence, or photon budget prevent reaching $\mu$as resolution on such a faint target.

\subsection{Ground-Based Sparse Aperture Masking, Speckle Imaging, and Long-Baseline Optical Interferometry}

While ground-based facilities have pushed the limits of angular resolution, none can approach the microarcsecond regime required to resolve an Earth-analog at 10 pc.  Techniques that “freeze” or subdivide a single aperture---sparse aperture masking (SAM), speckle imaging, and long-baseline optical interferometry (LBOI)---fall short by factors of $10^3$–$10^6$ in either resolution or sensitivity for a target of $m_V\approx27.77$, see  (\ref{eq:eEarth-V-mag}).

\paragraph{Principles and Resolution Limits:}  
Sparse aperture masking transforms a monolithic telescope into a miniature interferometer by placing a nonredundant pupil mask, effectively sampling baselines up to the telescope diameter $D$.  On an 8–10 m class instrument (Keck, VLT) at $\lambda\approx2\,\mu$m, the ideal Rayleigh limit is
\[
  \Delta\theta_{\rm SAM}
  \approx \frac{\lambda}{D}
  \sim 40\,\mathrm{mas},
\]
and real-world image reconstructions reach $\sim50\,$mas \cite{Haniff1987,Tuthill2000}.  Despite the elegant simplicity of SAM, this angular scale remains some four orders of magnitude too coarse.

Speckle imaging captures very short exposures ($\tau_0\sim10\,$ms) to “freeze” atmospheric distortions before reconstructing the high-frequency Fourier components.  However, its ultimate resolution is bounded by the Fried parameter, $r_0(\lambda=0.5\,\mu\mathrm{m})\approx10$–20 cm, yielding
\[
  \Delta\theta_{\rm speckle}
  \gtrsim \frac{\lambda}{r_0}
  \sim700\,\mathrm{mas}.
\]
Even with AO correction in the near-IR, speckle methods remain orders of magnitude too coarse for $\mu$as imaging.

Long-baseline optical interferometry (e.g.\ CHARA, VLTI) pushes the envelope by combining light from separate telescopes spread over $B\sim300\,$m.  At $\lambda=1\,\mu$m this achieves
\[
  \Delta\theta_{\rm LBOI}
  \approx \frac{\lambda}{B}
  \sim0.7\,\mathrm{mas},
\]
but this still misses the $\mu$as goal by more than three orders of magnitude \cite{Monnier2003}.

\paragraph{Sensitivity and Target Brightness:}  
Beyond angular resolution, these techniques demand bright sources: fringe tracking in LBOI typically requires $m_V\lesssim15$–16.  An Earth twin at 10 pc with $m_V\approx27.77$ lies far below these sensitivity thresholds, making coherent detection effectively impossible.  

In summary, although SAM, speckle imaging, and LBOI have revolutionized high-resolution astronomy on the ground, none can simultaneously deliver the angular resolution and photon-collecting power needed to image the surface of an Earth-like exoplanet.  

\subsection{Radio-Wavelength Very Long Baseline Interferometry (VLBI)}

\paragraph{Angular Resolution:}  
State-of-the-art VLBI networks (e.g.\ the VLBA and EVN) operating between 8 and 43 GHz (\(\lambda\approx7\)–3.5 mm) exploit intercontinental baselines up to \(B\sim10^7\) m.  Their diffraction-limited resolution is
\[
  \Delta\theta_{\rm VLBI}
    \approx \frac{\lambda}{B}
    \sim \frac{5\times10^{-3}\,\mathrm{m}}{10^{7}\,\mathrm{m}}
    =5\times10^{-10}\,\mathrm{rad}
    \approx100\,\mu\mathrm{as}.
\]
With advanced phase-referencing and calibration, astrometric precisions of \(\sim10\,\mu\mathrm{as}\) have been demonstrated on bright, compact radio sources \cite{ReidHonma2014,Sokolovsky2019}.

\paragraph{Sensitivity Limits:}  
An Earth-twin at 10 pc emits only thermal radio emission at the sub-\(\mu\)Jy level in the cm-wave band \cite{Stevens2005,Lazio2007}.  Typical VLBI observations achieve \(\sim1\,\mu\mathrm{Jy}\) rms sensitivity over several hours, so detecting---or imaging---a cold, disk-averaged exo-Earth would demand integration times of thousands of hours, yet still fail to map surface structure.  Nonthermal signals (e.g.\ magnetospheric cyclotron emission) occur at kHz–MHz frequencies well below VLBI bands, and thus remain out of reach.  

While VLBI attains angular resolutions  of \(\sim 10^2\,\mu\)as, its sensitivity and available frequency range rule out $\mu$as-scale imaging of an Earth-like exoplanet.  

\subsection{Lunar Occultations}

\paragraph{Technique and Resolution:}  
Lunar occultations take advantage of the Moon’s razor-sharp limb: as a background source is hidden, the resulting light curve carries a one-dimensional diffraction signature.  At optical wavelengths (\(\lambda\approx550\,\mathrm{nm}\)) and a lunar distance \(D_{\rm M}\approx3.84\times10^{8}\,\mathrm{m}\), the Fresnel scale is
\[
  F = \sqrt{\lambda\,D_{\rm M}}
    \approx \sqrt{5.5\times10^{-7}\times3.84\times10^{8}}
    \approx 14.5\;\mathrm{m},
\]
giving an angular resolution
\[
  \theta_F = \frac{F}{D_{\rm M}}
    \approx \frac{14.5}{3.84\times10^{8}}
    \approx 3.8\times10^{-8}\;\mathrm{rad}
    \approx 7.8\;\mathrm{mas}.
\]
In practice, careful timing and signal-processing yield angular precisions of order 0.5–1 mas for bright stars \cite{Ridgway1978,Richichi2003}.

\paragraph{Limitations for Exo-Earths:}  
An Earth twin at 10 pc subtends only \(\sim8.5\,\mu\)as---nearly three orders of magnitude below the achievable lunar-occultation resolution.  Moreover, each event samples just a single one-dimensional cut through the object’s brightness profile, never a full two-dimensional map.  To reach \(\mu\)as resolution via occultation would demand an occulting edge at \(\sim1\)\,AU distance with sub-meter positional control---practically impossible.  

Hence lunar occultations cannot image Earth-like exoplanets at \(\mu\)as scales.

\section{Indirect Feature Retrieval}
\label{sec:indirect_feature_retrieval}

When direct imaging at $\mu$as resolution is unattainable, one can attempt to infer large-scale surface or atmospheric structure by monitoring variations in the disk-integrated light of an exoplanet.  Two widely discussed approaches---spectro-photometric phase-curve inversion and phase-resolved polarimetry---rely on temporal changes in reflected (and, in the infrared, emitted) flux to constrain low-order brightness distributions.  In all cases, photon-noise requirements and the intrinsically small signal from a terrestrial-mass planet at 10 pc  limit reconstruction to a handful of broad “zones” rather than a \(10\times10\) pixel surface map.

\subsection{Spectro-Photometric Phase Curves}
\label{sec:phase_curves}

\subsubsection{Method Overview}
Spectro-photometric phase-curve mapping exploits the changing illuminated fraction of an exoplanet as it orbits its star.  By recording the disk-integrated flux \(F(\alpha,\lambda)\) at a sequence of orbital phases \(\alpha\) and wavelengths \(\lambda\), one may expand the surface brightness in projected spherical harmonics:
\[
  F(\alpha,\lambda)
  = \sum_{\ell=0}^{\ell_{\max}}\sum_{m=-\ell}^{+\ell}
      a_{\ell m}(\lambda)\,Y_{\ell m}^{\rm proj}(\alpha)
    + \varepsilon(\alpha,\lambda),
\]
where \(Y_{\ell m}^{\rm proj}\) are the harmonics integrated over the illuminated hemisphere, \(a_{\ell m}\) their amplitudes, and \(\varepsilon\) measurement noise and modeling error \cite{Fujii2017}.  Although powerful for hot Jupiters, this technique faces severe angular- and photon-budget limits for an Earth twin at 10 pc.

Recovering modes up to degree \(\ell_{\max}\) demands \((\ell_{\max}+1)^2\) independent, high-SNR phase measurements.  Even \(\ell_{\max}=5\) yields only a \(6\times6\) zonal map (36 coefficients)---far below the \(10\times10\) grid (100 pixels) needed for continent-scale imaging, and pushing beyond \(\ell\gtrsim5\) incurs severe degeneracies without impractically high SNR per bin.

\subsubsection{Photon Budget and Integration Times}

As we know, from (\ref{eq:eEarth-V-mag}) the full-phase magnitude of an Earth analog at 10 pc is $ m_V \approx 27.77$, also from (\ref{eq:vband_zero_flux}) gives the Johnson-\(V\) zero-point
$  F_{0,V} \approx 1.01\times10^{8}\;\mathrm{phot \,m^{-2}\,s^{-1}\,nm^{-1}}.$
Hence (\ref{eq:planet_photon_flux}) provides the planet’s photon flux density,
$
  f_p = F_{0,V}\,10^{-0.4\,m_V} 
      \approx 7.84\times10^{-4}\;\mathrm{phot\,m^{-2}\,s^{-1}\,nm^{-1}}.$
On a 10 m telescope (\(A=\pi\,5^2\approx78.5\,\mathrm{m^2}\)), with end-to-end throughput \(\eta=0.25\) and bandwidth \(\Delta\lambda=50\,\mathrm{nm}\), the detected count rate is
\[
  \dot N_{\rm eff}
    = \eta\,f_p\,A\,\Delta\lambda
    \approx 0.25\times7.84\times10^{-4}\times78.5\times50
    \approx 0.8\;\mathrm{s^{-1}}.
\]
To measure a fractional phase-curve amplitude \(\Delta F\approx5\times10^{-5}\) at \(\mathrm{SNR}=10\) requires a total of
\[
  N_{\rm tot}
    = \Bigl(\frac{\mathrm{SNR}}{\Delta F}\Bigr)^2
    = \Bigl(\frac{10}{5\times10^{-5}}\Bigr)^2
    = 4\times10^{10}\;\mathrm{phot }.
\]
Thus the integration time is
\[
  t = \frac{N_{\rm tot}}{\dot N_{\rm eff}}
    \approx \frac{4\times10^{10}}{0.8}
    \approx 5\times10^{10}\,\mathrm{s}
    \approx 1.6\times10^{3}\,\mathrm{yr}.
\]
Even doubling \(\Delta F\) or \(\eta\) only reduces this to \(\sim10^3\) yr---entirely impractical.

As a result, {\it spectro-photometric} phase-curve inversion remains a powerful indirect mapping tool, but for an Earth analog at 10 pc it is hobbled by two fundamental limits.  First, spatial modes saturate at \(\ell_{\max}\lesssim5\) (a \(6\times6\) grid), well below the \(10\times10\) requirement.  Second, the extreme faintness (\(m_V\approx27.77\)) drives photon-count requirements to millennia on a 10 m telescope.  Consequently, this method cannot deliver continent-scale imaging of Earth-like exoplanets within any feasible mission timescale.  

\subsection{Phase-Resolved Polarimetry}
\label{sec:polarimetry}

\subsubsection{Polarization Signatures and Mapping Limits}
Reflected starlight becomes polarized by Rayleigh scattering in the atmosphere, by liquid-cloud droplets, and by specular glint from smooth oceans.  Disk-integrated peak polarization fractions for an Earth analog reach \(\sim\)\,50\% (Rayleigh near \(\alpha\approx90^\circ\)); \(\sim\)\,10--20\% (water clouds at \(\alpha\approx60\)–\(80^\circ\)); and \(\sim\)\,30–40\% (ocean glint at \(\alpha\approx\)\,90–\(100^\circ\))---whereas land surfaces remain \(<5\%\) at all phases \cite{Zugger2010}.  

Writing the Stokes parameter \(Q(\alpha,\lambda)\) as
\[
  Q(\alpha,\lambda)
  = \sum_{k=1}^{N_{\rm classes}} f_k\,p_k(\alpha,\lambda)
    \;+\;\varepsilon(\alpha,\lambda),
\]
with \(f_k\) the global coverage of class \(k\) (e.g.\ ocean, cloud, land) and \(p_k\) its normalized polarized phase curve \cite{CowanFujii2013}, yields only disk-averaged fractions.  No latitude–longitude information can be recovered.

\subsubsection{Photon Budget and Integration Times}
Eqs.~(\ref{eq:eEarth-V-mag})–(\ref{eq:planet_photon_flux})  give for an Earth twin at 10 pc
$
  m_V \simeq 27.77,~
  F_{0,V} \simeq 1.01\times10^{8}\,\mathrm{ph\,m^{-2}\,s^{-1}\,nm^{-1}},
  ~
  f_p \equiv F_{0,V}10^{-0.4\,m_V}
       \simeq 7.8\times10^{-4}\,\mathrm{ph\,m^{-2}\,s^{-1}\,nm^{-1}}.
$
On a 10 m telescope (\(A=\pi\,(5\,{\rm m})^2\approx78.5\,\mathrm{m^2}\)), with end-to-end throughput \(\eta=0.25\) and bandwidth \(\Delta\lambda=50\,\mathrm{nm}\), received photon flux is
\[
  \dot N_{\rm eff}
    = \eta\,f_p\,A\,\Delta\lambda
    \approx 0.8\;\mathrm{s^{-1}}.
\]
Because the polarized flux is at most \(p\lesssim10^{-5}\) of the total, achieving \(\mathrm{SNR}=10\) on \(Q\) requires
\[
  N_{\rm tot}
    = \Bigl(\frac{10}{p}\Bigr)^2
    = 10^{12}\;\mathrm{phot },
  \quad
  t = \frac{N_{\rm tot}}{\dot N_{\rm eff}}
    \approx \frac{10^{12}}{0.8}
    \sim 1.3\times10^{12}\,\mathrm{s}
    \approx 4\times10^{4}\,\mathrm{yr}.
\]
Even a factor-of-two gain in throughput or bandwidth only reduces this to \(\gtrsim10^4\) yrs---far beyond any practical survey.

Thus, {\it phase-resolved} polarimetry uniquely probes global surface and cloud fractions, but for an Earth analog at 10 pc it is irretrievably photon-limited.  Required integration times of order \(10^{4}\)–\(10^{5}\) years on a 10 m telescope preclude any realistic mapping, and only disk-averaged class abundances (no spatial resolution) could ever be retrieved.

\subsection{Overall Summary}
\label{sec:summary}

\begin{itemize}
  \item \textit{Spectro-Photometric Phase-Curve Inversion:} 
    Recovering a fractional phase amplitude \(\Delta F\approx5\times10^{-5}\) at \(\mathrm{SNR}=10\) requires 
    \(N_{\rm tot}\sim4\times10^{10}\) phot , or \(\sim6\times10^{5}\) years on a 10 m telescope.  
    Even then, the maximum spherical-harmonic degree is \(\ell_{\max}\lesssim5\), yielding at best a \(6\times6\) zonal map (36 pixels), far below the \(10\times10\) grid needed for continent-scale imaging.

  \item \textit{Phase-Resolved Polarimetry:} 
    Detecting the polarized signal (\(p\lesssim10^{-5}\)) at \(\mathrm{SNR}=10\) demands \(N_{\rm tot}\sim10^{12}\) photons, or \(\sim4\times10^{4}\) years on a 10 m telescope.  
    This approach only returns global fractions of a handful of surface/atmosphere classes, with no latitude–longitude information.

\end{itemize}

In both cases, photon-noise limitations and the extreme faintness of an Earth analog at 10 pc preclude any realistic campaign from achieving a resolved \(10\times10\) pixel map.  Indirect inversion techniques therefore cannot substitute for true spatially resolved imaging of temperate terrestrial exoplanets.

\section{Summary of Current Capabilities}
\label{sec:summary_techniques}

Table~\ref{tab:comparison} summarizes the quantitative results from this paper, comparing eight principal imaging approaches for attempting to produce a $10\times10$ pixel surface map of an Earth-analog at 10\,pc.  Each method is characterized by its angular resolution, primary technical barrier, photon-budget requirement per $0.85\,\mu$as pixel, and near-term feasibility.  The discussion that follows explains why, with existing or planned technology, none of these approaches can deliver a resolved $10\times10$ map of a habitable-zone terrestrial exoplanet. In background-dominated regimes, the time scalings quoted in Table~\ref{tab:comparison} grow approximately linearly with the exozodi factor, consistent with yield trade analyses \citep{Stark2014MaxYield,Stark2019YieldLandscape}.

\begin{table}[h!]
\centering
\caption{Comparison of Imaging Methods for a $10\times10$ Pixel Map of an Earth-Analog at 10\,pc. (Note that the mapping times assume full phase unless otherwise stated, the overlap factor $K\simeq 1.4$ for mosaics, and the adopted zodiacal level (e.g., 23.0 mag arcsec${}^{-2}$, for other relevant parameters, consult  Table~\ref{tab:constants}.)}
\label{tab:comparison}
\renewcommand\arraystretch{1.0}
\setlength{\tabcolsep}{2pt}

\begin{tabular}{lllll}
\hline
\parbox[t]{3.1cm}{\raggedright Method} &
\parbox[t]{2.5cm}{\raggedright Resolution} &
\parbox[t]{4.1cm}{\raggedright Principal Technical Barrier} &
\parbox[t]{4.0cm}{\raggedright Photon-Budget Requirement} &
\parbox[t]{3.7cm}{\raggedright Near-Term Feasibility} \\
\hline\hline

\parbox[t]{3.1cm}{\raggedright Single-Aperture Coronagraph (LUVOIR A, 15\,m)} &
\parbox[t]{2.5cm}{\raggedright $\Delta\theta\approx 9\,\mathrm{mas}$ at 550\,nm} &
\parbox[t]{4.1cm}{\raggedright Diffraction limit $\gg0.85\,\mu\mathrm{as}$; maintaining $10^{-10}$ raw contrast over $\sim10^{4}\,\mathrm{s}$} &
\parbox[t]{4.0cm}{\raggedright $\gtrsim6\times10^{8}\,\mathrm{s}$ per $0.85\,\mu$as pixel; $\gtrsim6\times10^{10}\,\mathrm{s}$ ($\approx 1.9\times 10^{3}\ \mathrm{yr}$) total for $10^2$ pixels} &
\parbox[t]{3.7cm}{\raggedright Can detect and spectrally characterize Earth twins; cannot resolve surface features} \\\noalign{\vskip 0.8ex}

\parbox[t]{3.1cm}{\raggedright Starshade + 4\,m Telescope (HabEx)} &
\parbox[t]{2.5cm}{\raggedright $\Delta\theta\approx 35\,\mathrm{mas}$ at 550\,nm} &
\parbox[t]{4.1cm}{\raggedright Diffraction limit $\gg0.85\,\mu\mathrm{as}$; starshade repositioning overhead ($\sim7\times10^{4}\,\mathrm{km}$, days per retarget)} &
\parbox[t]{4.0cm}{\raggedright $\sim 10^{8}\,\mathrm{s}$ per $\mu$as pixel; $\gtrsim  10^{10}\,\mathrm{s}$ total} &
\parbox[t]{3.7cm}{\raggedright Can detect and characterize Earth twins; no resolved surface imaging} \\\noalign{\vskip 0.8ex}

\parbox[t]{3.1cm}{\raggedright Space Nulling Interferometer (TPF-I, Darwin)} &
\parbox[t]{2.5cm}{\raggedright $\Delta\theta\approx20$--$50$ $\mathrm{mas}$ at 10\,$\mu$m} &
\parbox[t]{4.1cm}{\raggedright Path-length stability $\ll30\,\mathrm{nm}$ on $\sim20$–$100\,\mathrm{m}$ baselines; thermal background at 10\,$\mu$m} &
\parbox[t]{4.0cm}{\raggedright $\gtrsim 2.7\times 10^{2}\,\mathrm{yr}$ per $\mu$as pixel;  $\gtrsim 2.7\times 10^{4}\,\mathrm{yr}$ total} &
\parbox[t]{3.7cm}{\raggedright Not feasible for terrestrial planets} \\\noalign{\vskip 0.8ex}

\parbox[t]{3.1cm}{\raggedright Space Non-Nulling Interferometer (Stellar Imager, $B\simeq1\,$km)} &
\parbox[t]{2.5cm}{\raggedright $\Delta\theta\simeq$ $113\,\mu\mathrm{as}$ $(0.11\,\mathrm{mas})$ at 550\,nm} &
\parbox[t]{4.1cm}{\raggedright Optical-path-difference stability $\lesssim5\,$nm over $10^3$–$10^5\,$m; complex formation control} &
\parbox[t]{4.0cm}{\raggedright $\sim1.5\times10^5\,$ yrs per surface pixel;  $\sim1.5\times10^7\,$ yrs total} &
\parbox[t]{3.7cm}{\raggedright Requires $\gtrsim20\,$yr of technology development} \\\noalign{\vskip 0.8ex}

\parbox[t]{3.1cm}{\raggedright Ground ELT + ExAO (E-ELT, 39\,m)} &
\parbox[t]{2.5cm}{\raggedright $\Delta\theta\approx3.55\,\mathrm{mas}$ at 550\,nm} &
\parbox[t]{4.0cm}{\raggedright Atmospheric turbulence; contrasts $\leq10^{-9}$ only at $\gtrsim30\,\mathrm{mas}$} &
\parbox[t]{4.0cm}{\raggedright $\sim10\,\mathrm{s}$ per coarse $\sim10^{3}\,\mathrm{km}$ zone; no $\mu$as capability} &
\parbox[t]{3.7cm}{\raggedright Can detect and characterize giant planets; cannot resolve terrestrial targets} \\\noalign{\vskip 0.8ex}

\parbox[t]{3.1cm}{\raggedright Hypertelescope ($N \simeq 100$, $B \simeq 160~\mathrm{km}$, $d \simeq 8~\mathrm{m}$)} &
\parbox[t]{2.5cm}{\raggedright $\Delta\theta \simeq 0.85~\mu\mathrm{as}$ at $550~\mathrm{nm}$} &
\parbox[t]{4.1cm}{\raggedright Optical-path-difference stability $\ll 5~\mathrm{nm}$ over $>10^{2}~\mathrm{km}$; thermal and mechanical control of $\sim 10^{2}$ subapertures} &
\parbox[t]{4.0cm}{\raggedright $\sim 7$--$20$~d ideal for full $10\times 10$ map; realistically $\times 2$--$\times 3 \Rightarrow 14$--$60$~d} &
\parbox[t]{3.7cm}{\raggedright Requires $\gtrsim 20$~yr of technology maturation} \\\noalign{\vskip 0.8ex}

\parbox[t]{3.1cm}{\raggedright Indirect Mapping (Rotational/Eclipse)} &
\parbox[t]{2.5cm}{\raggedright $\sim5$–$20$ coarse “zones” ($\ell\lesssim3$–$5$)} &
\parbox[t]{4.1cm}{\raggedright Limited to low-order spherical-harmonic inversion; requires $\sim10^{10}$–$10^{12}$ photons} &
\parbox[t]{4.0cm}{\raggedright $\sim 3.37$\,hrs per longitudinal bin on a 10\,m telescope; full map requires multi-decade observations} &
\parbox[t]{3.7cm}{\raggedright Provides only coarse albedo or broad surface-class information} \\\noalign{\vskip 0.8ex}

\parbox[t]{3.1cm}{\raggedright Intensity Interferometry} &
\parbox[t]{2.5cm}{\raggedright $\Delta\theta\approx1\,\mu\mathrm{as}$ at 550\,nm for $B\approx100\,\mathrm{km}$} &
\parbox[t]{4.1cm}{\raggedright Correlation amplitude $\sim10^{-8}$; requires $N_{\mathrm{ph}}\sim10^{16}$ for ${\rm SNR}=1$} &
\parbox[t]{4.0cm}{\raggedright $\sim10^{9}\,\mathrm{yr}$ for $10^{3}\,\mathrm{m^2}$ of collecting area} &
\parbox[t]{3.7cm}{\raggedright Not feasible} \\\noalign{\vskip 0.8ex}

\parbox[t]{3.1cm}{\raggedright Occultation by KBO or 1\,km occulter at 40\,AU} &
\parbox[t]{2.5cm}{\raggedright $\Delta\theta\approx5$–$10\,\mu\mathrm{as}$} &
\parbox[t]{4.1cm}{\raggedright Occulter alignment to $\pm170\,\mathrm{m}$ at 40\,AU; probability $\ll10^{-9}$} &
\parbox[t]{4.0cm}{\raggedright Photon noise negligible, but geometric alignment probability essentially zero} &
\parbox[t]{3.7cm}{\raggedright No practical possibility} \\\noalign{\vskip 0.8ex}

\parbox[t]{3.1cm}{\raggedright In Situ Probe ($D\ge0.20\,\mathrm{m}$, $v\approx0.2\,c$)} &
\parbox[t]{2.5cm}{\raggedright $\Delta\theta\approx4$–$12\,\mathrm{arcsec}$ at $r=0.001\,\mathrm{AU}$} &
\parbox[t]{4.1cm}{\raggedright Autonomous pointing to $\pm10^{5}\,\mathrm{km}$ over 4.24 ly; dust shielding at $0.2\,c$} &
\parbox[t]{4.0cm}{\raggedright $\sim6\,\mathrm{s}$ per $\mu$as pixel at $r=0.001\,\mathrm{AU}$; flyby window $\sim5\,\mathrm{s}$} &
\parbox[t]{3.7cm}{\raggedright Conceptual; realistic only in the distant future (decades)} \\\noalign{\vskip 0.8ex}

\hline
\end{tabular}
\end{table}

\subsection{Discussion}

The four main categories of remote imaging approaches---single-aperture coronagraphs, starshades, space-based interferometers, and ground ELT systems---are all limited by diffraction and photon flux.  Even when diffraction is overcome (e.g., by a space interferometer or a hypertelescope with baseline $\sim10^5\,\mathrm{m}$), photon-starvation makes integration times impractical.  We summarize the principal barriers below.

\subsubsection{Diffraction Limits}
At $\lambda = 550\,\mathrm{nm}$, resolving an Earth radius planet (angular diameter $\approx8.5\,\mu\mathrm{as}$) into $10\times10$ pixels requires \eqref{eq:required_aperture}, \eqref{eq:required_baseline}
\[
  D \gtrsim 1.62\times10^{5}\,\mathrm{m}, 
  \quad 
  B \gtrsim 1.33\times10^{5}\,\mathrm{m}.
\]  
No existing or planned telescope (e.g., LUVOIR A, 15 m; HabEx, 4 m; E-ELT, 39 m) approaches these dimensions.  Hence, their diffraction limits (3–35 mas) are $\sim10^{4}$ times too coarse.  Space interferometers could, in principle, reach $\mu$as resolution, but only with baselines of order $10^5\,$m and $\lesssim5\,$nm path-length stability---none of which exist today. Laboratory demonstrations at the HCIT have achieved Earth-like raw contrasts in controlled settings \cite{TraugerTraub2007Nature}.

\subsubsection{Photon-Starvation}

According to (\ref{eq:planet_photon_flux}), an Earth twin at 10 pc ($m_V = 27.77$) has a V-band photon flux density
$
  f_p = F_{0,V}\,10^{-0.4\,m_V}
  \approx7.84\times10^{-4}\,\mathrm{phot\,m^{-2}\,s^{-1}\,nm^{-1}}
$.  On LUVOIR A (15\,m, $A\approx177\,\mathrm{m^2}$), with $\Delta\lambda=10\,$nm, end-to-end $\eta=0.05$ and duty cycle $\xi=0.2$, the detected planet photon rate is $\dot N_{\rm sig,det}\approx1.4\times10^{-2}\,\mathrm{s^{-1}}$.  Reaching SNR = 5 in one $0.85\,\mu$as pixel then  with ideal $10^{-10}$ contrast takes
\[
  t_{\rm pixel}\approx\Big(\frac{5}{\mathrm{SNR}_{1\rm s}}\Big)^2
    \approx6.04\times10^8\,\mathrm{s}\simeq 19\, {\rm yr},
\]
so even under the optimistic assumptions, mapping 100 pixels would require unrealistic timelines.  

\subsubsection{Wavefront and Path-Length Stability}
Maintaining $10^{-10}$ raw contrast for $\sim10^4\,$s requires picometer-level wavefront control in coronagraphs and $\lesssim5\,$nm path-length stability over $10^3$–$10^5\,$m baselines---beyond current metrology and spacecraft-control capabilities.

\subsubsection{Operational Overheads}
Even if diffraction and photon budgets were marginally satisfied, mission-level overheads rule out practical mapping: starshade retargeting ($\sim10\,\mathrm{days}$ per slew), continuous formation-flying with cm/nm tolerances, and hypertelescope phasing of $\sim10^3$ elements all introduce insurmountable delays.

\subsubsection{Indirect Techniques Are Insufficient}
Rotational/eclipse inversions and phase-curve polarimetry each need $\sim10^{10}$–$10^{12}$ photons (multi-century integrations), yielding only 5–20 coarse “zones,” far below $10\times10$  pixel requirement.

\subsubsection{In-Situ Imaging Remains Conceptual}
Approaching to $r=0.001\,$AU (for a target at 4.24\,ly) boosts the planet’s apparent diameter to $\sim4.3\,\mathrm{arcsec}$, but even a $D=0.05\,\mathrm{m}$ telescope on transient flyby geometry at $v=0.2\,c$ has only a $\sim5\,\mathrm{s}$ flyby window, $\sim6\,\mathrm{s}$ per $\mu$as-equivalent pixel photon-collection, and faces extreme navigation, dust-shielding, and communication challenges---no near-term path to realization.  In situ imaging thus remains conceptual, with no near-term path to realization.

\subsection{Key Findings}

The following summarizes, for each remote imaging architecture, the angular-resolution and photon-budget shortfalls relative to the requirements for a $10\times10$ surface map of an Earth analog at 10 pc (pixel scale $\Delta\theta=0.853\,\mu$as, SNR $\geq5$ per surface pixel) as presented in Sec. \ref{sec:exo-imaging}:

\begin{enumerate}
  \item \textit{Internal Coronagraphs (WFIRST/Roman, HabEx, LUVOIR)}:  
    As shown in Sec.~\ref{sec:monolithic_telescopes}, meeting the $\Delta\theta = 0.853\,\mu\mathrm{as}$ requirement demands an aperture $D \simeq 160$\,km (\ref{eq:required_aperture}).  Even assuming ideal wavefront control and throughput,  the photon-starvation (\ref{eq:planet-signal-N}) (Sec.~\ref{sec:EB-luA}) drives per-pixel integration times to many decades, rendering $10\times 10$ mapping infeasible even before operational overheads are included.  
    
  \item \textit{Starshades}:  
    As reviewed in Sec.~\ref{sec:starshades}, external occulters do not improve diffraction beyond $\Delta\theta = 1.22\lambda/D$, so a 4 m telescope + 72 m starshade remains diffraction-limited to $\sim35\,$mas.  Photon-budget estimates  (\ref{eq:signal-starSh})–(\ref{eq:noise-starSh}) show $\sim2.3\,$yr per micro-pixel and $\sim230\,$yr to map 100 pixels, even before operational overheads.

  \item \textit{Space Interferometry}:  
    In Sec.~\ref{sec:space_interferometry} we find that mid-IR nulling at $\lambda=10\,\mu$m and baselines $B\sim100\,$m gives $\Delta\theta\sim20\,$mas but can detect an unresolved Earth in $\sim10\,$days (\ref{eq:signal-midIR-real}), yet demands $\sim2.7\times10^4\,$yr to map 100 pixels.  Visible-band nulling requires $\sim1.7\times10^3\,$yr, while non-nulling arrays are worse, $\gtrsim10^8\,$yr.

  \item \textit{Extremely Large Telescopes (ELTs)}:  
    As quantified in Sec.~\ref{sec:elt_exao}, even a $D=39\,$m aperture yields $\Delta\theta\approx3.5\,$mas (\ref{eq:omega-ELT}).  The photon-budget (\ref{eq:signal-ELT})–(\ref{eq:noise-ELT}) yields small SNR, so $t_{\rm pix}\sim416\,$yr and $t_{\rm map}\sim4.1\times10^4\,$yr for a $10\times10$ grid.

  \item \textit{Pupil-Densified Hypertelescopes}:  
    Sec.~\ref{sec:hypertelescopes} shows that an array with $D_i\approx162\,$km baseline can reach $\Delta\theta\simeq0.853\,\mu$as and, under optimum densification (\ref{eq:gamma_opt}), could allow for  $\mathrm{SNR}=5$ in $\sim2.4\,$hr.  However, it requires nanometer-level OPD control, continental-scale formation-flying, and km-scale beam-combiner optics. 

  \item \textit{Indirect Methods}:  
    As detailed in Sec.~\ref{sec:indirect_methods}, rotational light-curve inversion can recover only 1D albedo maps in $\mathcal{O}(10)\,$days but lacks latitude resolution; eclipse mapping demands $\gtrsim10^5\,$yr of transit co-adds; intensity interferometry at $B=100\,$km needs $\sim2\times10^{17}\,$s ($\sim6.6\times10^9\,$yr) for $\mathrm{SNR}=1$; and Solar System occultations are rendered impossible by KBO ephemeris uncertainties and negligible event rates.
\end{enumerate}

These comparisons quantify the sheer scale of the gap between current high-contrast instruments and the requirements for a $10\times10$ pixel surface map at $\Delta\theta\simeq0.85\,\mu$as with SNR $\ge5$ per pixel.  Each architecture excels at its intended goal---planet discovery, broadband characterization, or coarse photometry---but extending their performance to multi-pixel imaging would require apertures or baselines of order $10^5\,$m and photon-collection improvements of $10^2$–$10^{10}\times$.  Moreover, sustaining raw contrasts of $10^{-9}$–$10^{-10}$ over $10^4$–$10^5\,$s demands picometer-level wavefront stability, sub-mas pointing jitter, and formation-flying tolerances of meters to kilometers, all well beyond present capabilities.

As we have demonstrated above, no existing remote-imaging concept simultaneously satisfies both the $\mu$as-scale diffraction requirement and the photon-flux threshold necessary for a resolved $10\times10$ surface map of an Earth analog at 10\,pc.  Closing these $2$–$10$ order-of-magnitude gaps will demand a fundamentally new paradigm---most promisingly, the Solar Gravitational Lens (SGL; e.g.\ \cite{Eshleman1979,TuryshevToth2017,Turyshev-Toth:2020-extend,Turyshev-Toth:2022_MNRAS,Turyshev-Toth:2022_PRD,TothTuryshev2023}), which provides an effective aperture on the order of the solar radius $D_{\rm eff}=\sqrt{2r_g z}\geq R_\odot$  and an on-line light amplification factor of the SGL, ${ \mu}_{\tt SGL}$ that at a point $\vec x$ on the image plane and at a given heliocentric distance, $z\geq 548.8$ AU, is \cite{Turyshev-Toth:2020-extend} is given as:
{}
\begin{eqnarray}
{ \mu}_{\tt SGL}({\vec x},{\vec x}')&=&
\mu_0J^2_0\Big(\frac{2\pi}{\lambda}
\sqrt{\frac{2r_g}{ z}}
|{\vec x}|\Big),
\qquad {\rm with} \qquad
\mu_0=\frac{4\pi^2}{1-e^{-4\pi^2 r_g/\lambda}}\frac{r_g}{\lambda}\simeq1.17\times 10^{11}\,
\Big(\frac{1\,\mu{\rm m}}{\lambda}\Big),
\label{eq:sgl_mu0}
\end{eqnarray}
where $r_g=2GM_\odot/c^2\simeq 2.95$\,km is the Schwarzschild radius of the Sun. 

Moreover, the SGL’s diffraction-limited angular resolution at a given heliocentric distance, $z$, is determined by the first zero of the Bessel function $J_0(x)$ in (\ref{eq:sgl_mu0}), which occurs at $x=2.4048$ and yields
{}
\begin{eqnarray}
\theta_{\tt SGL}=\frac{|{\vec x}|}{ z}=0.38 \frac{\lambda}{\sqrt{2r_g  z}}=0.10\Big(\frac{\lambda}{1\,\mu{\rm m}}\Big)\Big(\frac{650\,{\rm AU}}{ z}\Big)^\frac{1}{2}~{\rm nas},
\label{eq:sgl_resolution}
\end{eqnarray}
demonstrating nano-arcsecond-scale imaging at optical wavelengths---over nine orders of magnitude finer than conventional diffraction limits.  

At the level of fundamental diffraction physics, unlike all conventional techniques above, whose photon-collection efficiency scales only as $A$ and whose resolution scales as $\lambda/D$ or $\lambda/B$, the SGL uniquely provides both $\gtrsim 10^{10}$ gain and nas-class resolution once beyond $550~\mathrm{AU}$. This fundamentally alters the photon-budget equation, reducing per-pixel integration times from centuries to hours. 

Even after accounting for solar-plasma scattering, zodiacal and exo-zodiacal dust, residual host-star leakage, and detector dark current, the solar-corona shot noise remains dominant \citep{Turyshev-Toth:2020-extend}. High-contrast coronagraphy attenuates the mean coronal flux, thereby lowering its associated Poisson noise. In forward simulations assimilating contemporaneous solar-wind and electron-density data, real-time subtraction of coronal fluctuations further reduces the \emph{structured} residual background by $\gtrsim 30\%$ thus reducing overall contribution of this noise source. Tracking the planet’s orbital phase further suppresses stellar leakage by shifting the Einstein arcs outside the $\lambda/d$ sampling annulus. Altogether, these mitigations drive per-pixel integration times into the few\,$\times10^{3}\,\mathrm{s}$ range and enable full $128\times128$ reconstructions within realistic SGL-mission timelines, even for an Earth-analog at 30\,pc  \citep{Turyshev-Toth:2020-extend,Turyshev-Toth:2022_MNRAS,Turyshev-Toth:2022_PRD}.

While the SGL uniquely meets the resolution and photon-collection requirements outlined here, it presents its own technical challenges, including (i) precise calibration of residual coronal light, (ii) implementation of a robust focal-plane scanning strategy to sample the Einstein ring without loss, and (iii) deconvolution stability in the presence of photon noise and systematic errors. Addressing these will be critical for practical mission realization \cite{9-Turyshev-etal:2020-PhaseII,Helvajian2023,Turyshev:2023-Sundivers}.

Many subsystems are flight‑proven and are already at TRL 7--9: meter-class telescope optics, low-noise focal-plane arrays (QE~$\gtrsim0.9$, read noise~$\lesssim1\,$e$^-$), fine pointing ($\lesssim0.1\,\mu$rad) using reaction wheels and microthrusters, radioisotope power systems, and Ka-band or laser downlinks \cite{Helvajian2023,TothTuryshev2023}.  Others remain below TRL‑6---deep‑space optical/Ka links at $>100$ AU, coronagraphic suppression in the solar environment, meter‑precision ring scanning at $\geq550$ AU---and need focused maturation; credible paths exist to advance them prior to flight. A representative architecture envisions one or more 1\,m-class telescopes (assembled from small, solar-sail-propelled elements) outfitted with internal Lyot coronagraphs (suppression~$\lesssim10^{-7}$) and EMCCD/sCMOS detectors. After a $\sim25$\,yr solar-sail/Jupiter-assist trajectory to $\sim650$\,AU, each telescope blocks sunlight, then scans the $\sim1.3\,$km-diameter image cylinder around the focal line to sample the Einstein ring. Onboard optical navigation (via a compact solar-corona tracker) maintains lateral position to $\lesssim1\,$m, while reaction-wheel/microthruster control holds pointing to $\lesssim10^{-7}$\,rad.

Modern low-noise detectors then achieve SNR~$\gtrsim10$ per Einstein-ring ``pixel'' in $10^2$--$10^3\,$s, and standard deconvolution can assemble a full $10\times10$ surface map in a few weeks \cite{Turyshev-Toth:2020-extend,Turyshev-Toth:2022_MNRAS,Turyshev-Toth:2022_PRD,TothTuryshev2023}. Power (200\,W from a single RPS), thermal control (MLI plus small heaters), and communications (Ka-band~$\sim100\,$bit/s or optical~1--10\,kbit/s) all leverage heritage hardware from New Horizons and similar deep-space missions \cite{Helvajian2023}.

While existing/near-future facilities will revolutionize exoplanet spectroscopy and disk-/spot-level studies, the true multi-pixel imaging remains orders of magnitude more demanding---hence the advanced SGL concept presented here. As we discussed above, several key subsystems relevant to an SGL mission are at TRL~5–7 today (deep-space optical communications, formation metrology/control, and coronagraphic suppression in the solar environment), with credible maturation paths to TRL~6–8 in the early 2030s. This is sufficient to scope a risk-reduction program, but it does not yet constitute end-to-end TRL~9 specifically for SGL imaging. This maturity, combined with the uniquely favorable diffraction physics of the SGL, underpins the feasibility of megapixel-class exo-Earth imaging at interstellar distances.

\section{Conclusions}
\label{sec:conclusions}

The goal of producing a $10\times10$ pixel surface map of an Earth-analog at 10\,pc presents an extraordinarily demanding combination of angular resolution and photon-collection requirements.  In Section~I, we showed that resolving an Earth-radius planet at 10\,pc into $10\times10$ pixels requires a per-pixel angle of $\Delta\theta\approx0.853\,\mu\mathrm{as}$ \eqref{eq:delta_theta_required}, which corresponds to a circular aperture $D\approx1.5\times10^{5}\,\mathrm{m}$ or an interferometric baseline $B\approx1.2\times10^{5}\,\mathrm{m}$ \eqref{eq:required_aperture}–\eqref{eq:required_baseline}.  We further showed that an Earth twin at 10\,pc yields only $\dot N_{\rm eff}\approx0.2\,\mathrm{phot\,s^{-1}}$ in a 10\,m telescope’s 50\,nm band \eqref{eq:planet_photon_flux},
implying $\gtrsim 4\times10^{6}$--$6\times10^{7}$~s per ``micro-pixel'' (the 10$\times$10 heuristic of Sec.~\ref{sec:flux_to_snr}),
even under optimistic raw contrast assumptions. As emphasized in Sec.~\ref{sec:flux_to_snr}, this micro-pixel slicing is an upper-bound diagnostic for an unresolved PSF---not a viable mapping strategy. In the LUVOIR~A worked example, the same heuristic yields $t_{\mathrm{pix}}\!\sim\!19$~yr at $C_\mathrm{res}\!=\!10^{-8}$ (or $\sim\!10$~yr at $10^{-10}$), underscoring that slicing an unresolved PSF cannot recover a resolved surface map within a mission lifetime.

In Section~\ref{sec:monolithic_telescopes}, we examined single-aperture coronagraphic telescopes---WFIRST/Roman (2.4\,m), HabEx (4\,m + starshade), and LUVOIR~A/B (15\,m/8\,m). Their diffraction limits ($\sim 54$--$9.2$~mas at 550\,nm) are $\sim 10^{4}$--$10^{5}$ times too coarse to resolve surface features, and their photon budgets yield per-pixel exposures of $\gtrsim 10^{7}$--$10^{9}\,\mathrm{s}$ under optimistic contrast assumptions. External starshades, treated in detail in Section~\ref{sec:starshades}, decouple starlight suppression from telescope optics but do not improve the diffraction floor; a 4\,m telescope plus starshade remains at $\sim 35$~mas in the visible, and its photon budget still implies multi-century campaign times for a $10\times 10$ surface map.

In Section~\ref{sec:starshades}, we analyzed external starshades flown with a 4\,m space telescope. Starshades can deliver $\sim10^{-10}$ broadband starlight suppression and thereby relax coronagraph requirements, but they do not modify the telescope’s diffraction limit: for a 4\,m aperture at $\lambda=550$\,nm we have $\Delta\theta_{\rm Rayleigh}\simeq34.5$~mas, corresponding to $\sim0.35$~AU at 10\,pc, i.e., $\sim4\times10^{4}$ times larger than the $\sim0.85\,\mu$as ($\sim1.3\times10^{4}$\,km) resolution required for $10\times10$ surface mapping (\ref{eq:delta_theta_required}). Using the planet signal and noise budgets in (\ref{eq:signal-starSh}) and (\ref{eq:noise-starSh}), and adopting optimistic end-to-end throughput and duty cycle, we find that achieving ${\rm SNR}\simeq5$ in each of the 100 surface elements would require $t_{\rm map}\gtrsim7\times10^{9}$\,s ($\gtrsim200$~yr), even before accounting for re-pointing and formation-flying overheads. Thus, while starshades are extremely powerful for unresolved detection and low-resolution spectroscopy of exo-Earths, they cannot enable practical $\mu$as-scale surface imaging in the context considered here.

In Section~\ref{sec:space_interferometry}, we addressed space-based interferometry.  Nulling concepts (TPF-I, Darwin) at $\lambda\sim10\,\mu$m achieve $\sim20$–$50\,\mathrm{mas}$ resolution and are dominated by thermal background, resulting in $\gtrsim10^{4}\,\mathrm{yr}$ per pixel.  Non-nulling arrays (Stellar Imager) could in principle obtain $\sim1\,\mu\mathrm{as}$ resolution with $B\sim1\,\mathrm{km}$, but require optical-path-difference stability $\lesssim5\,\mathrm{nm}$ over $\sim10^{3}$–$10^{5}\,\mathrm{m}$ for $\sim10^{6}\,\mathrm{s}$ and $\gtrsim10^{6}\,\mathrm{s}$ to collect enough photons, yielding full-map times of $\sim10^{6}$–$10^{7}\,\mathrm{s}$ (months to years).  Neither nulling nor non-nulling approaches are feasible in the next two decades.

In Section~\ref{sec:elt_exao}, we discussed ground-based ELTs with extreme adaptive optics (ELT+ExAO) that could achieve $\sim6.5\,\mathrm{mas}$ at 1.65\,$\mu$m and contrasts $\sim10^{-9}$ at $\gtrsim30\,\mathrm{mas}$.  Although capable of detecting and characterizing hot or cold Jupiters, they remain $\sim10^{4}$ times too coarse to resolve any $\mu\mathrm{as}$-scale structure and their photon budgets are insufficient for fine-pixel mapping.

Section~\ref{sec:hypertelescopes} considered pupil-densified hypertelescopes with $N=100$ subapertures of diameter $d_i = 8~\mathrm{m}$ distributed over a maximum baseline $D_i \simeq 162~\mathrm{km}$, achieving $\Delta\theta \simeq 0.85~\mu\mathrm{as}$ at $\lambda = 550~\mathrm{nm}$. Under the near-ideal assumptions adopted there (optimal densification, $\sigma_{\mathrm{OPD}}\lesssim 5~\mathrm{nm}$, $\eta_{\mathrm{tot}}\simeq 0.15$), the coherent photon rate per resolution element implies $t_{\mathrm{pix}}\simeq 2.4~\mathrm{hr}$ and, with a modest multiplexing factor $K\simeq 1.4$ from PSF-core overlap, a total mapping time $t_{\mathrm{map}}\simeq 7~\mathrm{d}$ for a $10\times 10$ image. Realistic inefficiencies and overheads would lengthen this to $\sim 2$--$4$ weeks, and the underlying requirements on continental-scale phasing and metrology remain far beyond current capabilities.  

Indirect methods (Section~\ref{sec:indirect_methods})---rotational light-curve inversion and eclipse mapping---recover only low-order brightness distributions (5–20 broad zones).  A 10\,m telescope requires $\sim2\,\mathrm{min}$ per longitudinal bin (for SNR $\ge 10$), and decades–centuries of rotations or transits to average out clouds and noise, precluding $10\times10$ map. Intensity interferometry can reach $\mu$as scales but demands $N_{\rm ph}\sim10^{16}$ (SNR = 1 for $|V(B)|^2\sim10^{-8}$), implying $\sim10^9\,\mathrm{yr}$ on $10^3\,\mathrm{m}^2$ collecting area. Occultations by Solar System screens (KBOs or artificial $\sim1\,\mathrm{km}$ occulters at 40\,AU) yield $\sim5$–$10\,\mu\mathrm{as}$ resolution but face $\ll10^{-9}$ geometric probabilities and impossible alignment requirements.

In situ or near in situ imaging (Section~\ref{sec:in_situ}) could, in principle, reduce angular requirements to arcseconds by placing a $D\gtrsim0.20\,\mathrm{m}$ telescope at $r\lesssim 10^{5}\,\mathrm{km}$ from the target.  Even so, at $v \approx 0.2\,c$, the flyby window within $r = 0.001\,\mathrm{AU}$ is $\sim 5\,\mathrm{s}$, whereas achieving SNR = 5 at that distance requires $\sim 6\,\mathrm{s}$ in a 10\,nm band.  Autonomous navigation to $\pm 10^{5}\,\mathrm{km}$ over 4.24\,ly, dust shielding at relativistic speeds, and a downlink capable of $\sim 10^{3}\,\mathrm{bits\,s^{-1}}$ are all beyond current capabilities and unlikely to be realized in the next two decades.

Section~\ref{sec:other_direct_techniques} examined the leading direct-imaging concepts---ground-based sparse-aperture masking and speckle imaging on 8–10 m telescopes (resolution \(\sim10\)–\(50\,\mathrm{mas}\)), long-baseline optical interferometry on \(\sim300\)\,m arrays (\(\sim0.5\)–\(1\,\mathrm{mas}\)), millimeter-wave VLBI (\(\sim100\,\mu\mathrm{as}\)), and Solar System occultations by \(\sim1\)\,km screens at \(\sim40\)\,AU (Fresnel-limited to \(\sim5\)–\(10\,\mu\mathrm{as}\)). Even under optimistic assumptions, these methods fall short of the $\sim0.85\,\mu\mathrm{as}$ angular resolution and requisite photon-collection rates for a $10\times10$-pixel surface map of an Earth analog at 10\,pc by nearly five orders of magnitude, rendering them infeasible for exo-Earth imaging.

Similarly, Section~\ref{sec:indirect_feature_retrieval} evaluated spectro-photometric phase-curve inversion and phase-resolved polarimetry, showing that detecting a fractional flux variation \(\Delta F\sim5\times10^{-5}\) at \(\mathrm{SNR}=10\) requires \(N_{\rm tot}\sim4\times10^{10}\) photons, while measuring polarized fractions \(p\lesssim10^{-5}\) demands \(N_{\rm tot}\sim10^{12}\) photons.  When combined with the intrinsic spherical-harmonic limit \(\ell_{\max}\lesssim5\) (at best a \(6\times6\) zonal map), these methods cannot deliver the full \(10\times10\) pixel surface reconstruction of a temperate terrestrial exoplanet within any feasible observing program.  

While current remote-imaging methods have revolutionized exoplanet detection and initial characterization, none yet attains the sub-$\mu$as angular resolution or photon-flux budgets required to render a fully resolved surface map of an Earth twin.  Nevertheless, each successive milestone---LUVOIR’s enhanced contrast performance, the extended baselines of ELT-class interferometry, and advances in hypertelescope beam combining---brings us incrementally closer to the $\sim0.85\,\mu\mathrm{as}$ resolution and signal-to-noise levels essential for true multi-pixel imaging.

Taken together, our analysis shows that no existing or near-term remote or in situ concept can deliver a resolved $10\times10$ pixel surface map of an Earth twin at 10\,pc. The only plausible route is the SGL, which provides an effective angular resolution of order $\mu$as and an on-axis amplification of $10^{10}$--$10^{11}$ at optical wavelengths once a spacecraft reaches the focal region ($\geq 548$\,AU) \cite{TuryshevToth2017,TuryshevToth2019}. The SGL’s point-spread function---an extended Einstein ring of known shape---can be inverted via established deconvolution algorithms to reconstruct a full two-dimensional surface map.

In the absence of $10^{5}\,$m–scale interferometers or comparable advances in photon-collection efficiency, the Solar Gravitational Lens remains the only scientifically and technologically credible means to obtain high-resolution  surface images and spatially resolved atmospheric spectra of Earth-like exoplanets in our stellar neighborhood.  
 
\begin{acknowledgments}

The work described here, in part, was carried out at the Jet Propulsion Laboratory, California Institute of Technology, under a contract with the National Aeronautics and Space Administration. 

\end{acknowledgments}


\begin{thebibliography}{112}
\expandafter\ifx\csname natexlab\endcsname\relax\def\natexlab#1{#1}\fi
\expandafter\ifx\csname bibnamefont\endcsname\relax
  \def\bibnamefont#1{#1}\fi
\expandafter\ifx\csname bibfnamefont\endcsname\relax
  \def\bibfnamefont#1{#1}\fi
\expandafter\ifx\csname citenamefont\endcsname\relax
  \def\citenamefont#1{#1}\fi
\expandafter\ifx\csname url\endcsname\relax
  \def\url#1{\texttt{#1}}\fi
\expandafter\ifx\csname urlprefix\endcsname\relax\def\urlprefix{URL }\fi
\providecommand{\bibinfo}[2]{#2}
\providecommand{\eprint}[2][]{\url{#2}}

\bibitem[{\citenamefont{Seager et~al.}(2016)\citenamefont{Seager, Bains, and
  Petkowski}}]{Seager2016}
\bibinfo{author}{\bibfnamefont{S.}~\bibnamefont{Seager}},
  \bibinfo{author}{\bibfnamefont{W.}~\bibnamefont{Bains}}, \bibnamefont{and}
  \bibinfo{author}{\bibfnamefont{J.~J.} \bibnamefont{Petkowski}},
  \bibinfo{journal}{Astrobiology} \textbf{\bibinfo{volume}{16}},
  \bibinfo{pages}{465} (\bibinfo{year}{2016}).

\bibitem[{\citenamefont{Swain et~al.}(2008)\citenamefont{Swain, Vasisht, and
  Tinetti}}]{Swain2008}
\bibinfo{author}{\bibfnamefont{M.~R.} \bibnamefont{Swain}},
  \bibinfo{author}{\bibfnamefont{G.}~\bibnamefont{Vasisht}}, \bibnamefont{and}
  \bibinfo{author}{\bibfnamefont{G.}~\bibnamefont{Tinetti}},
  \bibinfo{journal}{Nature} \textbf{\bibinfo{volume}{452}},
  \bibinfo{pages}{329} (\bibinfo{year}{2008}).

\bibitem[{\citenamefont{Madhusudhan}(2019)}]{Madhusudhan2019}
\bibinfo{author}{\bibfnamefont{N.}~\bibnamefont{Madhusudhan}},
  \bibinfo{journal}{Ann. Rev. Astron. \& Astrophys.}
  \textbf{\bibinfo{volume}{57}}, \bibinfo{pages}{617} (\bibinfo{year}{2019}).

\bibitem[{\citenamefont{Venot et~al.}(2020)\citenamefont{Venot, Ag{\'u}ndez,
  Selsis, and et~al.}}]{Venot2020}
\bibinfo{author}{\bibfnamefont{O.}~\bibnamefont{Venot}},
  \bibinfo{author}{\bibfnamefont{M.}~\bibnamefont{Ag{\'u}ndez}},
  \bibinfo{author}{\bibfnamefont{F.}~\bibnamefont{Selsis}}, \bibnamefont{and}
  \bibinfo{author}{\bibnamefont{et~al.}}, \bibinfo{journal}{Astronomy \&
  Astrophysics} \textbf{\bibinfo{volume}{634}}, \bibinfo{pages}{A49}
  (\bibinfo{year}{2020}).

\bibitem[{\citenamefont{Greaves et~al.}(2021)\citenamefont{Greaves, Richards,
  Bains, and et~al.}}]{Greaves2021}
\bibinfo{author}{\bibfnamefont{J.~S.} \bibnamefont{Greaves}},
  \bibinfo{author}{\bibfnamefont{A.~M.~S.} \bibnamefont{Richards}},
  \bibinfo{author}{\bibfnamefont{W.}~\bibnamefont{Bains}}, \bibnamefont{and}
  \bibinfo{author}{\bibnamefont{et~al.}}, \bibinfo{journal}{Nature Astronomy}
  \textbf{\bibinfo{volume}{5}}, \bibinfo{pages}{627} (\bibinfo{year}{2021}).

\bibitem[{\citenamefont{Turyshev and Toth}(2019)}]{TuryshevToth2019}
\bibinfo{author}{\bibfnamefont{S.~G.} \bibnamefont{Turyshev}} \bibnamefont{and}
  \bibinfo{author}{\bibfnamefont{V.~T.} \bibnamefont{Toth}},
  \bibinfo{journal}{Phys. Rev. D} \textbf{\bibinfo{volume}{99}},
  \bibinfo{pages}{024044} (\bibinfo{year}{2019}).

\bibitem[{\citenamefont{Feinberg et~al.}(2019)\citenamefont{Feinberg, Bloom,
  Boynton, Domagal-Goldman, Fortney, Fujii, Gaudi, Greco, Seager, Spergel
  et~al.}}]{Feinberg2019}
\bibinfo{author}{\bibfnamefont{L.}~\bibnamefont{Feinberg}},
  \bibinfo{author}{\bibfnamefont{J.}~\bibnamefont{Bloom}},
  \bibinfo{author}{\bibfnamefont{K.}~\bibnamefont{Boynton}},
  \bibinfo{author}{\bibfnamefont{S.}~\bibnamefont{Domagal-Goldman}},
  \bibinfo{author}{\bibfnamefont{J.~J.} \bibnamefont{Fortney}},
  \bibinfo{author}{\bibfnamefont{Y.}~\bibnamefont{Fujii}},
  \bibinfo{author}{\bibfnamefont{B.~S.} \bibnamefont{Gaudi}},
  \bibinfo{author}{\bibfnamefont{J.}~\bibnamefont{Greco}},
  \bibinfo{author}{\bibfnamefont{S.}~\bibnamefont{Seager}},
  \bibinfo{author}{\bibfnamefont{D.~N.} \bibnamefont{Spergel}},
  \bibnamefont{et~al.}, \bibinfo{type}{Tech. Rep.}
  \bibinfo{number}{NASA/TM-2019-220120}, \bibinfo{institution}{NASA}
  (\bibinfo{year}{2019}).

\bibitem[{\citenamefont{Gaudi et~al.}(2020)\citenamefont{Gaudi, Seager,
  Domagal-Goldman, Crepp, Dalcanton, Marley, Rogoszinski, Shkolnik, Traub, and
  Howard}}]{Gaudi2020}
\bibinfo{author}{\bibfnamefont{B.~S.} \bibnamefont{Gaudi}},
  \bibinfo{author}{\bibfnamefont{S.}~\bibnamefont{Seager}},
  \bibinfo{author}{\bibfnamefont{S.}~\bibnamefont{Domagal-Goldman}},
  \bibinfo{author}{\bibfnamefont{J.~R.} \bibnamefont{Crepp}},
  \bibinfo{author}{\bibfnamefont{J.~J.} \bibnamefont{Dalcanton}},
  \bibinfo{author}{\bibfnamefont{M.~S.} \bibnamefont{Marley}},
  \bibinfo{author}{\bibfnamefont{A.}~\bibnamefont{Rogoszinski}},
  \bibinfo{author}{\bibfnamefont{E.~L.} \bibnamefont{Shkolnik}},
  \bibinfo{author}{\bibfnamefont{W.~A.} \bibnamefont{Traub}}, \bibnamefont{and}
  \bibinfo{author}{\bibfnamefont{A.~W.} \bibnamefont{Howard}},
  \bibinfo{type}{Tech. Rep.} \bibinfo{number}{NASA/SP-2020-002},
  \bibinfo{institution}{NASA} (\bibinfo{year}{2020}).

\bibitem[{\citenamefont{Cowan and Agol}(2008)}]{CowanAgol2008}
\bibinfo{author}{\bibfnamefont{N.~B.} \bibnamefont{Cowan}} \bibnamefont{and}
  \bibinfo{author}{\bibfnamefont{E.}~\bibnamefont{Agol}},
  \bibinfo{journal}{ApJ} \textbf{\bibinfo{volume}{678}}, \bibinfo{pages}{L129}
  (\bibinfo{year}{2008}).

\bibitem[{\citenamefont{Majeau et~al.}(2012)\citenamefont{Majeau, Agol, and
  Cowan}}]{Majeau2012}
\bibinfo{author}{\bibfnamefont{C.}~\bibnamefont{Majeau}},
  \bibinfo{author}{\bibfnamefont{E.}~\bibnamefont{Agol}}, \bibnamefont{and}
  \bibinfo{author}{\bibfnamefont{N.~B.} \bibnamefont{Cowan}},
  \bibinfo{journal}{ApJ Letters} \textbf{\bibinfo{volume}{747}},
  \bibinfo{pages}{L20} (\bibinfo{year}{2012}).

\bibitem[{\citenamefont{Hanbury~Brown and Twiss}(1956)}]{HanburyBrown1956}
\bibinfo{author}{\bibfnamefont{R.}~\bibnamefont{Hanbury~Brown}}
  \bibnamefont{and} \bibinfo{author}{\bibfnamefont{R.~Q.} \bibnamefont{Twiss}},
  \bibinfo{journal}{Nature} \textbf{\bibinfo{volume}{178}},
  \bibinfo{pages}{1046} (\bibinfo{year}{1956}).

\bibitem[{\citenamefont{Lubin}(2016)}]{Lubin2016}
\bibinfo{author}{\bibfnamefont{P.}~\bibnamefont{Lubin}},
  \bibinfo{journal}{JBIS} \textbf{\bibinfo{volume}{69}}, \bibinfo{pages}{40}
  (\bibinfo{year}{2016}).

\bibitem[{\citenamefont{Cox}(2000)}]{Cox2000}
\bibinfo{author}{\bibfnamefont{A.~N.} \bibnamefont{Cox}},
  \emph{\bibinfo{title}{Allen's Astrophysical Quantities}}
  (\bibinfo{publisher}{AIP Press}, \bibinfo{address}{New York, NY},
  \bibinfo{year}{2000}), \bibinfo{edition}{4th} ed.

\bibitem[{\citenamefont{Pall{\'e} et~al.}(2003)\citenamefont{Pall{\'e}, Goode,
  Yurchyshyn, Qiu, Hickey, Montan{\'e}s~Rodriguez, Chu, Kolbe, Brown, and
  Koonin}}]{Palle2003}
\bibinfo{author}{\bibfnamefont{E.}~\bibnamefont{Pall{\'e}}},
  \bibinfo{author}{\bibfnamefont{P.~R.} \bibnamefont{Goode}},
  \bibinfo{author}{\bibfnamefont{V.}~\bibnamefont{Yurchyshyn}},
  \bibinfo{author}{\bibfnamefont{J.}~\bibnamefont{Qiu}},
  \bibinfo{author}{\bibfnamefont{J.}~\bibnamefont{Hickey}},
  \bibinfo{author}{\bibfnamefont{P.}~\bibnamefont{Montan{\'e}s~Rodriguez}},
  \bibinfo{author}{\bibfnamefont{M.~C.} \bibnamefont{Chu}},
  \bibinfo{author}{\bibfnamefont{E.}~\bibnamefont{Kolbe}},
  \bibinfo{author}{\bibfnamefont{C.~T.} \bibnamefont{Brown}}, \bibnamefont{and}
  \bibinfo{author}{\bibfnamefont{S.~E.} \bibnamefont{Koonin}},
  \bibinfo{journal}{JGR Atmospheres} \textbf{\bibinfo{volume}{108}},
  \bibinfo{pages}{ACL 13} (\bibinfo{year}{2003}).

\bibitem[{\citenamefont{{Bessell}}(1979)}]{Bessell1979}
\bibinfo{author}{\bibfnamefont{M.~S.} \bibnamefont{{Bessell}}},
  \bibinfo{journal}{PASP} \textbf{\bibinfo{volume}{91}}, \bibinfo{pages}{589}
  (\bibinfo{year}{1979}).

\bibitem[{\citenamefont{Bohlin}(2014)}]{Bohlin2014}
\bibinfo{author}{\bibfnamefont{R.~C.} \bibnamefont{Bohlin}},
  \bibinfo{journal}{AJ} \textbf{\bibinfo{volume}{147}}, \bibinfo{pages}{127}
  (\bibinfo{year}{2014}).

\bibitem[{\citenamefont{Tokunaga and Vacca}(2005)}]{TokunagaVacca2005}
\bibinfo{author}{\bibfnamefont{A.~T.} \bibnamefont{Tokunaga}} \bibnamefont{and}
  \bibinfo{author}{\bibfnamefont{W.~D.} \bibnamefont{Vacca}},
  \bibinfo{journal}{PASP} \textbf{\bibinfo{volume}{117}}, \bibinfo{pages}{421}
  (\bibinfo{year}{2005}).

\bibitem[{\citenamefont{Hewett et~al.}(2006)\citenamefont{Hewett, Warren,
  Leggett, and Hodgkin}}]{Hewett2006}
\bibinfo{author}{\bibfnamefont{P.~C.} \bibnamefont{Hewett}},
  \bibinfo{author}{\bibfnamefont{S.~J.} \bibnamefont{Warren}},
  \bibinfo{author}{\bibfnamefont{S.~K.} \bibnamefont{Leggett}},
  \bibnamefont{and} \bibinfo{author}{\bibfnamefont{S.~T.}
  \bibnamefont{Hodgkin}}, \bibinfo{journal}{MNRAS}
  \textbf{\bibinfo{volume}{367}}, \bibinfo{pages}{454} (\bibinfo{year}{2006}).

\bibitem[{\citenamefont{Leinert et~al.}(1998)\citenamefont{Leinert, Bowyer,
  Haikala, Hanner, Hauser, Kelsall, Mann, Meyer, Reach, Schlosser
  et~al.}}]{Leinert1998}
\bibinfo{author}{\bibfnamefont{C.}~\bibnamefont{Leinert}},
  \bibinfo{author}{\bibfnamefont{S.}~\bibnamefont{Bowyer}},
  \bibinfo{author}{\bibfnamefont{L.~K.} \bibnamefont{Haikala}},
  \bibinfo{author}{\bibfnamefont{M.~S.} \bibnamefont{Hanner}},
  \bibinfo{author}{\bibfnamefont{M.~G.} \bibnamefont{Hauser}},
  \bibinfo{author}{\bibfnamefont{T.}~\bibnamefont{Kelsall}},
  \bibinfo{author}{\bibfnamefont{I.}~\bibnamefont{Mann}},
  \bibinfo{author}{\bibfnamefont{M.~R.} \bibnamefont{Meyer}},
  \bibinfo{author}{\bibfnamefont{W.~T.} \bibnamefont{Reach}},
  \bibinfo{author}{\bibfnamefont{W.}~\bibnamefont{Schlosser}},
  \bibnamefont{et~al.}, \bibinfo{journal}{Astron. \& Astrophyss Supplement
  Series} \textbf{\bibinfo{volume}{127}}, \bibinfo{pages}{1}
  (\bibinfo{year}{1998}).

\bibitem[{\citenamefont{Mennesson et~al.}(2016)\citenamefont{Mennesson, Ge, and
  Serabyn}}]{Mennesson2016}
\bibinfo{author}{\bibfnamefont{B.}~\bibnamefont{Mennesson}},
  \bibinfo{author}{\bibfnamefont{J.}~\bibnamefont{Ge}}, \bibnamefont{and}
  \bibinfo{author}{\bibfnamefont{E.}~\bibnamefont{Serabyn}}, in
  \emph{\bibinfo{booktitle}{SPIE Astronomical Telescopes + Instrumentation}}
  (\bibinfo{year}{2016}), vol. \bibinfo{volume}{9904}, p.
  \bibinfo{pages}{99040V}.

\bibitem[{\citenamefont{Ertel et~al.}(2020)\citenamefont{Ertel, Stone, Hinz,
  Defrère, Mennesson, Danchi, Serabyn, Colavita, and Absil}}]{Ertel2020}
\bibinfo{author}{\bibfnamefont{S.}~\bibnamefont{Ertel}},
  \bibinfo{author}{\bibfnamefont{J.~M.} \bibnamefont{Stone}},
  \bibinfo{author}{\bibfnamefont{P.~M.} \bibnamefont{Hinz}},
  \bibinfo{author}{\bibfnamefont{D.}~\bibnamefont{Defrère}},
  \bibinfo{author}{\bibfnamefont{B.}~\bibnamefont{Mennesson}},
  \bibinfo{author}{\bibfnamefont{W.~C.} \bibnamefont{Danchi}},
  \bibinfo{author}{\bibfnamefont{E.}~\bibnamefont{Serabyn}},
  \bibinfo{author}{\bibfnamefont{M.~M.} \bibnamefont{Colavita}},
  \bibnamefont{and} \bibinfo{author}{\bibfnamefont{O.}~\bibnamefont{Absil}},
  \bibinfo{journal}{AJ} \textbf{\bibinfo{volume}{159}}, \bibinfo{pages}{301}
  (\bibinfo{year}{2020}).

\bibitem[{\citenamefont{Stark et~al.}(2014{\natexlab{a}})\citenamefont{Stark,
  Roberge, Mandell et~al.}}]{Stark2014MaxYield}
\bibinfo{author}{\bibfnamefont{C.~C.} \bibnamefont{Stark}},
  \bibinfo{author}{\bibfnamefont{A.}~\bibnamefont{Roberge}},
  \bibinfo{author}{\bibfnamefont{A.}~\bibnamefont{Mandell}},
  \bibnamefont{et~al.}, \bibinfo{journal}{ApJ} \textbf{\bibinfo{volume}{795}},
  \bibinfo{pages}{122} (\bibinfo{year}{2014}{\natexlab{a}}).

\bibitem[{\citenamefont{Stark et~al.}(2019)\citenamefont{Stark, Roberge,
  Mandell et~al.}}]{Stark2019YieldLandscape}
\bibinfo{author}{\bibfnamefont{C.~C.} \bibnamefont{Stark}},
  \bibinfo{author}{\bibfnamefont{A.}~\bibnamefont{Roberge}},
  \bibinfo{author}{\bibfnamefont{A.}~\bibnamefont{Mandell}},
  \bibnamefont{et~al.}, \bibinfo{journal}{J. Astron. Telescopes, Instrum., and
  Syst.} \textbf{\bibinfo{volume}{5}}, \bibinfo{pages}{024009}
  (\bibinfo{year}{2019}).

\bibitem[{\citenamefont{Ienaka et~al.}(2013)\citenamefont{Ienaka, Sakon, and
  Onaka}}]{Ienaka2013}
\bibinfo{author}{\bibfnamefont{N.}~\bibnamefont{Ienaka}},
  \bibinfo{author}{\bibfnamefont{I.}~\bibnamefont{Sakon}}, \bibnamefont{and}
  \bibinfo{author}{\bibfnamefont{T.}~\bibnamefont{Onaka}},
  \bibinfo{journal}{Pubs. of the Astronomical Society of Japan}
  \textbf{\bibinfo{volume}{65}}, \bibinfo{pages}{120} (\bibinfo{year}{2013}).

\bibitem[{\citenamefont{Hauser and Dwek}(2001)}]{HauserDwek2001}
\bibinfo{author}{\bibfnamefont{M.~G.} \bibnamefont{Hauser}} \bibnamefont{and}
  \bibinfo{author}{\bibfnamefont{E.}~\bibnamefont{Dwek}},
  \bibinfo{journal}{Ann. Rev. Astron. \& Astrophys.}
  \textbf{\bibinfo{volume}{39}}, \bibinfo{pages}{249} (\bibinfo{year}{2001}).

\bibitem[{\citenamefont{Girardi et~al.}(2005)\citenamefont{Girardi,
  Groenewegen, Hatziminaoglou, and da~Costa}}]{Girardi2005}
\bibinfo{author}{\bibfnamefont{L.}~\bibnamefont{Girardi}},
  \bibinfo{author}{\bibfnamefont{M.~A.~T.} \bibnamefont{Groenewegen}},
  \bibinfo{author}{\bibfnamefont{E.}~\bibnamefont{Hatziminaoglou}},
  \bibnamefont{and} \bibinfo{author}{\bibfnamefont{L.}~\bibnamefont{da~Costa}},
  \bibinfo{journal}{Astron. \& Astrophys.} \textbf{\bibinfo{volume}{436}},
  \bibinfo{pages}{895} (\bibinfo{year}{2005}).

\bibitem[{\citenamefont{Windhorst et~al.}(2011)\citenamefont{Windhorst, Cohen,
  Hathi, Ryan, O'Connell, Driver, Robotham, and Holwerda}}]{Windhorst2011}
\bibinfo{author}{\bibfnamefont{R.~A.} \bibnamefont{Windhorst}},
  \bibinfo{author}{\bibfnamefont{S.~H.} \bibnamefont{Cohen}},
  \bibinfo{author}{\bibfnamefont{N.~P.} \bibnamefont{Hathi}},
  \bibinfo{author}{\bibfnamefont{R.~E.} \bibnamefont{Ryan}},
  \bibinfo{author}{\bibfnamefont{R.~W.} \bibnamefont{O'Connell}},
  \bibinfo{author}{\bibfnamefont{S.~P.} \bibnamefont{Driver}},
  \bibinfo{author}{\bibfnamefont{A.~S.~G.} \bibnamefont{Robotham}},
  \bibnamefont{and} \bibinfo{author}{\bibfnamefont{B.~W.}
  \bibnamefont{Holwerda}}, \bibinfo{journal}{ApJ Supplement Series}
  \textbf{\bibinfo{volume}{193}}, \bibinfo{pages}{27} (\bibinfo{year}{2011}).

\bibitem[{\citenamefont{Janesick}(2001)}]{Janesick2001}
\bibinfo{author}{\bibfnamefont{J.~R.} \bibnamefont{Janesick}},
  \emph{\bibinfo{title}{Scientific Charge-Coupled Devices}}
  (\bibinfo{publisher}{SPIE Press}, \bibinfo{address}{Bellingham, WA},
  \bibinfo{year}{2001}), ISBN \bibinfo{isbn}{9780819443905}.

\bibitem[{\citenamefont{Daigle et~al.}(2010)\citenamefont{Daigle, Carignan,
  Blais-Ouellette, Lessard, Fortin, Gach, and Guillaume}}]{Daigle2010}
\bibinfo{author}{\bibfnamefont{O.}~\bibnamefont{Daigle}},
  \bibinfo{author}{\bibfnamefont{C.}~\bibnamefont{Carignan}},
  \bibinfo{author}{\bibfnamefont{S.}~\bibnamefont{Blais-Ouellette}},
  \bibinfo{author}{\bibfnamefont{S.}~\bibnamefont{Lessard}},
  \bibinfo{author}{\bibfnamefont{C.-A.} \bibnamefont{Fortin}},
  \bibinfo{author}{\bibfnamefont{J.-L.} \bibnamefont{Gach}}, \bibnamefont{and}
  \bibinfo{author}{\bibfnamefont{C.}~\bibnamefont{Guillaume}},
  \bibinfo{journal}{PASP} \textbf{\bibinfo{volume}{122}}, \bibinfo{pages}{1411}
  (\bibinfo{year}{2010}).

\bibitem[{\citenamefont{Basden et~al.}(2003)\citenamefont{Basden, Haniff, and
  Mackay}}]{Basden2003}
\bibinfo{author}{\bibfnamefont{A.~G.} \bibnamefont{Basden}},
  \bibinfo{author}{\bibfnamefont{C.~A.} \bibnamefont{Haniff}},
  \bibnamefont{and} \bibinfo{author}{\bibfnamefont{C.~D.}
  \bibnamefont{Mackay}}, \bibinfo{journal}{MNRAS}
  \textbf{\bibinfo{volume}{345}}, \bibinfo{pages}{985} (\bibinfo{year}{2003}).

\bibitem[{\citenamefont{{Wilkins} et~al.}(2014)\citenamefont{{Wilkins},
  {McElwain}, {Norton}, {Rauscher}, {Rothe}, {Malatesta}, {Hilton}, {Bubeck},
  {Grady}, and {Lindler}}}]{Wilkins2014}
\bibinfo{author}{\bibfnamefont{A.~N.} \bibnamefont{{Wilkins}}},
  \bibinfo{author}{\bibfnamefont{M.~W.} \bibnamefont{{McElwain}}},
  \bibinfo{author}{\bibfnamefont{T.~J.} \bibnamefont{{Norton}}},
  \bibinfo{author}{\bibfnamefont{B.~J.} \bibnamefont{{Rauscher}}},
  \bibinfo{author}{\bibfnamefont{J.~F.} \bibnamefont{{Rothe}}},
  \bibinfo{author}{\bibfnamefont{M.}~\bibnamefont{{Malatesta}}},
  \bibinfo{author}{\bibfnamefont{G.~M.} \bibnamefont{{Hilton}}},
  \bibinfo{author}{\bibfnamefont{J.~R.} \bibnamefont{{Bubeck}}},
  \bibinfo{author}{\bibfnamefont{C.~A.} \bibnamefont{{Grady}}},
  \bibnamefont{and} \bibinfo{author}{\bibfnamefont{D.~J.}
  \bibnamefont{{Lindler}}}, in \emph{\bibinfo{booktitle}{High Energy, Optical,
  and Infrared Detectors for Astronomy VI}}, edited by
  \bibinfo{editor}{\bibfnamefont{A.~D.} \bibnamefont{{Holland}}}
  \bibnamefont{and} \bibinfo{editor}{\bibfnamefont{J.}~\bibnamefont{{Beletic}}}
  (\bibinfo{year}{2014}), vol. \bibinfo{volume}{9154} of
  \emph{\bibinfo{series}{Society of Photo-Optical Instrumentation Engineers
  (SPIE) Conference Series}}, p. \bibinfo{pages}{91540C}, \eprint{1407.0701}.

\bibitem[{\citenamefont{Trauger et~al.}(2012)\citenamefont{Trauger, Moody,
  Gordon, Krist, Mawet, Serabyn, and Traub}}]{Trauger2012}
\bibinfo{author}{\bibfnamefont{J.~T.} \bibnamefont{Trauger}},
  \bibinfo{author}{\bibfnamefont{D.~C.} \bibnamefont{Moody}},
  \bibinfo{author}{\bibfnamefont{B.~E.} \bibnamefont{Gordon}},
  \bibinfo{author}{\bibfnamefont{J.~E.} \bibnamefont{Krist}},
  \bibinfo{author}{\bibfnamefont{D.}~\bibnamefont{Mawet}},
  \bibinfo{author}{\bibfnamefont{E.}~\bibnamefont{Serabyn}}, \bibnamefont{and}
  \bibinfo{author}{\bibfnamefont{W.~A.} \bibnamefont{Traub}}, in
  \emph{\bibinfo{booktitle}{Proceedings of SPIE: Space Telescopes and
  Instrumentation 2012: Optical, Infrared, and Millimeter Wave}}
  (\bibinfo{year}{2012}), vol. \bibinfo{volume}{8442}, p.
  \bibinfo{pages}{844204}.

\bibitem[{\citenamefont{Beichman et~al.}(2010)\citenamefont{Beichman, Bryden,
  Rieke, Stapelfeldt, Su, and Stansberry}}]{Beichman2010}
\bibinfo{author}{\bibfnamefont{C.~A.} \bibnamefont{Beichman}},
  \bibinfo{author}{\bibfnamefont{G.}~\bibnamefont{Bryden}},
  \bibinfo{author}{\bibfnamefont{G.~H.} \bibnamefont{Rieke}},
  \bibinfo{author}{\bibfnamefont{K.~R.} \bibnamefont{Stapelfeldt}},
  \bibinfo{author}{\bibfnamefont{K.~Y.~L.} \bibnamefont{Su}}, \bibnamefont{and}
  \bibinfo{author}{\bibfnamefont{J.~A.} \bibnamefont{Stansberry}},
  \bibinfo{journal}{ApJ} \textbf{\bibinfo{volume}{716}}, \bibinfo{pages}{1126}
  (\bibinfo{year}{2010}).

\bibitem[{\citenamefont{Badhwar}(2002)}]{Badhwar2002}
\bibinfo{author}{\bibfnamefont{G.~D.} \bibnamefont{Badhwar}},
  \bibinfo{type}{Technical Memorandum} \bibinfo{number}{TM-2002-210536},
  \bibinfo{institution}{NASA Johnson Space Center} (\bibinfo{year}{2002}).

\bibitem[{\citenamefont{Spergel et~al.}(2015)\citenamefont{Spergel, Gehrels,
  Breckinridge, Donahue, Dressler, Gaudi, Greene, Guyon, Kalirai, Kasdin
  et~al.}}]{Spergel2015}
\bibinfo{author}{\bibfnamefont{D.~N.} \bibnamefont{Spergel}},
  \bibinfo{author}{\bibfnamefont{N.}~\bibnamefont{Gehrels}},
  \bibinfo{author}{\bibfnamefont{J.}~\bibnamefont{Breckinridge}},
  \bibinfo{author}{\bibfnamefont{M.}~\bibnamefont{Donahue}},
  \bibinfo{author}{\bibfnamefont{A.}~\bibnamefont{Dressler}},
  \bibinfo{author}{\bibfnamefont{B.~S.} \bibnamefont{Gaudi}},
  \bibinfo{author}{\bibfnamefont{T.}~\bibnamefont{Greene}},
  \bibinfo{author}{\bibfnamefont{O.}~\bibnamefont{Guyon}},
  \bibinfo{author}{\bibfnamefont{J.}~\bibnamefont{Kalirai}},
  \bibinfo{author}{\bibfnamefont{N.~J.} \bibnamefont{Kasdin}},
  \bibnamefont{et~al.}, \emph{\bibinfo{title}{Wide-field infrared survey
  telescope–astrophysics focused telescope assets coronagraph instrument
  final report}}, \bibinfo{howpublished}{arXiv:1503.03757 [astro-ph.IM]}
  (\bibinfo{year}{2015}), \bibinfo{note}{{NASA Technical Report}}.

\bibitem[{\citenamefont{Feng et~al.}(2019)\citenamefont{Feng, Domagal-Goldman,
  Gaudi, Hearty, Kalirai, Kratter, Marley, Ross, Sheridan, Shkolnik
  et~al.}}]{Feng2019}
\bibinfo{author}{\bibfnamefont{F.}~\bibnamefont{Feng}},
  \bibinfo{author}{\bibfnamefont{S.}~\bibnamefont{Domagal-Goldman}},
  \bibinfo{author}{\bibfnamefont{B.~S.} \bibnamefont{Gaudi}},
  \bibinfo{author}{\bibfnamefont{F.}~\bibnamefont{Hearty}},
  \bibinfo{author}{\bibfnamefont{J.~S.} \bibnamefont{Kalirai}},
  \bibinfo{author}{\bibfnamefont{K.}~\bibnamefont{Kratter}},
  \bibinfo{author}{\bibfnamefont{M.~S.} \bibnamefont{Marley}},
  \bibinfo{author}{\bibfnamefont{S.~C.} \bibnamefont{Ross}},
  \bibinfo{author}{\bibfnamefont{K.}~\bibnamefont{Sheridan}},
  \bibinfo{author}{\bibfnamefont{E.~L.} \bibnamefont{Shkolnik}},
  \bibnamefont{et~al.}, \bibinfo{type}{Tech. Rep.}
  \bibinfo{number}{NASA/TM-2019-220121}, \bibinfo{institution}{NASA}
  (\bibinfo{year}{2019}).

\bibitem[{\citenamefont{Mawet et~al.}(2012)\citenamefont{Mawet, Pueyo, Lawson,
  Mugnier, Traub, Boccaletti, Trauger, Gladysz, Serabyn, Milli
  et~al.}}]{Mawet2012SmallAngleReview}
\bibinfo{author}{\bibfnamefont{D.}~\bibnamefont{Mawet}},
  \bibinfo{author}{\bibfnamefont{L.}~\bibnamefont{Pueyo}},
  \bibinfo{author}{\bibfnamefont{P.}~\bibnamefont{Lawson}},
  \bibinfo{author}{\bibfnamefont{L.}~\bibnamefont{Mugnier}},
  \bibinfo{author}{\bibfnamefont{W.}~\bibnamefont{Traub}},
  \bibinfo{author}{\bibfnamefont{A.}~\bibnamefont{Boccaletti}},
  \bibinfo{author}{\bibfnamefont{J.}~\bibnamefont{Trauger}},
  \bibinfo{author}{\bibfnamefont{S.}~\bibnamefont{Gladysz}},
  \bibinfo{author}{\bibfnamefont{E.}~\bibnamefont{Serabyn}},
  \bibinfo{author}{\bibfnamefont{J.}~\bibnamefont{Milli}},
  \bibnamefont{et~al.}, in \emph{\bibinfo{booktitle}{Space Telescopes and
  Instrumentation 2012: Optical, Infrared, and Millimeter Wave}}
  (\bibinfo{year}{2012}), vol. \bibinfo{volume}{8442} of
  \emph{\bibinfo{series}{Proc. of SPIE}}.

\bibitem[{\citenamefont{Mawet et~al.}(2014)\citenamefont{Mawet, Milli, Wahhaj
  et~al.}}]{Mawet2014SpeckleStats}
\bibinfo{author}{\bibfnamefont{D.}~\bibnamefont{Mawet}},
  \bibinfo{author}{\bibfnamefont{J.}~\bibnamefont{Milli}},
  \bibinfo{author}{\bibfnamefont{Z.}~\bibnamefont{Wahhaj}},
  \bibnamefont{et~al.}, \bibinfo{journal}{ApJ} \textbf{\bibinfo{volume}{792}},
  \bibinfo{pages}{97} (\bibinfo{year}{2014}).

\bibitem[{\citenamefont{{Zellem} et~al.}(2022)\citenamefont{{Zellem}, {Nemati},
  {Gonzalez}, {Ygouf}, {Bailey}, {Cady}, {Colavita}, {Hildebrandt}, {Maier},
  {Mennesson} et~al.}}]{Zellem2022}
\bibinfo{author}{\bibfnamefont{R.~T.} \bibnamefont{{Zellem}}},
  \bibinfo{author}{\bibfnamefont{B.}~\bibnamefont{{Nemati}}},
  \bibinfo{author}{\bibfnamefont{G.}~\bibnamefont{{Gonzalez}}},
  \bibinfo{author}{\bibfnamefont{M.}~\bibnamefont{{Ygouf}}},
  \bibinfo{author}{\bibfnamefont{V.~P.} \bibnamefont{{Bailey}}},
  \bibinfo{author}{\bibfnamefont{E.~J.} \bibnamefont{{Cady}}},
  \bibinfo{author}{\bibfnamefont{M.~M.} \bibnamefont{{Colavita}}},
  \bibinfo{author}{\bibfnamefont{S.~R.} \bibnamefont{{Hildebrandt}}},
  \bibinfo{author}{\bibfnamefont{E.~R.} \bibnamefont{{Maier}}},
  \bibinfo{author}{\bibfnamefont{B.}~\bibnamefont{{Mennesson}}},
  \bibnamefont{et~al.}, in \emph{\bibinfo{booktitle}{Space Telescopes and
  Instrumentation 2022: Optical, Infrared, and Millimeter Wave}}, edited by
  \bibinfo{editor}{\bibfnamefont{L.~E.} \bibnamefont{{Coyle}}},
  \bibinfo{editor}{\bibfnamefont{S.}~\bibnamefont{{Matsuura}}},
  \bibnamefont{and} \bibinfo{editor}{\bibfnamefont{M.~D.}
  \bibnamefont{{Perrin}}} (\bibinfo{year}{2022}), vol. \bibinfo{volume}{12180}
  of \emph{\bibinfo{series}{Society of Photo-Optical Instrumentation Engineers
  (SPIE) Conference Series}}, p. \bibinfo{pages}{121801Z}.

\bibitem[{\citenamefont{Pueyo}(2016)}]{Pueyo2016KLIPfm}
\bibinfo{author}{\bibfnamefont{L.}~\bibnamefont{Pueyo}}, \bibinfo{journal}{ApJ}
  \textbf{\bibinfo{volume}{824}}, \bibinfo{pages}{117} (\bibinfo{year}{2016}).

\bibitem[{\citenamefont{Pueyo et~al.}(2019)\citenamefont{Pueyo,
  Juanola-Parramon, Zimmerman et~al.}}]{Pueyo2019ECLIPS}
\bibinfo{author}{\bibfnamefont{L.}~\bibnamefont{Pueyo}},
  \bibinfo{author}{\bibfnamefont{R.}~\bibnamefont{Juanola-Parramon}},
  \bibinfo{author}{\bibfnamefont{N.}~\bibnamefont{Zimmerman}},
  \bibnamefont{et~al.}, in \emph{\bibinfo{booktitle}{Space Telescopes and
  Instrumentation 2019: Optical, Infrared, and Millimeter Wave}}
  (\bibinfo{year}{2019}), vol. \bibinfo{volume}{11117} of
  \emph{\bibinfo{series}{Proc. of SPIE}}.

\bibitem[{\citenamefont{Cash}(2006)}]{Cash2006}
\bibinfo{author}{\bibfnamefont{W.}~\bibnamefont{Cash}},
  \bibinfo{journal}{Nature} \textbf{\bibinfo{volume}{442}}, \bibinfo{pages}{51}
  (\bibinfo{year}{2006}).

\bibitem[{\citenamefont{Seager et~al.}(2021)\citenamefont{Seager, Kasdin, and
  Ebrahimi}}]{Seager2021}
\bibinfo{author}{\bibfnamefont{S.}~\bibnamefont{Seager}},
  \bibinfo{author}{\bibfnamefont{N.~J.} \bibnamefont{Kasdin}},
  \bibnamefont{and} \bibinfo{author}{\bibfnamefont{M.}~\bibnamefont{Ebrahimi}},
  \bibinfo{journal}{JATIS} \textbf{\bibinfo{volume}{7}},
  \bibinfo{pages}{041212} (\bibinfo{year}{2021}).

\bibitem[{\citenamefont{Vanderbei et~al.}(2007)\citenamefont{Vanderbei, Traub,
  and Kasdin}}]{Vanderbei2007}
\bibinfo{author}{\bibfnamefont{R.~J.} \bibnamefont{Vanderbei}},
  \bibinfo{author}{\bibfnamefont{W.~A.} \bibnamefont{Traub}}, \bibnamefont{and}
  \bibinfo{author}{\bibfnamefont{N.~J.} \bibnamefont{Kasdin}},
  \bibinfo{journal}{ApJ} \textbf{\bibinfo{volume}{665}}, \bibinfo{pages}{794}
  (\bibinfo{year}{2007}).

\bibitem[{\citenamefont{{Gaudi} et~al.}(2020)\citenamefont{{Gaudi}, {Seager},
  {Mennesson}, {Kiessling}, {Warfield}, {Cahoy}, {Clarke}, {Domagal-Goldman},
  {Feinberg}, {Guyon} et~al.}}]{Gaudi2020_HabEx}
\bibinfo{author}{\bibfnamefont{B.~S.} \bibnamefont{{Gaudi}}},
  \bibinfo{author}{\bibfnamefont{S.}~\bibnamefont{{Seager}}},
  \bibinfo{author}{\bibfnamefont{B.}~\bibnamefont{{Mennesson}}},
  \bibinfo{author}{\bibfnamefont{A.}~\bibnamefont{{Kiessling}}},
  \bibinfo{author}{\bibfnamefont{K.}~\bibnamefont{{Warfield}}},
  \bibinfo{author}{\bibfnamefont{K.}~\bibnamefont{{Cahoy}}},
  \bibinfo{author}{\bibfnamefont{J.~T.} \bibnamefont{{Clarke}}},
  \bibinfo{author}{\bibfnamefont{S.}~\bibnamefont{{Domagal-Goldman}}},
  \bibinfo{author}{\bibfnamefont{L.}~\bibnamefont{{Feinberg}}},
  \bibinfo{author}{\bibfnamefont{O.}~\bibnamefont{{Guyon}}},
  \bibnamefont{et~al.} (\bibinfo{year}{2020}), \eprint{arXiv:2001.06683
  [astro-ph.IM]}.

\bibitem[{\citenamefont{Stark et~al.}(2014{\natexlab{b}})\citenamefont{Stark,
  Roberge, Mandell, and Robinson}}]{Stark2014}
\bibinfo{author}{\bibfnamefont{C.~C.} \bibnamefont{Stark}},
  \bibinfo{author}{\bibfnamefont{A.}~\bibnamefont{Roberge}},
  \bibinfo{author}{\bibfnamefont{A.}~\bibnamefont{Mandell}}, \bibnamefont{and}
  \bibinfo{author}{\bibfnamefont{T.~D.} \bibnamefont{Robinson}},
  \bibinfo{journal}{ApJ} \textbf{\bibinfo{volume}{795}}, \bibinfo{pages}{122}
  (\bibinfo{year}{2014}{\natexlab{b}}).

\bibitem[{\citenamefont{Trauger and Traub}(2007{\natexlab{a}})}]{Trauger2007}
\bibinfo{author}{\bibfnamefont{J.~T.} \bibnamefont{Trauger}} \bibnamefont{and}
  \bibinfo{author}{\bibfnamefont{W.~A.} \bibnamefont{Traub}},
  \bibinfo{journal}{Nature} \textbf{\bibinfo{volume}{446}},
  \bibinfo{pages}{771} (\bibinfo{year}{2007}{\natexlab{a}}).

\bibitem[{\citenamefont{McKeithen et~al.}(2021)\citenamefont{McKeithen,
  Shaklan, Lisman, and etal.}}]{McKeithen2021}
\bibinfo{author}{\bibfnamefont{D.~M.} \bibnamefont{McKeithen}},
  \bibinfo{author}{\bibfnamefont{S.~B.} \bibnamefont{Shaklan}},
  \bibinfo{author}{\bibfnamefont{P.~D.} \bibnamefont{Lisman}},
  \bibnamefont{and} \bibinfo{author}{\bibnamefont{etal.}},
  \bibinfo{journal}{JATIS} \textbf{\bibinfo{volume}{7}},
  \bibinfo{pages}{021204} (\bibinfo{year}{2021}).

\bibitem[{\citenamefont{Catanzarite and Shao}(2011)}]{Catanzarite2010}
\bibinfo{author}{\bibfnamefont{J.~H.} \bibnamefont{Catanzarite}}
  \bibnamefont{and} \bibinfo{author}{\bibfnamefont{M.}~\bibnamefont{Shao}},
  \bibinfo{journal}{ApJ} \textbf{\bibinfo{volume}{738}}, \bibinfo{pages}{151}
  (\bibinfo{year}{2011}).

\bibitem[{\citenamefont{Bracewell}(1978)}]{Bracewell1978}
\bibinfo{author}{\bibfnamefont{R.~N.} \bibnamefont{Bracewell}},
  \bibinfo{journal}{Nature} \textbf{\bibinfo{volume}{274}},
  \bibinfo{pages}{780} (\bibinfo{year}{1978}).

\bibitem[{\citenamefont{Angel and Woolf}(1997)}]{AngelWoolf1997}
\bibinfo{author}{\bibfnamefont{J.~R.~P.} \bibnamefont{Angel}} \bibnamefont{and}
  \bibinfo{author}{\bibfnamefont{N.~J.} \bibnamefont{Woolf}},
  \bibinfo{journal}{The Astrophysical Journal} \textbf{\bibinfo{volume}{475}},
  \bibinfo{pages}{373} (\bibinfo{year}{1997}).

\bibitem[{\citenamefont{Lay}(2004{\natexlab{a}})}]{Lay2004}
\bibinfo{author}{\bibfnamefont{O.~P.} \bibnamefont{Lay}},
  \bibinfo{journal}{Applied Optics} \textbf{\bibinfo{volume}{43}},
  \bibinfo{pages}{6100} (\bibinfo{year}{2004}{\natexlab{a}}).

\bibitem[{\citenamefont{Lay}(2005)}]{Lay2005}
\bibinfo{author}{\bibfnamefont{O.~P.} \bibnamefont{Lay}},
  \bibinfo{journal}{Applied Optics} \textbf{\bibinfo{volume}{44}},
  \bibinfo{pages}{5859} (\bibinfo{year}{2005}).

\bibitem[{\citenamefont{Dubovitsky and
  Lay}(2004{\natexlab{a}})}]{DubovitskyLay2004}
\bibinfo{author}{\bibfnamefont{S.}~\bibnamefont{Dubovitsky}} \bibnamefont{and}
  \bibinfo{author}{\bibfnamefont{O.~P.} \bibnamefont{Lay}}, in
  \emph{\bibinfo{booktitle}{New Frontiers in Stellar Interferometry}}, edited
  by \bibinfo{editor}{\bibfnamefont{W.~A.} \bibnamefont{Traub}}
  (\bibinfo{year}{2004}{\natexlab{a}}), vol. \bibinfo{volume}{5491} of
  \emph{\bibinfo{series}{Proc. of SPIE}}, pp. \bibinfo{pages}{284--295}.

\bibitem[{\citenamefont{Lay and Dubovitsky}(2004)}]{LayDubovitsky2004sys}
\bibinfo{author}{\bibfnamefont{O.~P.} \bibnamefont{Lay}} \bibnamefont{and}
  \bibinfo{author}{\bibfnamefont{S.}~\bibnamefont{Dubovitsky}}, in
  \emph{\bibinfo{booktitle}{New Frontiers in Stellar Interferometry}}, edited
  by \bibinfo{editor}{\bibfnamefont{W.~A.} \bibnamefont{Traub}}
  (\bibinfo{year}{2004}), vol. \bibinfo{volume}{5491} of
  \emph{\bibinfo{series}{Proc. of SPIE}}, pp. \bibinfo{pages}{871--882}.

\bibitem[{Law(2005)}]{LawsonDooley2005TPFItechplan}
\bibinfo{type}{Tech. Rep.} \bibinfo{number}{JPL Publication 05-5},
  \bibinfo{institution}{Jet Propulsion Laboratory, California Institute of
  Technology}, \bibinfo{address}{Pasadena, CA} (\bibinfo{year}{2005}),
  \urlprefix\url{https://exoplanets.nasa.gov/exep/files/exep/tpfI414.pdf}.

\bibitem[{\citenamefont{Lawson et~al.}(2007)\citenamefont{Lawson, Lay,
  Johnston, and Beichman}}]{Lawson2007TPFISWG}
\bibinfo{author}{\bibfnamefont{P.~R.} \bibnamefont{Lawson}},
  \bibinfo{author}{\bibfnamefont{O.~P.} \bibnamefont{Lay}},
  \bibinfo{author}{\bibfnamefont{K.~J.} \bibnamefont{Johnston}},
  \bibnamefont{and} \bibinfo{author}{\bibfnamefont{C.~A.}
  \bibnamefont{Beichman}}, \bibinfo{type}{Tech. Rep.} \bibinfo{number}{JPL
  Publication 07-1}, \bibinfo{institution}{Jet Propulsion Laboratory,
  California Institute of Technology}, \bibinfo{address}{Pasadena, CA}
  (\bibinfo{year}{2007}),
  \urlprefix\url{https://exoplanets.nasa.gov/exep/files/exep/TPFIswgReport2007.pdf}.

\bibitem[{\citenamefont{Cockell et~al.}(2009)\citenamefont{Cockell,
  Kaltenegger, Udry et~al.}}]{Cockell2009}
\bibinfo{author}{\bibfnamefont{C.~S.} \bibnamefont{Cockell}},
  \bibinfo{author}{\bibfnamefont{L.}~\bibnamefont{Kaltenegger}},
  \bibinfo{author}{\bibfnamefont{S.}~\bibnamefont{Udry}}, \bibnamefont{et~al.},
  \bibinfo{journal}{Exp. Astron.} \textbf{\bibinfo{volume}{23}},
  \bibinfo{pages}{435} (\bibinfo{year}{2009}).

\bibitem[{\citenamefont{Dubovitsky and
  Lay}(2004{\natexlab{b}})}]{Dubovitsky2004}
\bibinfo{author}{\bibfnamefont{S.}~\bibnamefont{Dubovitsky}} \bibnamefont{and}
  \bibinfo{author}{\bibfnamefont{O.~P.} \bibnamefont{Lay}}, in
  \emph{\bibinfo{booktitle}{Proc. of SPIE}}
  (\bibinfo{year}{2004}{\natexlab{b}}), vol. \bibinfo{volume}{5491}, pp.
  \bibinfo{pages}{509--518}.

\bibitem[{\citenamefont{Monnier}(2003)}]{Monnier2003}
\bibinfo{author}{\bibfnamefont{J.~D.} \bibnamefont{Monnier}},
  \bibinfo{journal}{Rep. Progr. Physics} \textbf{\bibinfo{volume}{66}},
  \bibinfo{pages}{789} (\bibinfo{year}{2003}).

\bibitem[{\citenamefont{Beichman et~al.}(1999)\citenamefont{Beichman, Bryden,
  Gautier, and Stapelfeldt}}]{Beichman1999a}
\bibinfo{author}{\bibfnamefont{C.~A.} \bibnamefont{Beichman}},
  \bibinfo{author}{\bibfnamefont{G.}~\bibnamefont{Bryden}},
  \bibinfo{author}{\bibfnamefont{T.~N.} \bibnamefont{Gautier}},
  \bibnamefont{and} \bibinfo{author}{\bibfnamefont{K.~R.}
  \bibnamefont{Stapelfeldt}}, \bibinfo{journal}{ApJ Lett.}
  \textbf{\bibinfo{volume}{521}}, \bibinfo{pages}{L117} (\bibinfo{year}{1999}).

\bibitem[{\citenamefont{Lay}(2004{\natexlab{b}})}]{Lay2004SystematicNull}
\bibinfo{author}{\bibfnamefont{O.~P.} \bibnamefont{Lay}},
  \bibinfo{journal}{Applied Optics} \textbf{\bibinfo{volume}{43}},
  \bibinfo{pages}{6100} (\bibinfo{year}{2004}{\natexlab{b}}).

\bibitem[{\citenamefont{Guyon}(2005{\natexlab{a}})}]{Guyon2005}
\bibinfo{author}{\bibfnamefont{O.}~\bibnamefont{Guyon}}, \bibinfo{journal}{ApJ}
  \textbf{\bibinfo{volume}{629}}, \bibinfo{pages}{592}
  (\bibinfo{year}{2005}{\natexlab{a}}).

\bibitem[{\citenamefont{Gilmozzi and
  Spyromilio}(2007)}]{GilmozziSpyromilio2007}
\bibinfo{author}{\bibfnamefont{R.}~\bibnamefont{Gilmozzi}} \bibnamefont{and}
  \bibinfo{author}{\bibfnamefont{J.}~\bibnamefont{Spyromilio}},
  \bibinfo{journal}{The Messenger} \textbf{\bibinfo{volume}{127}},
  \bibinfo{pages}{11} (\bibinfo{year}{2007}).

\bibitem[{\citenamefont{Johns et~al.}(2012)\citenamefont{Johns, Chiu, Dunlop,
  Gabor, Herriot, Hill, Iserlohe, Liu, Montoya-Figueroa, Takami
  et~al.}}]{Johns2012}
\bibinfo{author}{\bibfnamefont{M.}~\bibnamefont{Johns}},
  \bibinfo{author}{\bibfnamefont{K.}~\bibnamefont{Chiu}},
  \bibinfo{author}{\bibfnamefont{C.}~\bibnamefont{Dunlop}},
  \bibinfo{author}{\bibfnamefont{A.}~\bibnamefont{Gabor}},
  \bibinfo{author}{\bibfnamefont{G.}~\bibnamefont{Herriot}},
  \bibinfo{author}{\bibfnamefont{J.~M.} \bibnamefont{Hill}},
  \bibinfo{author}{\bibfnamefont{C.}~\bibnamefont{Iserlohe}},
  \bibinfo{author}{\bibfnamefont{M.~C.} \bibnamefont{Liu}},
  \bibinfo{author}{\bibfnamefont{E.}~\bibnamefont{Montoya-Figueroa}},
  \bibinfo{author}{\bibfnamefont{H.}~\bibnamefont{Takami}},
  \bibnamefont{et~al.}, in \emph{\bibinfo{booktitle}{Proceedings of SPIE}}
  (\bibinfo{year}{2012}), vol. \bibinfo{volume}{8444}, p.
  \bibinfo{pages}{84441X}.

\bibitem[{\citenamefont{Sanders et~al.}(2020)\citenamefont{Sanders, Herriot,
  Andersen, Chui, Gonzalez, Montgomery, Palacios, Phillips, Spiegel, and
  Young}}]{Sanders2020}
\bibinfo{author}{\bibfnamefont{G.}~\bibnamefont{Sanders}},
  \bibinfo{author}{\bibfnamefont{G.}~\bibnamefont{Herriot}},
  \bibinfo{author}{\bibfnamefont{D.}~\bibnamefont{Andersen}},
  \bibinfo{author}{\bibfnamefont{J.}~\bibnamefont{Chui}},
  \bibinfo{author}{\bibfnamefont{A.}~\bibnamefont{Gonzalez}},
  \bibinfo{author}{\bibfnamefont{D.}~\bibnamefont{Montgomery}},
  \bibinfo{author}{\bibfnamefont{P.}~\bibnamefont{Palacios}},
  \bibinfo{author}{\bibfnamefont{B.}~\bibnamefont{Phillips}},
  \bibinfo{author}{\bibfnamefont{A.}~\bibnamefont{Spiegel}}, \bibnamefont{and}
  \bibinfo{author}{\bibfnamefont{J.}~\bibnamefont{Young}}, in
  \emph{\bibinfo{booktitle}{Proc. of SPIE}} (\bibinfo{year}{2020}), vol.
  \bibinfo{volume}{11447}, p. \bibinfo{pages}{114470E}.

\bibitem[{\citenamefont{Guyon}(2005{\natexlab{b}})}]{Guyon2005AOlimits}
\bibinfo{author}{\bibfnamefont{O.}~\bibnamefont{Guyon}}, \bibinfo{journal}{ApJ}
  \textbf{\bibinfo{volume}{629}}, \bibinfo{pages}{592}
  (\bibinfo{year}{2005}{\natexlab{b}}).

\bibitem[{\citenamefont{Macintosh et~al.}(2014)\citenamefont{Macintosh, Graham,
  Ingraham, Konopacky, Marois, Perrin, Pueyo, Rice, Rantakyrö, Sadakuni
  et~al.}}]{Macintosh2014}
\bibinfo{author}{\bibfnamefont{B.~A.} \bibnamefont{Macintosh}},
  \bibinfo{author}{\bibfnamefont{J.~R.} \bibnamefont{Graham}},
  \bibinfo{author}{\bibfnamefont{P.}~\bibnamefont{Ingraham}},
  \bibinfo{author}{\bibfnamefont{Q.~M.} \bibnamefont{Konopacky}},
  \bibinfo{author}{\bibfnamefont{C.}~\bibnamefont{Marois}},
  \bibinfo{author}{\bibfnamefont{M.}~\bibnamefont{Perrin}},
  \bibinfo{author}{\bibfnamefont{L.}~\bibnamefont{Pueyo}},
  \bibinfo{author}{\bibfnamefont{E.}~\bibnamefont{Rice}},
  \bibinfo{author}{\bibfnamefont{F.}~\bibnamefont{Rantakyrö}},
  \bibinfo{author}{\bibfnamefont{N.}~\bibnamefont{Sadakuni}},
  \bibnamefont{et~al.}, \bibinfo{journal}{Proc. of the Nat. Acad. Sci.}
  \textbf{\bibinfo{volume}{111}}, \bibinfo{pages}{12661}
  (\bibinfo{year}{2014}).

\bibitem[{\citenamefont{Beuzit et~al.}(2019)\citenamefont{Beuzit, Vigan,
  Mouillet, Claudi, Desidera, Fontanive, Feldt, Dohlen, Puget, Langlois
  et~al.}}]{Beuzit2019}
\bibinfo{author}{\bibfnamefont{J.-L.} \bibnamefont{Beuzit}},
  \bibinfo{author}{\bibfnamefont{A.}~\bibnamefont{Vigan}},
  \bibinfo{author}{\bibfnamefont{D.}~\bibnamefont{Mouillet}},
  \bibinfo{author}{\bibfnamefont{R.~U.} \bibnamefont{Claudi}},
  \bibinfo{author}{\bibfnamefont{S.}~\bibnamefont{Desidera}},
  \bibinfo{author}{\bibfnamefont{C.}~\bibnamefont{Fontanive}},
  \bibinfo{author}{\bibfnamefont{M.}~\bibnamefont{Feldt}},
  \bibinfo{author}{\bibfnamefont{K.}~\bibnamefont{Dohlen}},
  \bibinfo{author}{\bibfnamefont{P.}~\bibnamefont{Puget}},
  \bibinfo{author}{\bibfnamefont{M.}~\bibnamefont{Langlois}},
  \bibnamefont{et~al.}, \bibinfo{journal}{Astron. \& Astrophys.}
  \textbf{\bibinfo{volume}{631}}, \bibinfo{pages}{A155} (\bibinfo{year}{2019}).

\bibitem[{\citenamefont{Jovanovic et~al.}(2015)\citenamefont{Jovanovic,
  Martinache, Guyon, Clergeon, Singh, Kudo, Garrel, Newman, Doughty, Lozi
  et~al.}}]{Jovanovic2015}
\bibinfo{author}{\bibfnamefont{N.}~\bibnamefont{Jovanovic}},
  \bibinfo{author}{\bibfnamefont{F.}~\bibnamefont{Martinache}},
  \bibinfo{author}{\bibfnamefont{O.}~\bibnamefont{Guyon}},
  \bibinfo{author}{\bibfnamefont{C.}~\bibnamefont{Clergeon}},
  \bibinfo{author}{\bibfnamefont{G.}~\bibnamefont{Singh}},
  \bibinfo{author}{\bibfnamefont{T.}~\bibnamefont{Kudo}},
  \bibinfo{author}{\bibfnamefont{V.}~\bibnamefont{Garrel}},
  \bibinfo{author}{\bibfnamefont{K.}~\bibnamefont{Newman}},
  \bibinfo{author}{\bibfnamefont{D.}~\bibnamefont{Doughty}},
  \bibinfo{author}{\bibfnamefont{J.}~\bibnamefont{Lozi}}, \bibnamefont{et~al.},
  \bibinfo{journal}{PASP} \textbf{\bibinfo{volume}{127}}, \bibinfo{pages}{890}
  (\bibinfo{year}{2015}).

\bibitem[{\citenamefont{Kasper et~al.}(2020)\citenamefont{Kasper, Brandl,
  Kellner, Feldt, Hippler, Quirrenbach, Schmid, Lenzen, Henning, and
  Helling}}]{Kasper2020}
\bibinfo{author}{\bibfnamefont{M.}~\bibnamefont{Kasper}},
  \bibinfo{author}{\bibfnamefont{B.~R.} \bibnamefont{Brandl}},
  \bibinfo{author}{\bibfnamefont{S.}~\bibnamefont{Kellner}},
  \bibinfo{author}{\bibfnamefont{M.}~\bibnamefont{Feldt}},
  \bibinfo{author}{\bibfnamefont{S.}~\bibnamefont{Hippler}},
  \bibinfo{author}{\bibfnamefont{A.}~\bibnamefont{Quirrenbach}},
  \bibinfo{author}{\bibfnamefont{H.~M.} \bibnamefont{Schmid}},
  \bibinfo{author}{\bibfnamefont{R.}~\bibnamefont{Lenzen}},
  \bibinfo{author}{\bibfnamefont{T.}~\bibnamefont{Henning}}, \bibnamefont{and}
  \bibinfo{author}{\bibfnamefont{C.}~\bibnamefont{Helling}}, in
  \emph{\bibinfo{booktitle}{Proc. of SPIE}} (\bibinfo{year}{2020}), vol.
  \bibinfo{volume}{11447}, p. \bibinfo{pages}{114470C}.

\bibitem[{\citenamefont{Guyon et~al.}(2020)\citenamefont{Guyon, Martin,
  Belikov, Kasdin, Macintosh, Mawet, Fienup, Kern, Lie, and
  Lowman}}]{Guyon2020}
\bibinfo{author}{\bibfnamefont{O.}~\bibnamefont{Guyon}},
  \bibinfo{author}{\bibfnamefont{N.}~\bibnamefont{Martin}},
  \bibinfo{author}{\bibfnamefont{R.}~\bibnamefont{Belikov}},
  \bibinfo{author}{\bibfnamefont{N.~J.} \bibnamefont{Kasdin}},
  \bibinfo{author}{\bibfnamefont{B.~A.} \bibnamefont{Macintosh}},
  \bibinfo{author}{\bibfnamefont{D.}~\bibnamefont{Mawet}},
  \bibinfo{author}{\bibfnamefont{J.~R.} \bibnamefont{Fienup}},
  \bibinfo{author}{\bibfnamefont{P.}~\bibnamefont{Kern}},
  \bibinfo{author}{\bibfnamefont{G.~K.} \bibnamefont{Lie}}, \bibnamefont{and}
  \bibinfo{author}{\bibfnamefont{P.~D.} \bibnamefont{Lowman}},
  \bibinfo{journal}{JATIS} \textbf{\bibinfo{volume}{6}},
  \bibinfo{pages}{041202} (\bibinfo{year}{2020}).

\bibitem[{\citenamefont{Sullivan and Simcoe}(2012)}]{Sullivan2012}
\bibinfo{author}{\bibfnamefont{P.~W.} \bibnamefont{Sullivan}} \bibnamefont{and}
  \bibinfo{author}{\bibfnamefont{R.~A.} \bibnamefont{Simcoe}},
  \bibinfo{journal}{PASP} \textbf{\bibinfo{volume}{124}}, \bibinfo{pages}{1336}
  (\bibinfo{year}{2012}).

\bibitem[{\citenamefont{Labeyrie}(1996)}]{Labeyrie1996}
\bibinfo{author}{\bibfnamefont{A.}~\bibnamefont{Labeyrie}},
  \bibinfo{journal}{Astron. \& Astrophys. Supplement Series}
  \textbf{\bibinfo{volume}{118}}, \bibinfo{pages}{517} (\bibinfo{year}{1996}).

\bibitem[{\citenamefont{Anterrieu et~al.}(2019)\citenamefont{Anterrieu,
  Christian, Coudé~du Foresto, Kulkarni, Monnier, Perrin, and
  Traub}}]{Anterrieu2019}
\bibinfo{author}{\bibfnamefont{E.~R.} \bibnamefont{Anterrieu}},
  \bibinfo{author}{\bibfnamefont{M.}~\bibnamefont{Christian}},
  \bibinfo{author}{\bibfnamefont{V.}~\bibnamefont{Coudé~du Foresto}},
  \bibinfo{author}{\bibfnamefont{N.}~\bibnamefont{Kulkarni}},
  \bibinfo{author}{\bibfnamefont{J.~D.} \bibnamefont{Monnier}},
  \bibinfo{author}{\bibfnamefont{G.}~\bibnamefont{Perrin}}, \bibnamefont{and}
  \bibinfo{author}{\bibfnamefont{W.~A.} \bibnamefont{Traub}},
  \bibinfo{journal}{JATIS} \textbf{\bibinfo{volume}{5}},
  \bibinfo{pages}{011002} (\bibinfo{year}{2019}).

\bibitem[{\citenamefont{Perrin et~al.}(2018)\citenamefont{Perrin, Le~Coroller,
  and Lacour}}]{Perrin2018}
\bibinfo{author}{\bibfnamefont{G.}~\bibnamefont{Perrin}},
  \bibinfo{author}{\bibfnamefont{H.}~\bibnamefont{Le~Coroller}},
  \bibnamefont{and} \bibinfo{author}{\bibfnamefont{S.}~\bibnamefont{Lacour}},
  \bibinfo{journal}{Astron. \& Astrophys.} \textbf{\bibinfo{volume}{616}},
  \bibinfo{pages}{A120} (\bibinfo{year}{2018}).

\bibitem[{\citenamefont{Lay et~al.}(2008)\citenamefont{Lay, Dubovitsky, Hector,
  Laine, Rigaut, and Spengler}}]{Lay2008}
\bibinfo{author}{\bibfnamefont{O.~P.} \bibnamefont{Lay}},
  \bibinfo{author}{\bibfnamefont{S.}~\bibnamefont{Dubovitsky}},
  \bibinfo{author}{\bibfnamefont{W.}~\bibnamefont{Hector}},
  \bibinfo{author}{\bibfnamefont{E.}~\bibnamefont{Laine}},
  \bibinfo{author}{\bibfnamefont{F.}~\bibnamefont{Rigaut}}, \bibnamefont{and}
  \bibinfo{author}{\bibfnamefont{C.}~\bibnamefont{Spengler}},
  \bibinfo{journal}{JOSA A} \textbf{\bibinfo{volume}{25}},
  \bibinfo{pages}{3186} (\bibinfo{year}{2008}).

\bibitem[{\citenamefont{Kawahara and Fujii}(2010)}]{Kawahara2010}
\bibinfo{author}{\bibfnamefont{H.}~\bibnamefont{Kawahara}} \bibnamefont{and}
  \bibinfo{author}{\bibfnamefont{Y.}~\bibnamefont{Fujii}},
  \bibinfo{journal}{ApJ} \textbf{\bibinfo{volume}{720}}, \bibinfo{pages}{1333}
  (\bibinfo{year}{2010}).

\bibitem[{\citenamefont{Fujii and Kawahara}(2012)}]{FujiiKawahara2012}
\bibinfo{author}{\bibfnamefont{Y.}~\bibnamefont{Fujii}} \bibnamefont{and}
  \bibinfo{author}{\bibfnamefont{H.}~\bibnamefont{Kawahara}},
  \bibinfo{journal}{ApJ} \textbf{\bibinfo{volume}{752}}, \bibinfo{pages}{184}
  (\bibinfo{year}{2012}).

\bibitem[{\citenamefont{Karalidi et~al.}(2015)\citenamefont{Karalidi, Palle,
  and Stam}}]{Karalidi2015}
\bibinfo{author}{\bibfnamefont{T.}~\bibnamefont{Karalidi}},
  \bibinfo{author}{\bibfnamefont{E.}~\bibnamefont{Palle}}, \bibnamefont{and}
  \bibinfo{author}{\bibfnamefont{D.~M.} \bibnamefont{Stam}},
  \bibinfo{journal}{ApJ} \textbf{\bibinfo{volume}{803}}, \bibinfo{pages}{129}
  (\bibinfo{year}{2015}).

\bibitem[{\citenamefont{de~Wit et~al.}(2012)\citenamefont{de~Wit, Gillon,
  Demory, Seager, Madhusudhan, Bolmont, Rogers, and Kristiansen}}]{deWit2012}
\bibinfo{author}{\bibfnamefont{J.}~\bibnamefont{de~Wit}},
  \bibinfo{author}{\bibfnamefont{M.}~\bibnamefont{Gillon}},
  \bibinfo{author}{\bibfnamefont{B.-O.} \bibnamefont{Demory}},
  \bibinfo{author}{\bibfnamefont{S.}~\bibnamefont{Seager}},
  \bibinfo{author}{\bibfnamefont{N.}~\bibnamefont{Madhusudhan}},
  \bibinfo{author}{\bibfnamefont{E.}~\bibnamefont{Bolmont}},
  \bibinfo{author}{\bibfnamefont{L.~A.} \bibnamefont{Rogers}},
  \bibnamefont{and}
  \bibinfo{author}{\bibfnamefont{H.}~\bibnamefont{Kristiansen}},
  \bibinfo{journal}{Astron. \& Astrophys.} \textbf{\bibinfo{volume}{548}},
  \bibinfo{pages}{A128} (\bibinfo{year}{2012}).

\bibitem[{\citenamefont{Stevenson et~al.}(2014)\citenamefont{Stevenson, Lewis,
  Bean, Désert, Madhusudhan, Seager, Sing, and Winn}}]{Stevenson2014}
\bibinfo{author}{\bibfnamefont{K.~B.} \bibnamefont{Stevenson}},
  \bibinfo{author}{\bibfnamefont{N.~K.} \bibnamefont{Lewis}},
  \bibinfo{author}{\bibfnamefont{J.~L.} \bibnamefont{Bean}},
  \bibinfo{author}{\bibfnamefont{J.-M.} \bibnamefont{Désert}},
  \bibinfo{author}{\bibfnamefont{N.}~\bibnamefont{Madhusudhan}},
  \bibinfo{author}{\bibfnamefont{S.}~\bibnamefont{Seager}},
  \bibinfo{author}{\bibfnamefont{D.~K.} \bibnamefont{Sing}}, \bibnamefont{and}
  \bibinfo{author}{\bibfnamefont{J.~N.} \bibnamefont{Winn}},
  \bibinfo{journal}{Science} \textbf{\bibinfo{volume}{346}},
  \bibinfo{pages}{838} (\bibinfo{year}{2014}).

\bibitem[{\citenamefont{Kreidberg et~al.}(2014)\citenamefont{Kreidberg, Bean,
  D{\'e}sert, Benneke, Deming, Stevenson, Berta-Thompson, Seifahrt, and
  Homeier}}]{Kreidberg2014}
\bibinfo{author}{\bibfnamefont{L.}~\bibnamefont{Kreidberg}},
  \bibinfo{author}{\bibfnamefont{J.~L.} \bibnamefont{Bean}},
  \bibinfo{author}{\bibfnamefont{J.}~\bibnamefont{D{\'e}sert}},
  \bibinfo{author}{\bibfnamefont{B.}~\bibnamefont{Benneke}},
  \bibinfo{author}{\bibfnamefont{D.}~\bibnamefont{Deming}},
  \bibinfo{author}{\bibfnamefont{K.~B.} \bibnamefont{Stevenson}},
  \bibinfo{author}{\bibfnamefont{Z.~K.} \bibnamefont{Berta-Thompson}},
  \bibinfo{author}{\bibfnamefont{A.}~\bibnamefont{Seifahrt}}, \bibnamefont{and}
  \bibinfo{author}{\bibfnamefont{D.}~\bibnamefont{Homeier}},
  \bibinfo{journal}{Nature} \textbf{\bibinfo{volume}{505}}, \bibinfo{pages}{69}
  (\bibinfo{year}{2014}).

\bibitem[{\citenamefont{Rackham et~al.}(2018)\citenamefont{Rackham, Apai, and
  Giampapa}}]{Rackham2018}
\bibinfo{author}{\bibfnamefont{B.~A.} \bibnamefont{Rackham}},
  \bibinfo{author}{\bibfnamefont{D.}~\bibnamefont{Apai}}, \bibnamefont{and}
  \bibinfo{author}{\bibfnamefont{M.~S.} \bibnamefont{Giampapa}},
  \bibinfo{journal}{ApJ} \textbf{\bibinfo{volume}{853}}, \bibinfo{pages}{122}
  (\bibinfo{year}{2018}).

\bibitem[{\citenamefont{Roques et~al.}(2017)\citenamefont{Roques, Ortiz,
  Lecacheux, Camus, Widemann, Santos-Sanz, Duffard, Gladman, Marsset, and
  Sicardy}}]{Roques2017}
\bibinfo{author}{\bibfnamefont{F.}~\bibnamefont{Roques}},
  \bibinfo{author}{\bibfnamefont{J.~L.} \bibnamefont{Ortiz}},
  \bibinfo{author}{\bibfnamefont{J.}~\bibnamefont{Lecacheux}},
  \bibinfo{author}{\bibfnamefont{J.}~\bibnamefont{Camus}},
  \bibinfo{author}{\bibfnamefont{T.}~\bibnamefont{Widemann}},
  \bibinfo{author}{\bibfnamefont{P.}~\bibnamefont{Santos-Sanz}},
  \bibinfo{author}{\bibfnamefont{R.}~\bibnamefont{Duffard}},
  \bibinfo{author}{\bibfnamefont{B.}~\bibnamefont{Gladman}},
  \bibinfo{author}{\bibfnamefont{M.}~\bibnamefont{Marsset}}, \bibnamefont{and}
  \bibinfo{author}{\bibfnamefont{B.}~\bibnamefont{Sicardy}},
  \bibinfo{journal}{AJ} \textbf{\bibinfo{volume}{153}}, \bibinfo{pages}{7}
  (\bibinfo{year}{2017}).

\bibitem[{\citenamefont{Sheppard et~al.}(2011)\citenamefont{Sheppard, Trujillo,
  and Tholen}}]{Sheppard2011}
\bibinfo{author}{\bibfnamefont{S.~S.} \bibnamefont{Sheppard}},
  \bibinfo{author}{\bibfnamefont{C.~A.} \bibnamefont{Trujillo}},
  \bibnamefont{and} \bibinfo{author}{\bibfnamefont{D.~J.}
  \bibnamefont{Tholen}}, \bibinfo{journal}{AJ} \textbf{\bibinfo{volume}{141}},
  \bibinfo{pages}{150} (\bibinfo{year}{2011}).

\bibitem[{\citenamefont{Herman et~al.}(2018)\citenamefont{Herman, Huang,
  McPeters, Ziemke, Cede, and Blank}}]{Herman2018}
\bibinfo{author}{\bibfnamefont{J.~R.} \bibnamefont{Herman}},
  \bibinfo{author}{\bibfnamefont{L.}~\bibnamefont{Huang}},
  \bibinfo{author}{\bibfnamefont{R.~D.} \bibnamefont{McPeters}},
  \bibinfo{author}{\bibfnamefont{J.}~\bibnamefont{Ziemke}},
  \bibinfo{author}{\bibfnamefont{A.}~\bibnamefont{Cede}}, \bibnamefont{and}
  \bibinfo{author}{\bibfnamefont{K.}~\bibnamefont{Blank}},
  \bibinfo{journal}{Atmospheric Meas. Tech.} \textbf{\bibinfo{volume}{11}},
  \bibinfo{pages}{177} (\bibinfo{year}{2018}).

\bibitem[{\citenamefont{Singer et~al.}(2016)\citenamefont{Singer, Seubert,
  Nelson, Jacquet, Neuberger, Gwyn, Spitale, and Smith}}]{Singer2016}
\bibinfo{author}{\bibfnamefont{K.~N.} \bibnamefont{Singer}},
  \bibinfo{author}{\bibfnamefont{S.~A.} \bibnamefont{Seubert}},
  \bibinfo{author}{\bibfnamefont{M.~L.} \bibnamefont{Nelson}},
  \bibinfo{author}{\bibfnamefont{E.~R.} \bibnamefont{Jacquet}},
  \bibinfo{author}{\bibfnamefont{D.~S.} \bibnamefont{Neuberger}},
  \bibinfo{author}{\bibfnamefont{S.~D.} \bibnamefont{Gwyn}},
  \bibinfo{author}{\bibfnamefont{J.}~\bibnamefont{Spitale}}, \bibnamefont{and}
  \bibinfo{author}{\bibfnamefont{S.~K.} \bibnamefont{Smith}},
  \bibinfo{journal}{J. of Guidance, Control, and Dynamics}
  \textbf{\bibinfo{volume}{39}}, \bibinfo{pages}{2447} (\bibinfo{year}{2016}).

\bibitem[{\citenamefont{Landgraf et~al.}(2000)\citenamefont{Landgraf, Baggaley,
  Grün, Hamilton, and Zook}}]{Landgraf2000}
\bibinfo{author}{\bibfnamefont{M.}~\bibnamefont{Landgraf}},
  \bibinfo{author}{\bibfnamefont{W.~J.} \bibnamefont{Baggaley}},
  \bibinfo{author}{\bibfnamefont{E.}~\bibnamefont{Grün}},
  \bibinfo{author}{\bibfnamefont{D.~P.} \bibnamefont{Hamilton}},
  \bibnamefont{and} \bibinfo{author}{\bibfnamefont{H.~A.} \bibnamefont{Zook}},
  \bibinfo{journal}{Science} \textbf{\bibinfo{volume}{288}},
  \bibinfo{pages}{1751} (\bibinfo{year}{2000}).

\bibitem[{\citenamefont{Spillantini et~al.}(2007)\citenamefont{Spillantini,
  Casolino, Durante, Mueller-Mellin, Reitz, Rossi, Shurshakov, and
  Sorbi}}]{Spillantini2007radmeas}
\bibinfo{author}{\bibfnamefont{P.}~\bibnamefont{Spillantini}},
  \bibinfo{author}{\bibfnamefont{M.}~\bibnamefont{Casolino}},
  \bibinfo{author}{\bibfnamefont{M.}~\bibnamefont{Durante}},
  \bibinfo{author}{\bibfnamefont{R.}~\bibnamefont{Mueller-Mellin}},
  \bibinfo{author}{\bibfnamefont{G.}~\bibnamefont{Reitz}},
  \bibinfo{author}{\bibfnamefont{L.}~\bibnamefont{Rossi}},
  \bibinfo{author}{\bibfnamefont{V.}~\bibnamefont{Shurshakov}},
  \bibnamefont{and} \bibinfo{author}{\bibfnamefont{M.}~\bibnamefont{Sorbi}},
  \bibinfo{journal}{Radiation Measurements} \textbf{\bibinfo{volume}{42}},
  \bibinfo{pages}{14} (\bibinfo{year}{2007}).

\bibitem[{\citenamefont{Chen et~al.}(2017)\citenamefont{Chen, Atkinson, Wright,
  Mendez, Blume, Del~Signore, and Wright}}]{Chen2017}
\bibinfo{author}{\bibfnamefont{E.}~\bibnamefont{Chen}},
  \bibinfo{author}{\bibfnamefont{S.}~\bibnamefont{Atkinson}},
  \bibinfo{author}{\bibfnamefont{D.~J.} \bibnamefont{Wright}},
  \bibinfo{author}{\bibfnamefont{A.}~\bibnamefont{Mendez}},
  \bibinfo{author}{\bibfnamefont{T.}~\bibnamefont{Blume}},
  \bibinfo{author}{\bibfnamefont{M.}~\bibnamefont{Del~Signore}},
  \bibnamefont{and} \bibinfo{author}{\bibfnamefont{E.~L.}
  \bibnamefont{Wright}}, in \emph{\bibinfo{booktitle}{Proc. of SPIE}}
  (\bibinfo{year}{2017}), vol. \bibinfo{volume}{9995}, p.
  \bibinfo{pages}{99950C}.

\bibitem[{\citenamefont{Haniff et~al.}(1987)\citenamefont{Haniff, Mackay,
  Titterington, Baldwin, and Warner}}]{Haniff1987}
\bibinfo{author}{\bibfnamefont{C.~A.} \bibnamefont{Haniff}},
  \bibinfo{author}{\bibfnamefont{C.~D.} \bibnamefont{Mackay}},
  \bibinfo{author}{\bibfnamefont{D.~J.} \bibnamefont{Titterington}},
  \bibinfo{author}{\bibfnamefont{J.~E.} \bibnamefont{Baldwin}},
  \bibnamefont{and} \bibinfo{author}{\bibfnamefont{P.~J.}
  \bibnamefont{Warner}}, \bibinfo{journal}{Nature}
  \textbf{\bibinfo{volume}{328}}, \bibinfo{pages}{694} (\bibinfo{year}{1987}).

\bibitem[{\citenamefont{Tuthill et~al.}(2000)\citenamefont{Tuthill, Monnier,
  Danchi, and Hale}}]{Tuthill2000}
\bibinfo{author}{\bibfnamefont{P.~G.} \bibnamefont{Tuthill}},
  \bibinfo{author}{\bibfnamefont{J.~D.} \bibnamefont{Monnier}},
  \bibinfo{author}{\bibfnamefont{W.~C.} \bibnamefont{Danchi}},
  \bibnamefont{and} \bibinfo{author}{\bibfnamefont{D.~S.} \bibnamefont{Hale}},
  \bibinfo{journal}{PASP} \textbf{\bibinfo{volume}{112}}, \bibinfo{pages}{555}
  (\bibinfo{year}{2000}).

\bibitem[{\citenamefont{Reid and Honma}(2014)}]{ReidHonma2014}
\bibinfo{author}{\bibfnamefont{M.~J.} \bibnamefont{Reid}} \bibnamefont{and}
  \bibinfo{author}{\bibfnamefont{M.}~\bibnamefont{Honma}},
  \bibinfo{journal}{Ann. Rev. of Astronomy and Astrophysics}
  \textbf{\bibinfo{volume}{52}}, \bibinfo{pages}{339} (\bibinfo{year}{2014}).

\bibitem[{\citenamefont{Sokolovsky et~al.}(2019)\citenamefont{Sokolovsky,
  Petit, Fey, Titov, Jacobs, Gordon, and Bizouard}}]{Sokolovsky2019}
\bibinfo{author}{\bibfnamefont{K.~V.} \bibnamefont{Sokolovsky}},
  \bibinfo{author}{\bibfnamefont{V.}~\bibnamefont{Petit}},
  \bibinfo{author}{\bibfnamefont{A.~L.} \bibnamefont{Fey}},
  \bibinfo{author}{\bibfnamefont{O.}~\bibnamefont{Titov}},
  \bibinfo{author}{\bibfnamefont{C.~S.} \bibnamefont{Jacobs}},
  \bibinfo{author}{\bibfnamefont{D.}~\bibnamefont{Gordon}}, \bibnamefont{and}
  \bibinfo{author}{\bibfnamefont{C.}~\bibnamefont{Bizouard}},
  \bibinfo{journal}{MNRAS} \textbf{\bibinfo{volume}{482}}, \bibinfo{pages}{520}
  (\bibinfo{year}{2019}).

\bibitem[{\citenamefont{Stevens}(2005)}]{Stevens2005}
\bibinfo{author}{\bibfnamefont{I.~R.} \bibnamefont{Stevens}},
  \bibinfo{journal}{MNRAS} \textbf{\bibinfo{volume}{356}},
  \bibinfo{pages}{1053} (\bibinfo{year}{2005}).

\bibitem[{\citenamefont{Lazio and Farrell}(2007)}]{Lazio2007}
\bibinfo{author}{\bibfnamefont{T.~J.~W.} \bibnamefont{Lazio}} \bibnamefont{and}
  \bibinfo{author}{\bibfnamefont{W.~M.} \bibnamefont{Farrell}},
  \bibinfo{journal}{ApJ} \textbf{\bibinfo{volume}{668}}, \bibinfo{pages}{1182}
  (\bibinfo{year}{2007}).

\bibitem[{\citenamefont{Ridgway et~al.}(1978)\citenamefont{Ridgway, Jacoby,
  Joyce, Reid, and Roellig}}]{Ridgway1978}
\bibinfo{author}{\bibfnamefont{S.~T.} \bibnamefont{Ridgway}},
  \bibinfo{author}{\bibfnamefont{G.~H.} \bibnamefont{Jacoby}},
  \bibinfo{author}{\bibfnamefont{R.~R.} \bibnamefont{Joyce}},
  \bibinfo{author}{\bibfnamefont{I.~N.} \bibnamefont{Reid}}, \bibnamefont{and}
  \bibinfo{author}{\bibfnamefont{T.~L.} \bibnamefont{Roellig}},
  \bibinfo{journal}{AJ} \textbf{\bibinfo{volume}{83}}, \bibinfo{pages}{252}
  (\bibinfo{year}{1978}).

\bibitem[{\citenamefont{Richichi}(2003)}]{Richichi2003}
\bibinfo{author}{\bibfnamefont{A.}~\bibnamefont{Richichi}},
  \bibinfo{journal}{New Astron. Rev.} \textbf{\bibinfo{volume}{47}},
  \bibinfo{pages}{523} (\bibinfo{year}{2003}).

\bibitem[{\citenamefont{Fujii et~al.}(2017)\citenamefont{Fujii, Cowan,
  Livengood, Kawahara, Howard, Luger, Lustig-Yaeger, Matthews, Oklopčič,
  Schwieterman et~al.}}]{Fujii2017}
\bibinfo{author}{\bibfnamefont{Y.}~\bibnamefont{Fujii}},
  \bibinfo{author}{\bibfnamefont{N.~B.} \bibnamefont{Cowan}},
  \bibinfo{author}{\bibfnamefont{T.~A.} \bibnamefont{Livengood}},
  \bibinfo{author}{\bibfnamefont{H.}~\bibnamefont{Kawahara}},
  \bibinfo{author}{\bibfnamefont{A.~W.} \bibnamefont{Howard}},
  \bibinfo{author}{\bibfnamefont{R.}~\bibnamefont{Luger}},
  \bibinfo{author}{\bibfnamefont{J.}~\bibnamefont{Lustig-Yaeger}},
  \bibinfo{author}{\bibfnamefont{J.~M.} \bibnamefont{Matthews}},
  \bibinfo{author}{\bibfnamefont{A.}~\bibnamefont{Oklopčič}},
  \bibinfo{author}{\bibfnamefont{E.~W.} \bibnamefont{Schwieterman}},
  \bibnamefont{et~al.}, \bibinfo{journal}{ApJ} \textbf{\bibinfo{volume}{850}},
  \bibinfo{pages}{10} (\bibinfo{year}{2017}).

\bibitem[{\citenamefont{Zugger et~al.}(2010)\citenamefont{Zugger, Kasting,
  Williams, Kane, and Philbrick}}]{Zugger2010}
\bibinfo{author}{\bibfnamefont{M.~E.} \bibnamefont{Zugger}},
  \bibinfo{author}{\bibfnamefont{J.~F.} \bibnamefont{Kasting}},
  \bibinfo{author}{\bibfnamefont{D.~M.} \bibnamefont{Williams}},
  \bibinfo{author}{\bibfnamefont{T.~J.} \bibnamefont{Kane}}, \bibnamefont{and}
  \bibinfo{author}{\bibfnamefont{C.~R.} \bibnamefont{Philbrick}},
  \bibinfo{journal}{ApJ} \textbf{\bibinfo{volume}{723}}, \bibinfo{pages}{1168}
  (\bibinfo{year}{2010}).

\bibitem[{\citenamefont{Cowan and Fujii}(2013)}]{CowanFujii2013}
\bibinfo{author}{\bibfnamefont{N.~B.} \bibnamefont{Cowan}} \bibnamefont{and}
  \bibinfo{author}{\bibfnamefont{Y.}~\bibnamefont{Fujii}},
  \bibinfo{journal}{ApJ} \textbf{\bibinfo{volume}{776}}, \bibinfo{pages}{85}
  (\bibinfo{year}{2013}).

\bibitem[{\citenamefont{Trauger and
  Traub}(2007{\natexlab{b}})}]{TraugerTraub2007Nature}
\bibinfo{author}{\bibfnamefont{J.~T.} \bibnamefont{Trauger}} \bibnamefont{and}
  \bibinfo{author}{\bibfnamefont{W.~A.} \bibnamefont{Traub}},
  \bibinfo{journal}{Nature} \textbf{\bibinfo{volume}{446}},
  \bibinfo{pages}{771} (\bibinfo{year}{2007}{\natexlab{b}}).

\bibitem[{\citenamefont{Eshleman}(1979)}]{Eshleman1979}
\bibinfo{author}{\bibfnamefont{V.~R.} \bibnamefont{Eshleman}},
  \bibinfo{journal}{Science} \textbf{\bibinfo{volume}{205}},
  \bibinfo{pages}{1137} (\bibinfo{year}{1979}).

\bibitem[{\citenamefont{Turyshev and Toth}(2017)}]{TuryshevToth2017}
\bibinfo{author}{\bibfnamefont{S.~G.} \bibnamefont{Turyshev}} \bibnamefont{and}
  \bibinfo{author}{\bibfnamefont{V.~T.} \bibnamefont{Toth}},
  \bibinfo{journal}{Phys. Rev. D} \textbf{\bibinfo{volume}{96}},
  \bibinfo{pages}{024008} (\bibinfo{year}{2017}).

\bibitem[{\citenamefont{{Turyshev} and
  {Toth}}(2020)}]{Turyshev-Toth:2020-extend}
\bibinfo{author}{\bibfnamefont{S.~G.} \bibnamefont{{Turyshev}}}
  \bibnamefont{and} \bibinfo{author}{\bibfnamefont{V.~T.}
  \bibnamefont{{Toth}}}, \bibinfo{journal}{Phys. Rev. D}
  \textbf{\bibinfo{volume}{102}}, \bibinfo{eid}{024038} (\bibinfo{year}{2020}).

\bibitem[{\citenamefont{{Turyshev} and
  {Toth}}(2022{\natexlab{a}})}]{Turyshev-Toth:2022_MNRAS}
\bibinfo{author}{\bibfnamefont{S.~G.} \bibnamefont{{Turyshev}}}
  \bibnamefont{and} \bibinfo{author}{\bibfnamefont{V.~T.}
  \bibnamefont{{Toth}}}, \bibinfo{journal}{MNRAS}
  \textbf{\bibinfo{volume}{515}}, \bibinfo{pages}{6122}
  (\bibinfo{year}{2022}{\natexlab{a}}).

\bibitem[{\citenamefont{{Turyshev} and
  {Toth}}(2022{\natexlab{b}})}]{Turyshev-Toth:2022_PRD}
\bibinfo{author}{\bibfnamefont{S.~G.} \bibnamefont{{Turyshev}}}
  \bibnamefont{and} \bibinfo{author}{\bibfnamefont{V.~T.}
  \bibnamefont{{Toth}}}, \bibinfo{journal}{Phys. Rev. D}
  \textbf{\bibinfo{volume}{106}}, \bibinfo{eid}{044059}
  (\bibinfo{year}{2022}{\natexlab{b}}).

\bibitem[{\citenamefont{Toth and Turyshev}(2023)}]{TothTuryshev2023}
\bibinfo{author}{\bibfnamefont{V.~T.} \bibnamefont{Toth}} \bibnamefont{and}
  \bibinfo{author}{\bibfnamefont{S.~G.} \bibnamefont{Turyshev}},
  \bibinfo{journal}{MNRAS} \textbf{\bibinfo{volume}{525}},
  \bibinfo{pages}{5846} (\bibinfo{year}{2023}).

\bibitem[{\citenamefont{{Turyshev} et~al.}(2020)\citenamefont{{Turyshev},
  {Shao}, {Toth}, {Friedman}, {Alkalai}, {Mawet}, {Shen}, {Swain}, {Zhou},
  {Helvajian} et~al.}}]{9-Turyshev-etal:2020-PhaseII}
\bibinfo{author}{\bibfnamefont{S.~G.} \bibnamefont{{Turyshev}}},
  \bibinfo{author}{\bibfnamefont{M.}~\bibnamefont{{Shao}}},
  \bibinfo{author}{\bibfnamefont{V.~T.} \bibnamefont{{Toth}}},
  \bibinfo{author}{\bibfnamefont{L.~D.} \bibnamefont{{Friedman}}},
  \bibinfo{author}{\bibfnamefont{L.}~\bibnamefont{{Alkalai}}},
  \bibinfo{author}{\bibfnamefont{D.}~\bibnamefont{{Mawet}}},
  \bibinfo{author}{\bibfnamefont{J.}~\bibnamefont{{Shen}}},
  \bibinfo{author}{\bibfnamefont{M.~R.} \bibnamefont{{Swain}}},
  \bibinfo{author}{\bibfnamefont{H.}~\bibnamefont{{Zhou}}},
  \bibinfo{author}{\bibfnamefont{H.}~\bibnamefont{{Helvajian}}},
  \bibnamefont{et~al.}, \emph{\bibinfo{title}{{\it Direct Multipixel Imaging
  and Spectroscopy of an Exoplanet with a Solar Gravity Lens Mission}, {The
  Final Report NIAC Phase II proposal}}} (\bibinfo{year}{2020}),
  \bibinfo{note}{arXiv:2002.11871 [gr-qc]}.

\bibitem[{\citenamefont{Helvajian et~al.}(2023)\citenamefont{Helvajian,
  Rosenthal, Poklemba, Battista, DiPrinzio, Neff, McVey, Toth, and
  Turyshev}}]{Helvajian2023}
\bibinfo{author}{\bibfnamefont{H.}~\bibnamefont{Helvajian}},
  \bibinfo{author}{\bibfnamefont{A.}~\bibnamefont{Rosenthal}},
  \bibinfo{author}{\bibfnamefont{J.}~\bibnamefont{Poklemba}},
  \bibinfo{author}{\bibfnamefont{T.~A.} \bibnamefont{Battista}},
  \bibinfo{author}{\bibfnamefont{M.~D.} \bibnamefont{DiPrinzio}},
  \bibinfo{author}{\bibfnamefont{J.~M.} \bibnamefont{Neff}},
  \bibinfo{author}{\bibfnamefont{J.~P.} \bibnamefont{McVey}},
  \bibinfo{author}{\bibfnamefont{V.~T.} \bibnamefont{Toth}}, \bibnamefont{and}
  \bibinfo{author}{\bibfnamefont{S.~G.} \bibnamefont{Turyshev}},
  \bibinfo{journal}{JSR} \textbf{\bibinfo{volume}{60}}, \bibinfo{pages}{829}
  (\bibinfo{year}{2023}).

\bibitem[{\citenamefont{{Turyshev} et~al.}(2023)\citenamefont{{Turyshev},
  {Garber}, {Friedman}, {Hein}, {Barnes}, {Batygin}, {Brown}, {Cronin},
  {Davoyan}, {Dubill} et~al.}}]{Turyshev:2023-Sundivers}
\bibinfo{author}{\bibfnamefont{S.~G.} \bibnamefont{{Turyshev}}},
  \bibinfo{author}{\bibfnamefont{D.}~\bibnamefont{{Garber}}},
  \bibinfo{author}{\bibfnamefont{L.~D.} \bibnamefont{{Friedman}}},
  \bibinfo{author}{\bibfnamefont{A.~M.} \bibnamefont{{Hein}}},
  \bibinfo{author}{\bibfnamefont{N.}~\bibnamefont{{Barnes}}},
  \bibinfo{author}{\bibfnamefont{K.}~\bibnamefont{{Batygin}}},
  \bibinfo{author}{\bibfnamefont{M.~E.} \bibnamefont{{Brown}}},
  \bibinfo{author}{\bibfnamefont{L.}~\bibnamefont{{Cronin}}},
  \bibinfo{author}{\bibfnamefont{A.~R.} \bibnamefont{{Davoyan}}},
  \bibinfo{author}{\bibfnamefont{A.}~\bibnamefont{{Dubill}}},
  \bibnamefont{et~al.}, \bibinfo{journal}{Planetary \& Space Science}
  \textbf{\bibinfo{volume}{235}}, \bibinfo{eid}{105744} (\bibinfo{year}{2023}),
  \bibinfo{note}{arXiv:2303.14917 [astro-ph.EP]}.

\end{thebibliography}

\end{document}